\documentclass[preprint2]{aastex}

\def\sun{\hbox{$\odot$}}

\slugcomment{Accepted by the Astronomical Journal.}

\shorttitle{Spitzer Imaging of Interacting Galaxies}
\shortauthors{Smith et al.}

\begin{document}


\title{The Spitzer Spirals, Bridges, and Tails Interacting Galaxy Survey: 
Interaction-Induced Star Formation in the Mid-Infrared}


\author{Beverly J. Smith}
\affil{Department of Physics, Astronomy, and Geology, East Tennessee
State University, Johnson City TN  37614}
\email{smithbj@etsu.edu}

\author{Curtis Struck}
\affil{Department of Physics and Astronomy, Iowa State University, Ames IA  50011}
\email{curt@iastate.edu}

\author{Mark Hancock}
\affil{Department of Physics, Astronomy, and Geology, East Tennessee
State University, Johnson City TN  37614}
\email{hancockm@etsu.edu}

\author{Philip N. Appleton}
\affil{Spitzer Science Center,
California Institute of Technology, Pasadena CA  91125}
\email{apple@ipac.caltech.edu}
                                                                                                         
\author{Vassilis Charmandaris\footnote{Chercheur Associ\'e, Observatoire
de Paris, F-75014, Paris, France, and IESL/Foundation for Research
and Technology-Hellas, PO Box 1527, 71110, Heraklion, Greece}}
\affil{Department of Physics, University of Crete, GR-71003, Heraklion, Greece}
\email{vassilis@physics.uoc.gr}

\and
                                                                                                         
\author{William T. Reach}
\affil{Spitzer Science Center,
California Institute of Technology, Pasadena CA  91125}
\email{reach@ipac.caltech.edu}



\begin{abstract}

We present Spitzer mid-infrared imaging of a sample of 35 
tidally-distorted
pre-merger interacting galaxy pairs selected from the Arp Atlas. We compare
their global mid-infrared properties with those of normal
galaxies from the SINGS Spitzer Legacy survey,
and separate the disk emission from that of the tidal
features.  
The [8.0 $\mu$m] $-$ [24 $\mu$m], [3.6 $\mu$m] $-$ [24 $\mu$m], 
and 
[5.8 $\mu$m] $-$ [8.0 $\mu$m]
colors
of these optically-selected interacting galaxies
are redder on average than those of spirals,
implying enhancements to the mass-normalized star formation rates (SFRs)
of a factor of $\sim$2.
Furthermore, the 24 $\mu$m emission in the Arp galaxies is more 
centrally concentrated than that in the spirals, suggesting that gas is
being concentrated into the inner regions and fueling central star formation.
No significant differences can be discerned in
the shorter wavelength 
Spitzer colors of the Arp galaxies compared to the spirals, thus these
quantities are less sensitive to star formation enhancements.
No strong
trend of Spitzer 
color with pair separation is visible in our sample; this may be because
our sample was selected to be tidally disturbed.
The tidal features contribute $\le$10$\%$ of the total Spitzer fluxes
on average.  
The SFRs implied for the Arp galaxies by the Spitzer 24 $\mu$m luminosities are relatively
modest, $\sim$1 M$_{\sun}$~yr$^{-1}$ on average.

\end{abstract}




\section{Introduction}

Since the early work of \citet{larson78} and \citet{struckmarcell78},
there has been a great deal of interest in how star
formation in galaxies is affected by collisions with other
galaxies.  Infrared Astronomical Satellite (IRAS)
observations led to the discovery of galaxies with very high
far-infrared luminosities \citep{soifer87, smith87} that are the
result of mergers between equal-mass gas-rich progenitors
\citep{sanders88}.  Many studies since that time have shown that
spectacular and prolonged enhancement of star formation is the cause
of much of the infrared emission in major mergers (see reviews by
\citealp{sanders96} and \citealp{struck99}).

The question of how star formation is affected in interacting galaxies
before a merger occurs is subtle and more difficult to answer.  
Since dynamical models of interactions suggest that the pre-merger
stage is relatively long-lived, understanding the properties
of nearby interacting galaxies and their differences from field
spirals is important in interpreting the results of high redshift surveys
(e.g., \citealp{lefevre00, bundy04}).
Early studies
indicated that on average the H$\alpha$ equivalent widths and
far-infrared-to-blue luminosity ratios in both close pairs and in
morphologically-peculiar \citet{arp66} Atlas interacting galaxies are
enhanced by about a factor of two relative to those of spiral galaxies,
with more disturbed systems tending to have higher values
\citep{bushouse87, bushouse88, kennicutt87}.  There is, however, a
large scatter in these values, with many interacting galaxies having
SFRs
typical of normal spirals.  Star formation
enhancement appears to be a function of pair separation, with pairs
closer than $\sim$30 kpc having higher H$\alpha$ equivalent widths on
average than wider pairs
\citep{barton00, lambas03, nikolic04}.  
However,
there is much scatter in this
relationship, with many close pairs having low H$\alpha$
equivalent widths. 
In many
cases, star formation is enhanced in the nuclear regions compared
to the disks \citep{hummel90, nikolic04}.  Although luminous star formation
regions have been found in tidal tails and bridges (e.g.,
\citealp{schweizer78, mirabel91, mirabel92}), the optical broadband
colors of tidal features are often similar to those of outer disks,
but with a larger scatter \citep{schombert90}.

To date, most studies comparing interacting galaxies to normal
galaxies have been based on optical data, which suffer from
extinction, or IRAS far-infrared data. The limited spatial resolution
of IRAS, however, provided only a global measurement of the far-infrared
luminosities of all but the most nearby galaxies; furthermore,
dust heating by the older stellar population may also contribute
to the far-infrared luminosities (e.g., \citealp{persson87,
smith91, smith94}).
An alternative tracer of star formation
is the mid-infrared, given the close coupling between HII regions and
the surrounding photodissociation regions, which emit strongly
in the mid-infrared (see \citealp{hollenbach97}). 
Observations with 
the 
Infrared Space Observatory (ISO) 
revealed 
direct correlations between the H$\alpha$ luminosity, the mid-infrared
dust features, and the mid-infrared continuum fluxes
in galaxies
\citep{roussel01, helou01,
forster04}.

With the advent of the Spitzer Space Telescope \citep{werner04},
mid-infrared imaging of galaxies
entered a new era. The large format mid-infrared detector arrays of
Spitzer, nearly two orders of magnitude more sensitive than those 
of ISO, make feasible the detailed study of individual star formation
complexes in nearby galaxies (e.g.,
\citealp{calzetti05, smith05,
elmegreen06, hancock06}).
Spitzer is well-suited for studying
induced star formation in spiral arms and tidal bridges and tails, and
detecting knots of star formation (see \citealp{smith05}).

In the current paper, we 
present results from a Spitzer mid-infrared survey
of a well-defined sample of nearby interacting galaxy pairs.
We discuss the global infrared properties of the
galaxies in our sample, and compare them with those of normal spiral,
elliptical/S0, and irregular galaxies.
We also compare the infrared colors
of the main disks with those of the tidal features in these galaxies.
Detailed studies of clumps of star formation in 
two of the galaxies in our sample, Arp 107 and Arp 82,
are presented in \citet{smith05} and \citet{hancock06}.

\section{The Samples}


\subsection{The Interacting Galaxy Sample } 

Our interacting galaxy sample was selected from the Arp Atlas of
Peculiar Galaxies \citep{arp66}, based on the following criteria: 1)
They are relatively isolated binary systems; we eliminated merger
remnants and close triples and multiple systems in which the galaxies
have similar optical brightnesses (systems with additional smaller
angular size companions were not excluded).  2) They are tidally
disturbed.  3) They have radial velocities less than 11,000 km/s
(150 Mpc, for H$_{\rm o}$ = 75 km~s$^{-1}$~Mpc$^{-1}$).  4) Their
total angular size is $>$ 3$'$
and the angular sizes of the individual galaxies are
$\ge$ 30$''$, 
to allow for good spatial resolution
with Spitzer.  
A
total of 35 Arp systems fit these criteria.  One of these systems, Arp
297, consists of two pairs at different redshifts.  These two pairs
are included separately in our sample.  We also include the
interacting pair NGC 4567, which fits the above criteria but is not in
the Arp Atlas.  This brings the sample to 37. 

Of these 37 systems, 28 were included in our `Spirals, Bridges, and
Tails' (SB\&T) Guest Observer Cycle 1 Spitzer program.
The
remaining nine galaxies were reserved as part of various Guaranteed Time
or Guest Observer programs.  
For completeness, in our survey we included the seven of 
these additional galaxies for which the data were publicly-available.
This brings our final sample size to 35 pairs.

The 35 galaxy pairs in our sample are listed in Table 1, in order of
their number in the Arp Atlas.
Table 1 also includes alternative names for the galaxies,
notes on the optical morphologies of the systems based on the
\citet{arp66} photographs, as well as the nuclear optical spectral
type when available from \citet{keel85} or \citet{dahari85}.  The
interacting sample contains three known Seyferts and 13 Low Ionization
Emission Line (LINER) galaxies (4$\%$ and 19$\%$, respectively).  The
distances given in Table 1 were calculated using
velocities from the NASA Extragalactic Database (NED), a Hubble
constant of 75 km~s$^{-1}$~Mpc$^{-1}$, and the \citet{schechter80}
Virgocentric infall model with parameters as in \citet{heckman98}.  A
histogram of the distances is shown in the top panel of Figure 1.
The median distance is 47 Mpc and the maximum distance is 143 Mpc.

Table 1 also includes far-infrared luminosities (40 $-$ 120 $\mu$m), 
when available,
calculated\footnote{L$_{FIR}$(L$_{\sun}$) 
= 3.94 $\times$ 10$^{5}$D$^2$(2.58F$_{60}$ + F$_{100}$), where D
is the distance in Mpc and F$_{60}$
and L$_{100}$ are the IRAS 60 and 100 $\mu$m flux densities in Jy.}
as in
\citet{lonsdale85} and \citet{persson87} using IRAS fluxes obtained from 
\citet{sanders03}, the IRAS
Faint Source Catalog \citep{moshir92}, or the {\it scanpi} 
software\footnote{$http:$//$irsa.ipac.caltech.edu/IRASdocs/scanpi\_over.html$}.
A histogram of these luminosities is shown in the top panel of Figure 2.

\subsection{The `Normal' Galaxy Sample}

As a `control' sample of nearby `normal' galaxies, we started with
the 75 galaxies in the Spitzer Infrared Nearby Galaxies Survey (SINGS)
\citep{kennicutt03, dale05}.  The SINGS sample was selected to cover a
wide range in parameter space, with a wide range in Hubble type and
luminosity.  Most objects have angular sizes between 5$'$ and 15$'$.  

We used the NASA Extragalactic Database 
(NED\footnote{The NASA/IPAC Extragalactic Database (NED) (http$://$nedwww.ipac.caltech.edu) is operated by the 
Jet Propulsion Laboratory, California Institute of Technology, under contract 
with the National Aeronautics and Space Administration.}) to search for companions
to the SINGS galaxies.  
We eliminated from our `normal' sample galaxies
with companions 
whose velocities differ from that of the target galaxy by 
$\le$1000 km~s$^{-1}$,
have an optical luminosity brighter than 1/10th of that
of the target galaxy, and 
are separated from the target galaxy by less than 10 times
the optical diameter of the target galaxy or the companion, whichever 
is larger.
We emphasize that our `normal' galaxies are not necessarily completely
isolated; some have distant or low mass companions.  However, they are
not subject to the strong perturbing forces of the Arp sample.
For comparison, the Large Magellanic Cloud has an optical
luminosity $\sim$1/9th that of the Milky Way \citep{vandenbergh00},
thus our `normal' galaxies are less perturbed than the Milky Way.
We note that three of the galaxies eliminated from
the `normal' sample by
this method had previously been included in the interacting sample:
NGC 2798
(Arp 283), NGC 5194, and NGC 5195 (Arp 85 = M51).
A total of 42 SINGS galaxies remain in our `normal' galaxy
sample; of these,
26 are spirals, 4 are ellipticals or S0s,
and 12 are irregulars/Sm.  These galaxies are listed in Table 2.
These three groups are treated separately in our analysis below.  

Table 2 also includes the distances, Hubble types,
and far-infrared luminosities for the sample galaxies.
As shown
in Figure 1, these `normal' galaxies
are more nearby than our interacting sample,
with median and maximum distances of 9.6 Mpc and 25 Mpc, respectively.

The mean and median far-infrared luminosities L$_{FIR}$
for the various samples are given in Table 3.
Table 3 also includes the mean, median, and rms
of the
total 8 $-$ 1000 $\mu$m infrared luminosity
L$_{IR}$, calculated 
using the relationship in \citet{perault87},
\citet{sanders96},
and 
\citet{kennicutt98}.
A histogram of the far-infrared luminosities is given in Figure 2.
After dividing the Arp luminosities by a factor of two, to account
for the two disks in the pair, a Kolmogorov-Smirnov (KS) test cannot rule
out that the spiral and Arp luminosities come from the same parent
distribution.
Although the distributions are similar,
the median far-infrared luminosity
for the Arp galaxies, per disk, is slightly larger than 
that of the spirals, a factor of $\sim$2,
consistent with the results of \citet{bushouse88}
for a similar optically-selected sample.

\section{Observations and Data Reduction}

The galaxies in our SB\&T sample were observed between
November 2004 and August 2005
in the 3.6, 4.5, 5.8 and 8.0 $\mu$m
broadband filters of the Spitzer Infrared Array Camera (IRAC; 
\citealp{fazio04})
and the 24 $\mu$m band of the Multiband Imaging Photometer
for Spitzer (MIPS; \citealp{rieke04}).
For the IRAC observations,
we used 12 sec frames in a cycling
dither pattern, with between 4 and 23 positions per system.
Most galaxies had a small enough angular size to
fit within the
5\arcmin$\times$5\arcmin~ field of view of IRAC, or could be
imaged with a small map.
The exceptions were NGC 4567, Arp 82, Arp 89, and
Arp 279, for which two pointings were necessary.
For the MIPS observations,
we used fixed single observations with two cycles
of 10 sec integrations per frame, resulting in
an on-source time of 312.5 sec per target.  
The details of the
SB\&T
observations are presented in Table 4.

We started our analysis of the 3.6 $\mu$m and 4.5 $\mu$m interacting
galaxy data
with the version S13.2.0 
post-pipeline Basic Calibrated Data
(post-BCD) mosaicked images.
At 5.8 $\mu$m, 8.0 $\mu$m, and 24 $\mu$m,
the pipeline post-BCD 
mosaicked images show significant artifacts in the sky background,
in particular, `boxiness' due to imperfect matching of sky levels
between images during mosaicking.  Therefore, for these wavelengths
we started with the S13.2.0 BCD images and mosaicked them ourselves using
the MOPEX\footnote{$http$$:$$//$$ssc.spitzer.caltech.edu/postbcd/download\-mopex.html$}
software.  We first ran an overlap correction
using a drizzle interpolation method with a drizzle factor of 1,
and then ran the mosaicking module.
In addition,
at 5.8 $\mu$m and 8.0 $\mu$m, 
bright point sources in some of the images
(often the nuclei of the galaxies) cause horizonal `bands' in the
images (`muxbleed').  These were removed by interpolation
from nearby clean areas.
Furthermore, a strong `sky' gradient was present in some of the mosaicked 
images.  To remove this gradient, we used the 
Image Reduction and Analysis Facility (IRAF\footnote{IRAF is distributed by the 
National Optical Astronomy Observatories,
which are operated by the Association 
of Universities for Research
in Astronomy, Inc., under cooperative agreement with the National
Science Foundation.})
routine {\it imsurfit} to 
fit and subtract 
the background.
For the SINGS galaxies, we started with the SINGS Data Release 4 mosaicked
images\footnote{See
$http:$//$ssc.spitzer.caltech.edu/legacy$}.  When necessary, we
subtracted a gradient from the SINGS images.

The final 3.6 $\mu$m, 8.0 $\mu$m, and 24 $\mu$m images for the
interacting galaxy sample are shown in Figures 3 $-$ 11, ordered by
their Arp number.  From these and the SINGS
images, at each wavelength we determined 
flux densities for each disk and tidal feature.
We used multiple
rectangular boxes that cover the entire observed extent of these
features in the 3.6 $\mu$m images, 
avoiding bright foreground
stars and their residuals.
These same regions were then used for the other wavelength images.
Multiple blank sky fields away from the galaxies 
were
measured for sky subtraction.  We avoided the edges of the images
in selecting sky regions, due to edge effects in the final images.

The final flux densities for the Arp galaxies
are tabulated in Table 5, while those for the SINGS galaxies are given
in Table 6.  The quoted upper limits are 3$\sigma$.
The uncertainties given in Tables 5 and 6 and used
to calculate upper limits are a combination of 
statistical uncertainties and uncertainties in the sky subtraction
due to inhomogeneities
and residual sky gradients
in the final mosaicked images.
For the sky regions in an image, we calculated the mean rms
noise $<$$rms$$>$ and the rms of the mean 
sky
$rms_{<sky>}$,
where 
$rms_{<sky>}$ is the standard deviation
of the means of the different sky
regions in an image.
The quantity 
$rms_{<sky>}$ is a measure of the `flatness' of the final image,
i.e., the artifacts in the background.  The uncertainties, $\sigma$, quoted
in Tables 5 and 6 and used in the 
subsequent figures were calculated using $\sigma^2 = 
(<$$rms$$>*\sqrt{N})^2 + (rms_{<sky>}*N)^2$, where $N$ is the number of
pixels in the galaxy apertures in question.  The first term in this
equation is the standard
statistical uncertainty, while the second term is a measure of the
uncertainty due to variations in the sky level across the image.
The latter term typically dominates by a factor of $\sim$9.
The median uncertainty in Table 5 is $\sim$3$\%$, with
80$\%$ of the sources having uncertainties $\le$10$\%$.
The sky artifacts are worse at 5.8 $\mu$m and 24 $\mu$m than at
the other wavelengths, thus these two bands have the most upper limits quoted in
Tables 5 and 6.
Note that our upper limits are quite conservative; in some cases,
galaxies are visible in the image but sky problems prevent an accurate
flux from being determined.

The uncertainties given in Tables 5 and 6
do not include absolute calibration uncertainties, which are $\le$10$\%$ 
(IRAC Data Manual; MIPS Data Manual).  
Additional
uncertainties arise from ambiguities in selecting the boundaries of the
galaxy and sky regions, possible
contributions from foreground stars inadvertently included in the
galaxy region, and the mosaicking process.  
Our fluxes for the SINGS galaxies typically agree
within 10 $-$ 20$\%$ with those determined independently by
\citet{dale05}.  This provides an upper limit to these additional
uncertainties.

In the 24 $\mu$m images of some of the galaxies with very bright
galactic nuclei, the MIPS point spread function (see MIPS Data
Handbook) is visible as a ring around the nucleus, and six radial
spikes extending out from the nucleus (see Arp 84, 87,
104, 181, 242, 283, 
284, 295, 297,
and 298 in Figures 4 $-$ 10).  In these systems, accurate
measurement of 24 $\mu$m tidal flux densities are not possible.

\section{Global Spitzer Luminosities }

In Figures 12, 13, and 14, we present histograms of the 3.6 $\mu$m,
8.0 $\mu$m, and 24 $\mu$m luminosities, respectively,
for the various classes of galaxies.  In Table 7, we give the median,
mean, and rms 3.6 $\mu$m, 8.0 $\mu$m, and 24 $\mu$m luminosity for each class.
As with the far-infrared luminosities,
at all three Spitzer wavelengths
KS tests cannot rule out that the Arp and spiral luminosities originated
from the same parent distribution.

The 3.6 $\mu$m luminosity provides an 
approximate measure of the stellar mass
in the galaxy.
At this wavelength,
the distribution for the Arp disks extends
to lower luminosities than that of spirals,
and has a slightly lower mean luminosity than the spiral
galaxies.
This is likely because of the inclusion
of low mass companions, which in some cases may have
originally been irregular or dwarf elliptical galaxies.
At 8 $\mu$m and 24 $\mu$m, 
the Arp distributions are similar to those of spirals, however,
as at 3.6 $\mu$m,
there is a tail of lower luminosity,
low mass galaxies, compared to spirals.  There is also 
a slight excess of higher luminosity Arp disks at  
24 $\mu$m.  The median
24 $\mu$m luminosity for the Arp disks
is 
slightly higher ($\sim$1.8$\times$) that of the spirals.

As expected, the luminosities of the irregular galaxies are typically
less than those of the spirals.
The luminosities of the tails and bridges also
tend to be
lower than those of the Arp disks on 
average, as expected.  
The tidal features
tend to be more luminous than the irregular galaxies, 
overlapping in luminosities
with the spirals to some extent.  
This
is probably due in part to contributions from the underlying
disks, and in part due to a selection effect,
as these systems were selected 
for having prominent tidal features.
On average, the 3.6 $\mu$m luminosities (and therefore
the stellar masses) of the tidal features
are $\sim$7$\%$ that of the Arp disks (see Table 7).
The fractional fluxes in the tidal features are similar
at the other Spitzer wavelengths.
We note that
there is some uncertainty in defining `tidal' vs.\ `disk' regions;
for example, in some systems, large
star formation regions
lie at the base of a tidal feature.
Generally 
mid-infrared-bright
extra-nuclear star formation regions
in the vicinity of the main galaxies
are considered `disk' rather than `tidal', but this
separation is rather subjective.

\section{Global Spitzer Colors}

In Figure 15, we plot the Spitzer IRAC [3.6] $-$ [4.5] color against
the [4.5] $-$ [5.8] color for the main disks (black circles) and tidal
features (filled blue triangles) of the galaxies in the interacting
sample.  In Figure 16, we present the same plot for the SINGS galaxies
(spirals: black circles; irregular/Sm: open blue circles; E/S0: open
magenta diamonds).  In Figures 17 and 18, the [4.5] $-$ [5.8] vs.\
[5.8] $-$ [8.0] colors for the Arp and SINGS samples, respectively,
are plotted.  Figures 19 and 20 show the [5.8] $-$ [8.0] vs.\ [8.0]
$-$ [24] colors for the Arp and SINGS galaxies, respectively.  No
color corrections were included in making these plots.  The colors of
M0III stars (open magenta square; M. Cohen 2005, private
communication), field stars (magenta triangle; \citealp{whitney04}),
and predicted IRAC colors of interstellar dust
\citep{li01} are also included in these plots. 
The \citet{hatz05} colors of quasars are also plotted (red circles).
In calculating magnitudes at 3.6 $\mu$m,
4.5 $\mu$m, 5.8 $\mu$m, 8.0 $\mu$m, and 24 $\mu$m, 
we used zero points of 280.9 Jy, 179.7 Jy,
115.0 Jy, 64.1 Jy, and 7.14 Jy, respectively (IRAC Data Manual;
MIPS Data Manual).

Histograms showing the distribution of Spitzer colors for the different
samples are shown in Figures 21 $-$ 26.  
In addition to histograms for all of the Arp galaxies,
we plot separately the colors for the
eight M51-like systems identified in Table 1.  
Our results are
quantified in Table 8, where we provide the mean, median, and rms
spread in the Spitzer colors for the different samples.  
For comparison with Table 8, the mean [3.6] $-$ [4.5], [4.5] $-$ [5.8],
[5.8] $-$ [8.0], and [3.6] $-$ [8.0] colors in the \citet{whitney04}
field stars are $-$0.05, 0.1, 0.05, and 0.1, respectively,
and the predicted values
for interstellar dust are $-$0.35, 3.2, 2.1, and 4.95,
respectively \citep{li01}.
The expected [8.0] $-$ [24] color for dust varies from 2.6 $-$ 4.3 \citep{li01},
increasing with increasing interstellar radiation field intensity,
while
stars are expected to be at $\sim$0.0.
In Figures 21 $-$ 26, the expected colors for stars and dust
are indicated by vertical lines.

Results of KS tests for whether
these samples were drawn from the same parent sample are 
given in Table 9.
We ran up to three different KS tests per comparison.  First, we
did a KS test ignoring the upper/lower limits, if any.  Then, if
these results suggest that the samples are significantly
different, we ran another
test assuming the limits were detections.  Third, if the 
upper/lower limits were above/below the peak of the comparison sample,
to force the best possible match between the samples, we treated the
limits as detections near the peak of the comparison sample.

The [3.6] $-$ [4.5] colors for all of the samples tend to lie near
0.0, indicating that these bands are generally dominated by starlight.  
No significant differences can be discerned between the Arp galaxies
and the spirals in
this color. 
The [4.5] $-$ [5.8], [5.8] $-$ [8.0], and [3.6] $-$ [8.0] 
colors show a relatively wide range, 
between those expected for stellar photospheres
and those of interstellar dust (see Figures 22 $-$ 24).  
In the [4.5] $-$ [5.8] and [3.6] $-$ [8.0] colors,
the mean and median colors the spirals and Arp disks are
similar, and 
KS tests cannot rule out that the spirals and Arp 
galaxies were drawn from the same sample (see Tables 8 and 9).

In contrast to these shorter wavelength colors, there are
statistically significant differences between
the [8.0] $-$ [24], [3.6] $-$ [24], and [5.8] $-$ [8.0]
colors of the Arp galaxies 
and those of the spirals, with the Arp galaxies being redder (see
Figures 23, 25, and 26, and Tables 8 and 9).
The M51-like subset of the Arp galaxies also shows possible 
differences from the spirals in these colors, 
however, the M51 sample size is small so these
results are tentative.

In the histograms of the colors of the Arp galaxies (Figures 21 $-$ 26), 
we have distinguished
between the more and less massive galaxies in the pair.  No significant
difference is seen between these two groups.

In most cases, our sample sizes are too small and/or there
are too many upper limits to draw strong conclusions
about the colors of the irregular/Sm galaxies, the E/S0 galaxies, or
the
tidal features compared to
the spirals or Arp disks.
However, we do find a significant difference between the
[3.6] $-$ [8.0] colors of irregular/Sm galaxies compared
to the spirals, with the Irr/Sm galaxies being bluer.
Furthermore, the [3.6] $-$ [4.5] colors of the Irr/Sm galaxies
tend to be redder than those of the other samples.

\section{Separation vs. IR colors}

To search for correlations between star formation
rates and interaction stage, 
in Figures 27 and 28
we plotted pair separation
against the [3.6] $-$ [8.0] and 
[8.0] $-$ [24] colors.
To distinguish between equal mass galaxy pairs and unequal
mass pairs, using the 3.6 $\mu$m luminosity ratio as a proxy for mass
ratio we have separated the galaxies into three groups.  
The first group (filled black
triangles) represent pairs with approximately
equal mass (i.e., 3.6 $\mu$m luminosity ratios greater
or equal to 0.75).  The second group, with intermediate
mass ratios 
(3.6 $\mu$m luminosity ratios
between 0.1 $-$ 0.75) is plotted with red and blue open circles 
(more and less massive galaxy in the pair, respectively).
Pairs with 3.6
$\mu$m luminosity ratios less than 0.1 are marked as magenta and
green open diamonds, respectively, for the more and less massive
galaxy in the pair.  
These plots show large scatter, without a strong correlation.
Plots of the other colors (not shown) show similar scatter, for both
the disks and the tidal features.

As an alternative test for
whether separation correlates with Spitzer color, we have
binned the Arp galaxies into two groups: those with separations less
than 30 kpc, and wider pairs.  In Figures 29 $-$ 34, we display
histograms of the Spitzer colors for these two groups.  
Mean and median colors for these two groups are given in Table 8,
while the results of KS comparison tests 
are given in Table 9.
No significant difference between the wide and close pairs is seen.
Although the reddest disks tend to be closer pairs, there is a large
amount of scatter, and this result is not statistically significant.
The KS
test cannot rule out the possibility that the colors of 
the two samples come from
the same parent population.

\section{Central Concentration}

Unlike IRAS studies, which had limited spatial resolution,
Spitzer allows us to look for differences in the spatial distribution
of the infrared emission between our samples.
To investigate whether the Arp pairs have more centrally concentrated
star formation, we have evaluated the 24 $\mu$m nuclear emission in a region
of diameter 2~kpc centered on the position of the 2MASS K-band nucleus
of each galaxy. The aperture sizes used were calculated using the
assumed distances to each galaxy given in Tables 1 and 2.
We choose a diameter of 2~kpc so that the central part of
the point spread function
at 24 $\mu$m would subtend at least 2 pixels ($\sim$5$''$) for
even the most distant members of the sample. We define the nuclear
concentration C$_{24}$ to be the ratio of the flux in the aperture
S$_{2~kpc}$ to the total flux S$_{tot}$ from the galaxy.

Since the MIPS point spread function contains considerable power
outside the inner peak, we apply an aperture correction to the flux
measured in our small nuclear aperture to estimate the contribution
from an unresolved nuclear point source to the extended disk. For the
pixels scales in the Arp galaxy images (2.5$''$~pixel$^{-1}$) and the
SINGs images (1.5$''$~pixel$^{-1}$) we tabulated the correction factors
using a high dynamic range map provided by the MIPS team at the
Spitzer Science Center.
We note that because we assume a point source for the aperture
correction, the concentration C$_{24}$ will represent a theoretical
lower limit to the actual concentration. The values of C$_{24}$ should
therefore be treated as limiting cases where the emission is a true
point source at the galaxy center. We performed tests of our aperture
correction using some galaxies with pure point sources, and found that
it was capable of reproducing the correct flux in the extended PSF to
within 5$\%$, even for the smallest aperture. The correction factors
range from values of 2.96 in the most extreme cases, to more typically
1.5 for the Arp sample, and 1.1 for the normal galaxy sample. We carefully
investigated any redshift-dependent systematic trends of the
correction factor on the derived value of C$_{24}$, and found
none. Hence we believe that the values of C$_{24}$ are measured
reliably for the Arp and SINGs sample, despite differences in redshift
between the two samples. 

In Figure 35 we present the distribution of measured values of C$_{24}$ 
for the spirals and the Arp disks. We bin the value of C$_{24}$ 
in increments of 0.1 from 0 (no nuclear flux) to 1 (pure point
source). The values for the Arp primary and secondary pair members as
well as the combined Arp galaxies are shown as separate curves. The vertical
axis is the fraction of the total population that falls within each
concentration bin. 

Figure 35 shows a clear difference in the distribution of 24 $\mu$m nuclear
concentrations between the spirals and the Arp systems. The
spirals are predominantly diffuse objects with the median value
of C$_{24}$ being around 0.2 (i.e., 20$\%$ of the flux coming from the
inner 2kpc) and only tiny fraction of the systems having nuclear
concentrations, whereas the Arp systems show an almost flat (and
slowly declining) distribution of C$_{24}$. In particular, large
numbers of Arp systems (both primary and secondary galaxies) have
values of C$_{24}$ greater than 0.3 and in some cases are totally
nuclear dominated. We therefore conclude that Arp systems show a much
wider distribution of MIPS 24 $\mu$m nuclear concentrations, with a
large excess (over the spirals) extending up to objects which are
essentially point sources (e.g., the minor component of Arp 89).
No strong correlation is seen between the [3.6] - [24] and [8.0] - [24]
colors and the central concentration for the Arp
galaxies, although the reddest galaxies 
tend to be somewhat more centrally concentrated.

\section{Discussion}

\subsection{Spitzer Colors and Mass-Normalized Star Formation Rates}

To search for interaction-enhanced 
star formation in the mid-infrared, instead of comparing
luminosities, which measure the absolute SFR,
it is better to compare Spitzer colors, which measure
mass-normalized SFRs.
The [4.5] $-$ [5.8], [3.6] $-$ [8.0], [8.0] $-$ [24],
and [3.6] $-$ [24]
colors are all measures of the mass-normalized SFR,
with a redder color indicating a higher normalized SFR
(e.g., \citealp{smith05}).  The 3.6 $\mu$m and 4.5 $\mu$m
bands are dominated by the older stellar population,
while the other bands have significant contributions from interstellar
dust heated by young stars.

The 
[3.6] $-$ [4.5], 
[4.5] $-$ [5.8], and [3.6] $-$ [8.0]
colors of 
our optically-selected pre-merger interacting galaxies are not
strongly different from those of normal spirals.  However,
the [8.0] $-$ [24], [3.6] $-$ [24], and [5.8] $-$ [8.0] colors of the Arp
galaxies are significantly
redder on average than the spirals.
This is consistent with previous IRAS studies, which showed that
the IRAS 12 $\mu$m to 25 $\mu$m colors are redder in optically-selected
pre-merger interacting galaxies than in normal spirals
\citep{bushouse88, surace04}.
Reddening of these colors is expected as the 
interstellar radiation
field increases \citep{li01}.

The 
24 $\mu$m band is the most sensitive of our filters to the SFR,
as it is 
dominated by emission from very small
grains, which are heated by the ultraviolet radiation field.   
In contrast, 
the 8 $\mu$m band contains 
the dust emission features commonly attributed to 
polycylic aromatic hydrocarbons (PAHs);
these PAHs may be heated by the 
general interstellar radiation field.
For star forming clumps in M51, \citet{calzetti05} found
a direct proportionality between 24 $\mu$m and
extinction-corrected Paschen-$\alpha$ flux.
In contrast,
the 8 $\mu$m/Pa$\alpha$ 
ratio for these M51 clumps increases with decreasing luminosity,
implying some PAH heating by 
non-ionizing photons.
Thus to find weak enhancements in SFR,
colors involving 24 $\mu$m are best.  The [3.6] $-$ [24] color
is particularly valuable, as it is the 
cleanest measure of the ratio of the star formation
rate to the total stellar mass.

The mean difference in [3.6] $-$ [24] color between the Arp disks
and spirals is $\sim$0.8 magnitudes.  This corresponds to a mean
enhancement in normalized SFR of 2.1 in the Arp disks.
This is consistent with previous H$\alpha$ and far-infrared studies
of optically-selected interacting systems
\citep{bushouse87, bushouse88, kennicutt87}.
As with these previous studies, there is a lot of
scatter in this relation, with some Arp galaxies having little 
star formation enhancement.
Starbursts tend to be physically localized within interacting
galaxies, and have relatively short timescales compared to the
timescale of the interaction.  Thus they may not dominate the
global stellar population over the duration of the interaction.  An
optically-selected sample of galaxies like our sample would have a range of
interaction parameters and timescales, thus would tend to catch a lot of
systems in a non-dominating-burst stage.  For a strong burst to happen
before a merger and to dominate the global infrared colors of the
galaxy, the interaction parameters and timescale would have to be optimal.
It is also possible that some of our spiral galaxies may not be completely
isolated: they may be the result of recent minor mergers or
interactions with distant or small companions, so our set of spirals
may not be a perfect control sample.

As with the Arp galaxies as a whole, we see 
redder [8.0] $-$ [24], [3.6] $-$ [24], and [5.8] $-$ [8.0]
colors for the M51-like galaxies in 
our sample compared to spirals, and therefore enhanced star formation.
This is in agreement with the results of 
\citet{laurikainen98}, 
who found bluer optical colors in the central regions of
nine out of thirteen M51-like galaxies they studied.
As with their study,
our result is very tentative since we have 
a very small M51 sample size.
We note that 
strongly 
enhanced star formation is present
in localized regions within some of our M51-like galaxies. For examples,
see our detailed
studies on Arp 107 \citep{smith05} and Arp 82 \citep{hancock06}.

Examining the mid-infrared properties of the tails and bridges
in our galaxies has been challenging because of the low level of
mid-infrared emission from these regions.
On average, the tidal features contribute less than 10$\%$ to the 
total Spitzer light from these galaxies.  We cannot
distinguish strong differences in the colors of the disks
and tails, implying that the normalized SFR in the tails/bridges
is probably similar to that in the disks, however, this is uncertain.

In most cases, we also cannot make any definitive conclusions
about the colors of the non-spiral, non-Arp galaxy samples,
because of our small sample size and the number of upper limits.
We do, however, find
significantly bluer 
[3.6] $-$ [8.0] 
colors for the irregular/Sm galaxies than for the spirals.
This is likely
because of
weakness in the mid-infrared PAH features,
which contribute
strongly to the 8 $\mu$m band.  
Weak PAH features and low 8 $\mu$m
fluxes have been noted
before in low metallicity systems \citep{madden00, houck04, 
engelbracht05, dale05, madden06, ohalloran06}.

The irregular/Sm galaxies also have slightly redder [3.6] $-$ [4.5]
colors than spirals, as previously 
found by \citet{pahre04}.  This may be due to younger stars
on average
(main sequence A stars are 
redder in this color than K giants; \citealp{reach05})
or higher
interstellar-matter-to-stellar mass ratios; there may be larger
fractional
contributions from hot dust to the 4.5 $\mu$m band in irregular/Sm
galaxies compared to spirals.

\subsection{Spitzer and IRAS Luminosities
vs.\ Absolute Star Formation Rates}

Unlike the case with the observed distribution of Spitzer colors,
KS tests cannot rule out that the 
observed distributions of 
8.0 $\mu$m and 24 $\mu$m luminosities
for the Arp disks and the spirals came from the same parent
population.
Since the fluxes in these bands can be used as a proxy for
star formation (e.g., \citealp{forster04, calzetti05}),
this implies that the absolute rate of star formation in these Arp disks
(as opposed to the mass-normalized rate)
is similar to that in spirals.
From their Pa$\alpha$-24 $\mu$m correlation for star forming 
regions in M51, \citet{calzetti05} derived a relationship
between the 
SFR
(for
0.1 $-$ 100 M$_{\sun}$ stars)
and the
24 $\mu$m luminosity 
of
SFR(M$_{\sun}$~yr$^{-1}$) 
$\approx$ 4 $\times$ 10$^{-43}$ L$_{24}$(erg~s$^{-1}$).
\citet{wu05}
obtained the same relationship using 
global Spitzer, radio continuum, and H$\alpha$ fluxes for a sample
of non-dwarf infrared-selected galaxies
from the Spitzer First Look Survey.
Combined with our median luminosities (Table 7),
this relationship implies a median SFR
of $\sim$1.7 
M$_{\sun}$~yr$^{-1}$ for our Arp galaxies, and 
0.9
M$_{\sun}$~yr$^{-1}$ 
for the spirals.  
The possibly lower implied masses for the Arp galaxies compared
to the spirals (Section 4), 
combined with the relatively
small sample sizes, 
may account for the lack of a significant difference in the 
distribution of luminosities.

For dusty starburst galaxies, another measure of the star formation
rate can be obtained from the total 8 $-$ 1000 $\mu$m
infrared luminosity using
the relationship in \citet{kennicutt98}.  This gives, on average, a 
SFR of $\sim$2.6
M$_{\sun}$~yr$^{-1}$ per disk in our Arp galaxies, and 
$\sim$0.9
M$_{\sun}$~yr$^{-1}$ in the spirals.
These are similar to the values obtained using the 24 $\mu$m
luminosities.
As noted by \citet{kennicutt98} and many others (e.g., \citealp{persson87,
smith91, smith94, sauvage92}), in normal galaxies dust heating by non-OB
stars can contribute to the far-infrared luminosity, thus
the use of the Kennicutt relationship for non-starbursts may over-estimate
the SFR in some cases.  
\citet{calzetti05} found that for the M51 clumps, Pa$\alpha$ scales
directly with L$_{24}$ but not with total far-infrared luminosity;
there is an excess of far-infrared emission at low luminosities,
implying heating by non-OB stars.   
If the fraction of dust heated by young stars contributing to 
the FIR emission varies from galaxy to galaxy, then 
the L$_{FIR}$/SFR relationship may be different in interacting
galaxies than in spirals.
For example, in interacting galaxies the star formation is more
centrally concentrated (see Section 7), perhaps due to gas being driven
into the inner regions of the galaxy
by the interaction.
The dust associated with this central gas may be in closer proximity to
both OB and non-OB stars, on average, than in normal spirals, producing
a higher FIR luminosity for the same absolute star formation rate.
In spite of these possible problems, however, the median
SFRs that we obtain for the two samples from L$_{24}$ and L$_{IR}$
are similar.

To further investigate the question of SFR indicators
and whether the conversion from L$_{FIR}$ to SFR differs
from galaxy to galaxy and 
between samples, in Figure 36 we provide 
histograms of 
L$_{24 {\mu}m}$/L$_{FIR}$ for the Arp galaxies,
the spirals, the M51-like galaxies, the Irr/Sm galaxies,
and the Elliptical/S0 galaxies.  
For the Arp systems, the total
flux of both galaxies in the pairs is included.
For comparison, in Figure 37 we show the same histograms, but using
the total 8 $-$ 1000 $\mu$m infrared luminosity L$_{IR}$.
For both the 
L$_{24 {\mu}m}$/L$_{FIR}$ 
and 
L$_{24 {\mu}m}$/L$_{IR}$ ratios,
KS tests cannot rule out that the Arp and spiral samples originated
from the same parent distribution.
There is a spread in these two ratios from galaxy to
galaxy, with a total range in the 
L$_{24 {\mu}m}$/L$_{FIR}$ 
ratio of a factor
of $\sim$6 for the Arp galaxies.
The spread in the L$_{24}$/L$_{IR}$
ratio is slightly larger, with a total range
of 
a factor
of $\sim$10 for the Arp galaxies.
Note that L$_{24}$ and L$_{IR}$ are not completely independent,
in that L$_{IR}$ includes the 25 $\mu$m IRAS flux, thus an
excess in L$_{24}$ compared to L$_{FIR}$ increases L$_{IR}$.
Since the ratio of the star formation rates derived by
the two methods, SFR(24 $\mu$m)/SFR(IR), is simply proportional
to L$_{24}$/L$_{IR}$, then the derived SFR for an individual
galaxy can differ by 
up to a factor of $\sim$10, depending upon the method used.
As our galaxies do not have
strongly enhanced SFRs vs.\ normal spirals, the 24 $\mu$m luminosity
may be a better tracer of star formation than the far-infrared.
This is supported by models of the 
infrared spectral energy distribution in normal
spirals, which suggest that the 20 $-$ 42 $\mu$m range is
the best tracer of star formation in normal 
star forming galaxies \citep{dale01}.
Follow-up studies with 
bigger samples would be helpful
in further investigating this issue.

\subsection{Star Formation Enhancement vs.\ Pair Separation}

No strong correlation between the Spitzer mid-infrared colors of
the interacting systems and the projected pair separation is visible
in our sample.
A number of different factors likely contribute to weakening such a
correlation.  First, some scatter is introduced by the fact that we
have only projected separations.  Second, scatter is introduced
because the strength of tidal star formation triggering during an
interaction depends upon many other parameters in addition to
separation and mass ratio, such as orbital parameters and interstellar
gas content and distribution.  Third, the observed projected
separation is a complicated function of the orbital parameters and the
timescale of the interaction, which differ from system to system.
Some of our wider pairs may have already passed the point of closest
approach and may be in the process of separating, thus may be in a burst
phase.  Numerical modeling shows that when a flyby encounter triggers
a starburst, the burst often happens after the time of closest approach
(e.g., \citealp{mihos92}).

In contrast to our Spitzer results, previous optical studies
do 
measure a detectable difference in SFR between close pairs and 
wider pairs, though with a lot of scatter in the relationship
\citep{barton00, lambas03, nikolic04}.
Our lack of a detectable difference is likely due in part
to a selection effect; our pairs were selected based on tidal
distortion, while these other studies selected pairs based solely
on proximity.  Among our wider pairs, we probably include
a higher fraction that
are post-closest encounter.
Another likely 
factor is our relatively small sample size. 
In the histograms of the [8] $-$ [24] and [3.6] $-$ [24]
colors of the wide vs.\ close pairs (Figures 33 and 34),
the reddest galaxies are in close pairs.
However,
this result is not statistically significant (Table 9).  
With a larger sample size, a significant difference may be discernable.

We note that the so-called ultra-luminous infrared
galaxies (ULIRGs) (infrared-selected galaxies which have very high
SFRs) tend to be either mergers or very close pairs
(separation $<$ 2 kpc) \citep{veilleux02}.  Our pairs have much larger
separations, and so are in an earlier pre-merger state.

\subsection{The Central Concentration of Star Formation}

We have found that the 24 $\mu$m distributions in the Arp disks
are more centrally-concentrated than those of normal spirals.
This implies that, even before merger, 
interactions can modify the
radial distribution of the active star formation regions,
by driving gas into the central region and triggering nuclear
and circumnuclear star formation.  This result is consistent
with
previous optical, near-infrared, radio continuum,
and ground-based mid-infrared studies, which also implied
enhanced 
central 
star formation in interacting
galaxies \citep{joseph84, lonsdale84, keel85, hummel90, 
bartongillespie03, bergvall03,
nikolic04}.
Our observations show that,
even in relatively low-luminosity pre-mergers, 
where the total L$_{FIR}$ of a
typical system is less than that of a so-called luminous
infrared galaxy (i.e., $<$ 10$^{11}$
L$_{\odot}$), the effects of the tidal processes has already begun to
operate on the disks, stimulating more activity in the center
of the disk.
These results are consistent with numerical models of gas buildup in the
central regions of interacting galaxies, via
bar- or wave-induced angular momentum transport, mass transfer
between galaxies, and ring-like compression 
(e.g., \citealp{noguchi88, barnes96, struck03, iono04}).
Such activity
may be a prelude to dramatically increased nuclear activity when
the galaxies fully merge.

\section{Summary}

We compare the Spitzer infrared luminosities and colors for
a sample of 35 strongly interacting Arp galaxy pairs with those
of 26 isolated spirals.  
The 
[3.6] $-$ [24], 
[8.0] $-$ [24], and [5.8] $-$ [8.0] 
colors of the Arp disks 
are significantly
redder than those of the spirals.  
The [3.6] $-$ [24] colors of the Arp disks are 0.8 magnitudes redder,
on average, than the spirals, implying an enhancement to the mass-normalized
SFR of a factor of $\sim$2.
For the shorter wavelength Spitzer colors 
([3.6] $-$ [4.5], 
[4.5] $-$ [5.8], and [3.6] $-$ [8.0]) we cannot
detect a significant
difference between spirals and Arp galaxies.  
This is likely because these bands are less sensitive to star
formation, and include a larger fraction
of emission not related to star formation.

We see little statistical difference in
the 
3.6 $\mu$m, 8.0 $\mu$m, and 24 $\mu$m luminosities of the Arp and 
spiral
samples.  However, some of the small Arp companions have 
low 3.6 $\mu$m
luminosities, and therefore low masses, compared to spirals. 
The Spitzer colors are more sensitive to small interaction-induced
enhancements
in star formation rate than the Spitzer luminosities, because
they are normalized for mass.
The Arp disks have more centrally-concentrated 24 $\mu$m emission
than the normal spirals, implying that the interactions have 
driven gas into the central regions, triggering central star formation.
No trend of Spitzer colors with pair separation is visible in our sample.
This may be a selection effect, as our galaxies were chosen to have
strong tidal distortions.





\acknowledgments

We thank the Spitzer team for making this research possible.
We also thank Rob Kennicutt and the SINGS team for producing
the SINGS dataset.
We thank Edith Seier and Mark Giroux for helpful communications and
Amanda Moffett for help with the analysis.
This research was supported by NASA Spitzer grant 1263924, NSF
grant AST-0097616, and 
NASA LTSA grant NAG5-13079.
This research has made use of the NASA/IPAC Extragalactic Database (NED) which is operated by the Jet Propulsion Laboratory, California Institute of Technology, under contract with the National Aeronautics and Space Administration.

\clearpage



\clearpage

\begin{figure*}
\centerline{\includegraphics[width=\textwidth]{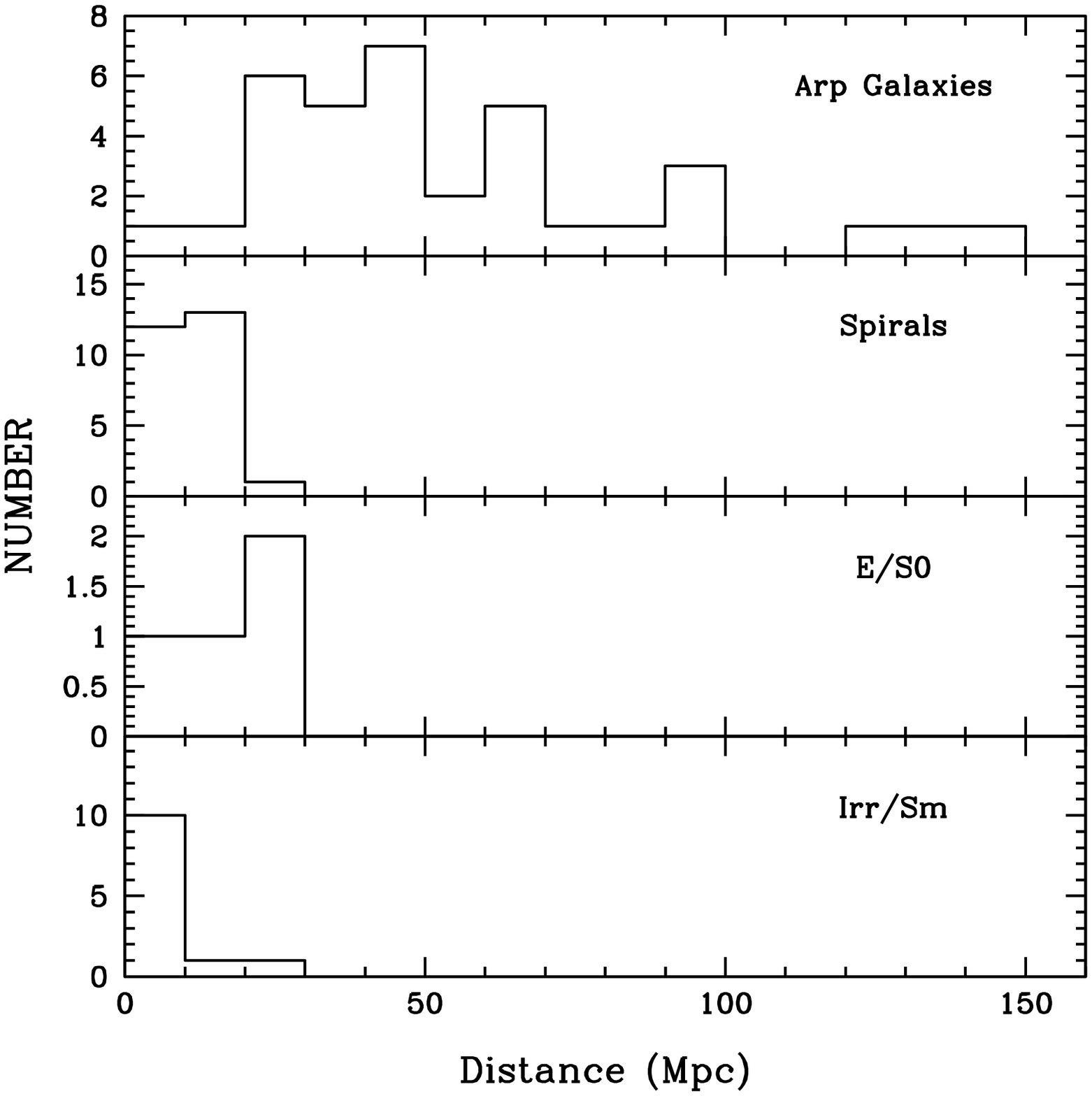}}
\caption{
\small 
Histograms of the distances to the four samples of galaxies.
Top panel: interacting sample.  
Middle panel: spiral sample.
Third panel: Elliptical/S0 sample.
Bottom panel: Irregular/Sm sample.
}
\end{figure*}

\clearpage

\begin{figure*}
\centerline{\includegraphics[width=\textwidth]{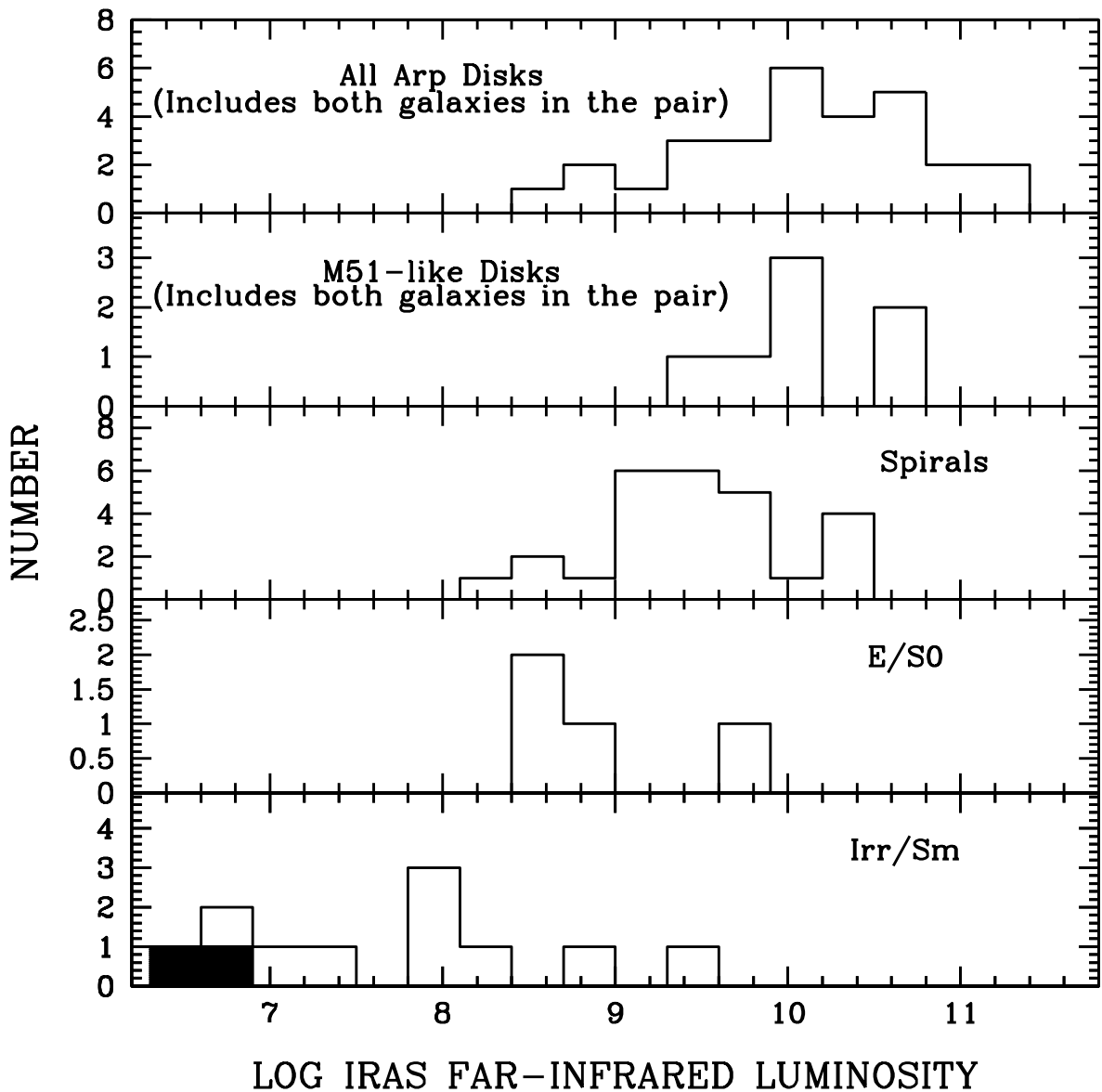}}
\caption{
\small 
Histograms of the far-infrared luminosities of the various samples.
Upper limits are indicated by shaded regions.
Note that the luminosities for the Arp galaxies and M51-like
galaxies include the combined
flux from both galaxies in the pair.
Top panel: interacting sample.  
Second panel: M51-like galaxies.
Third panel: spiral sample.
Fourth panel: Elliptical/S0 sample.
Bottom panel: Irregular/Sm sample.
}
\end{figure*}

\clearpage

\begin{figure*}[t]
\includegraphics[width=6.1in]{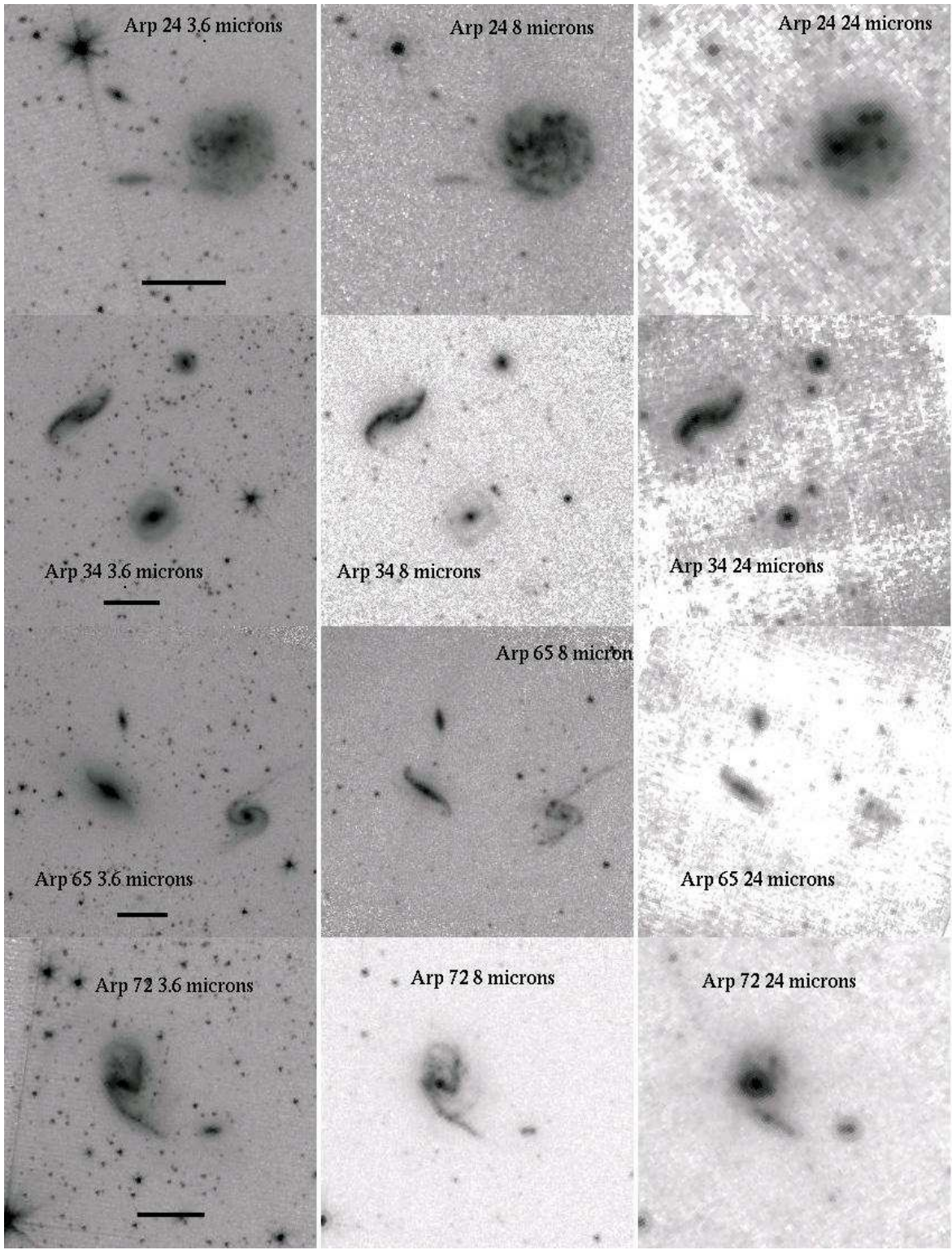}
\caption{
\small 
The 3.6 $\mu$m, 8.0 $\mu$m, and 24 $\mu$m 
images of galaxies in the interacting galaxy sample.
North is up and east to the left.  The scale bar is 60$''$ (1$'$).
The 8 and 24 $\mu$m images are displayed with the same spatial scale as the 
3.6 $\mu$m image.
In Figures 3 $-$ 11, the galaxies are plotted in order of Arp number.
For identification of individual galaxies in the pairs, see Table 5.
In the Arp 24 and 65 3.6 $\mu$m images, the small galaxies in the northeast
(upper left) are background galaxies, so are not included in Table 5.
}
\end{figure*}

\clearpage

\begin{figure*}
\includegraphics[width=6.1in]{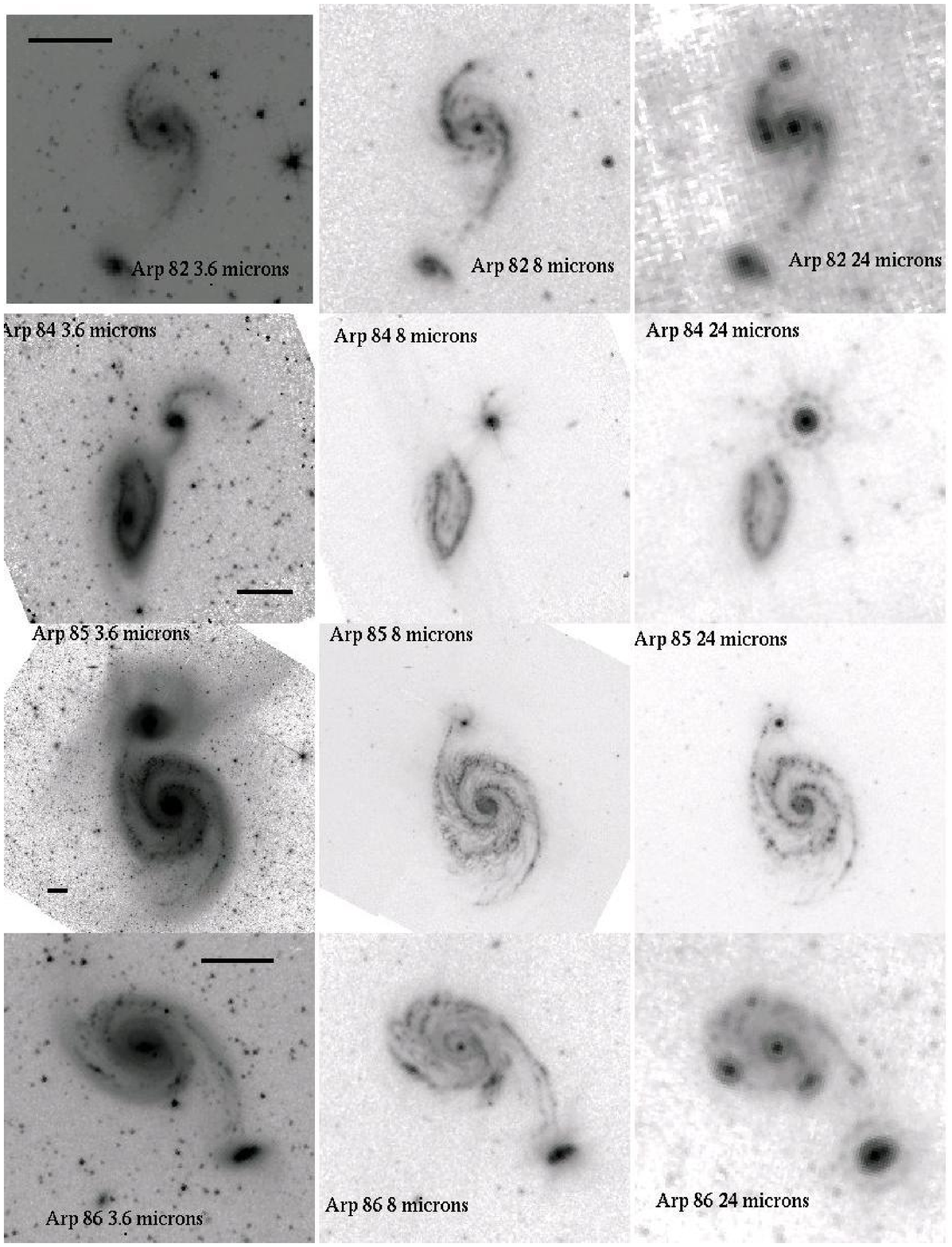}
\caption{
\small 
The 3.6 $\mu$m, 8.0 $\mu$m, and 24 $\mu$m 
images of galaxies in the interacting galaxy sample.
North is up and east to the left.  The scale bar is 60$''$ (1$'$).
In the 24 $\mu$m Arp 84 image, the MIPS point spread function is visible
as a `ring' and `spikes' around the very bright nucleus of the northern
galaxy.
}
\end{figure*}

\clearpage

\begin{figure*}
\includegraphics[width=6.1in]{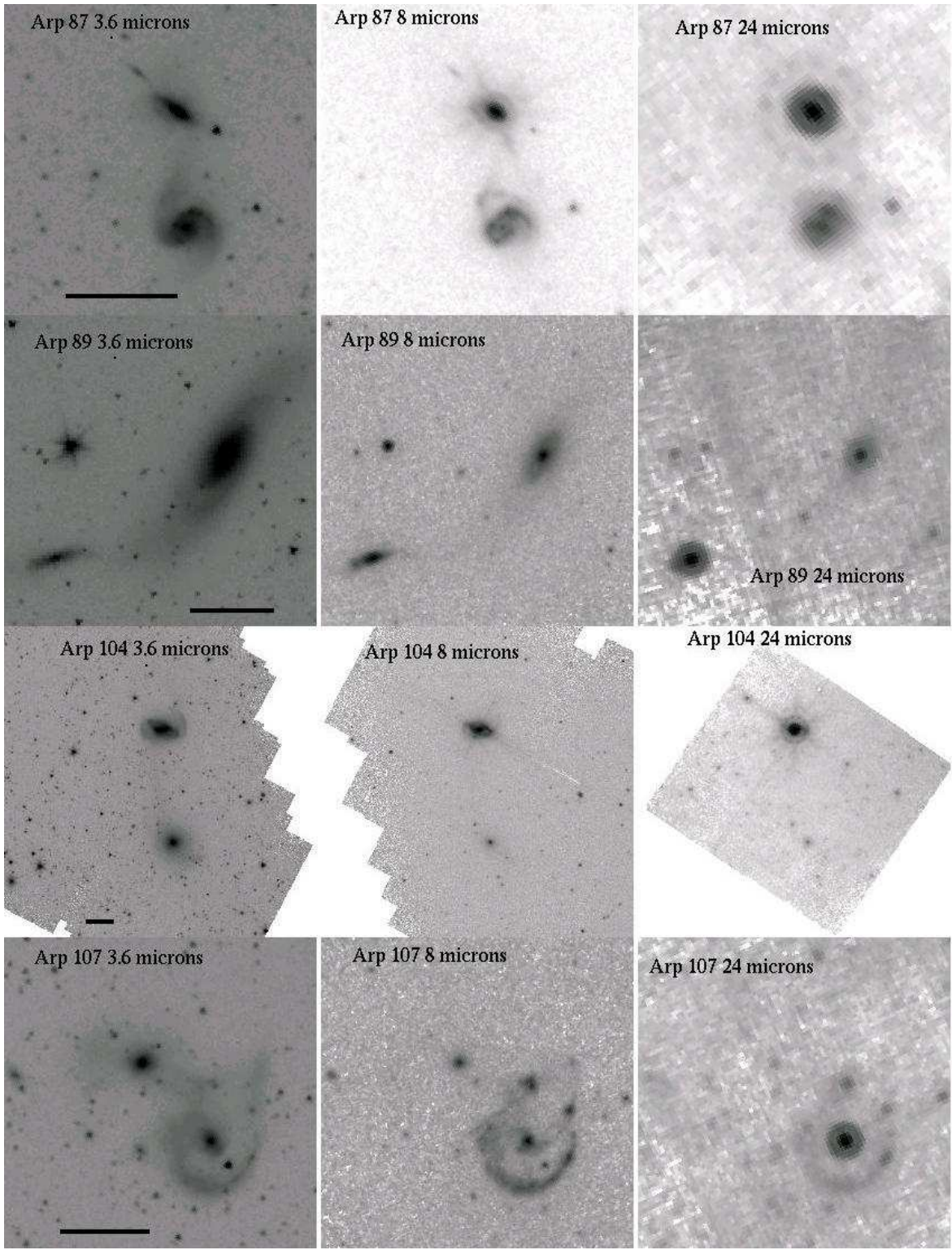}
\caption{
\small 
More the 3.6 $\mu$m, 8.0 $\mu$m, and 24 $\mu$m 
images of the sample galaxies.
North is up and east to the left.  The scale bar is 60$''$ (1$'$).
No redshift is available for the small feature seen to the northeast
of the northern galaxy in Arp 87, thus it is uncertain whether this
is associated with the Arp pair or not.  It is not included in the following
statistics.
In the 24 $\mu$m Arp 87 and Arp 104 images, 
the MIPS point spread function is visible
around the very bright nuclei of the northern
galaxies.  
The ring in Arp 107 is a real physical structure (see
\citet{smith05}
for a detailed study of the Spitzer Arp 107 images).
}
\end{figure*}

\clearpage

\begin{figure*}
\epsscale{.7}
\plotone{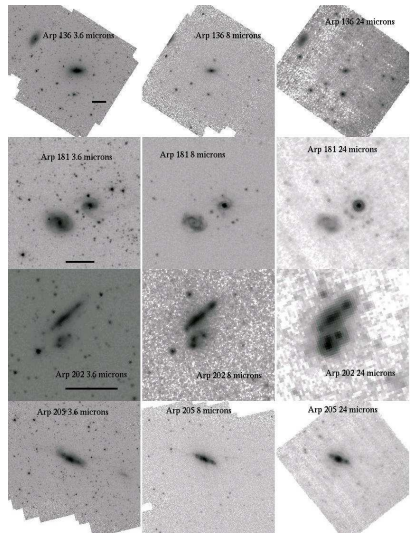}
\epsscale{1}
\caption{
\small 
The 3.6 $\mu$m, 8.0 $\mu$m, and 24 $\mu$m 
images of galaxies in the interacting galaxy sample.
North is up and east to the left.  The scale bar is 60$''$ (1$'$).
In the \citet{arp66} Atlas, Arp 136 consists of only NGC 5820, the
large galaxy near the center of the images shown above.  NGC 5820 has
a similar-mass companion outside of the Arp Atlas field of view, NGC 5821.
This second galaxy is visible in the northeast (upper left) of the 
images shown above.  Unfortunately NGC 5821 was missed by our 4.5 and 8.0 $\mu$m
Spitzer observations.  The small galaxy to the south of NGC 5820 is a
background galaxy.
In the 24 $\mu$m Arp 181 image, the MIPS point spread function is visible
as a `ring' and `spikes' around the very bright nucleus of the northern
galaxy.
}
\end{figure*}

\clearpage

\begin{figure*}
\includegraphics[width=6.1in]{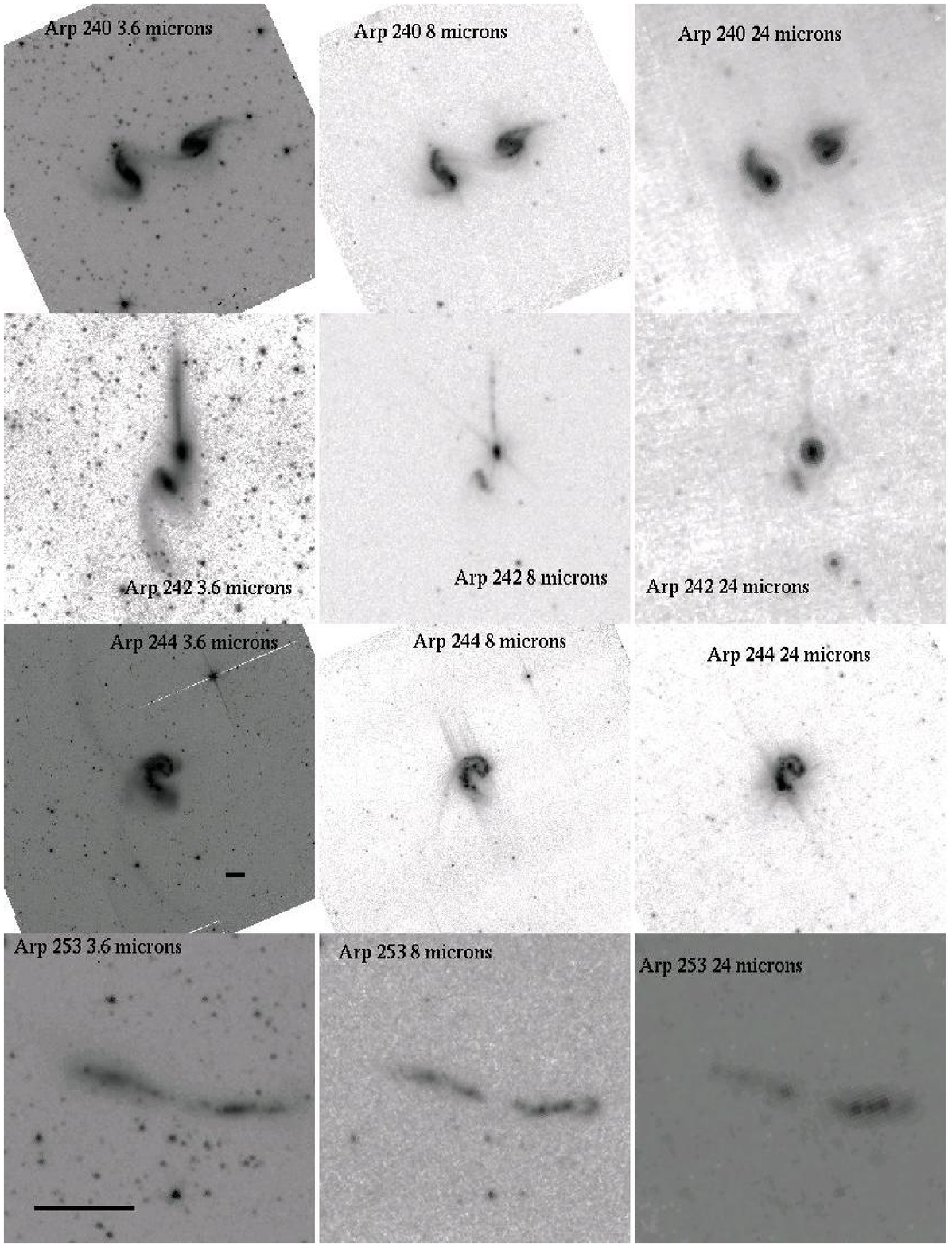}
\caption{
\small 
The 3.6 $\mu$m, 8.0 $\mu$m, and 24 $\mu$m 
images of galaxies in the interacting galaxy sample.
North is up and east to the left.  The scale bar is 60$''$ (1$'$).
In the 24 $\mu$m Arp 242 image, 
the MIPS point spread function is visible
around the nucleus of the northern
galaxy.  
}
\end{figure*}

\begin{figure*}
\includegraphics[width=6.1in]{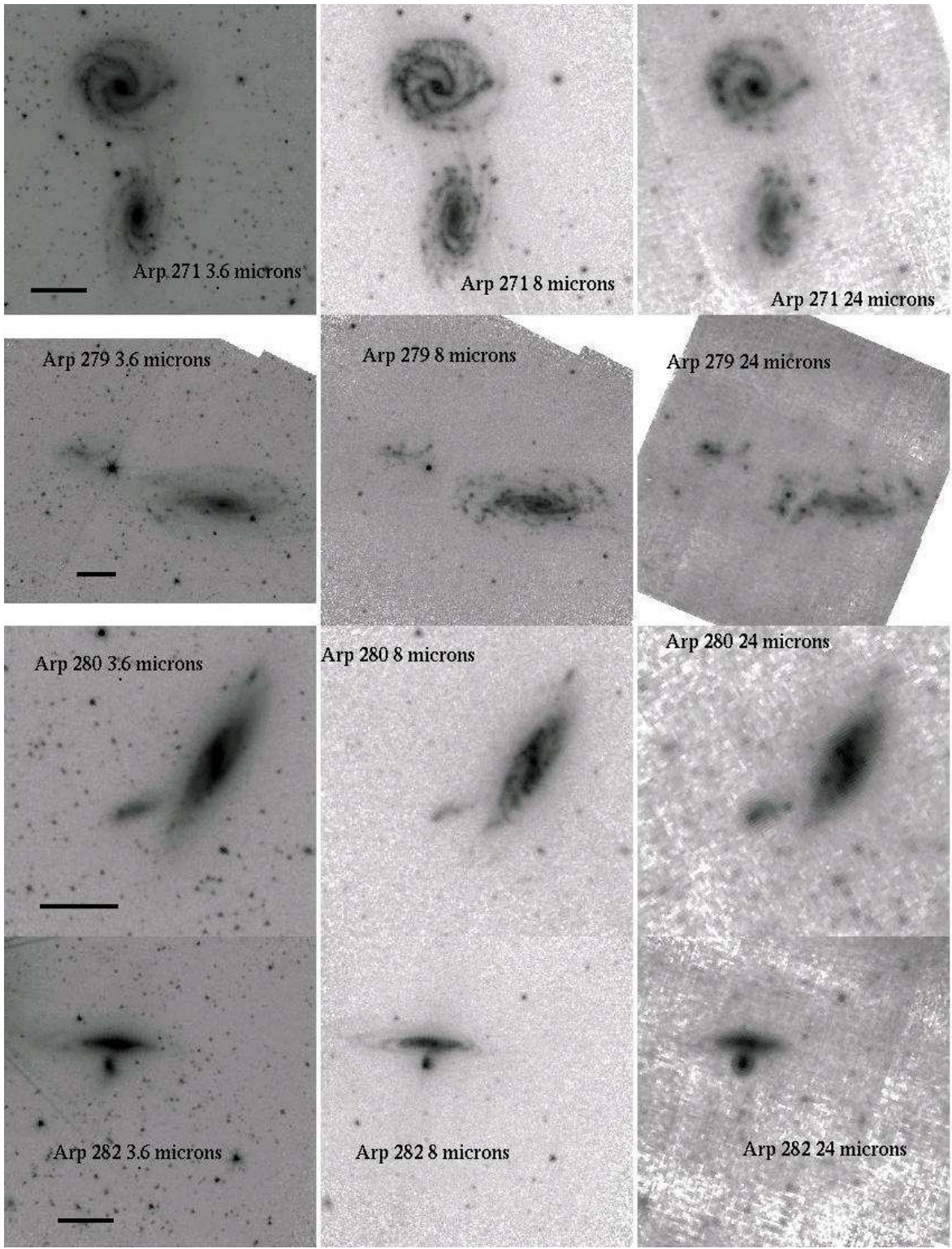}
\caption{
\small 
The 3.6 $\mu$m, 8.0 $\mu$m, and 24 $\mu$m 
images of galaxies in the interacting galaxy sample.
North is up and east to the left.  The scale bar is 60$''$ (1$'$).
}
\end{figure*}

\clearpage

\begin{figure*}
\includegraphics[width=6.1in]{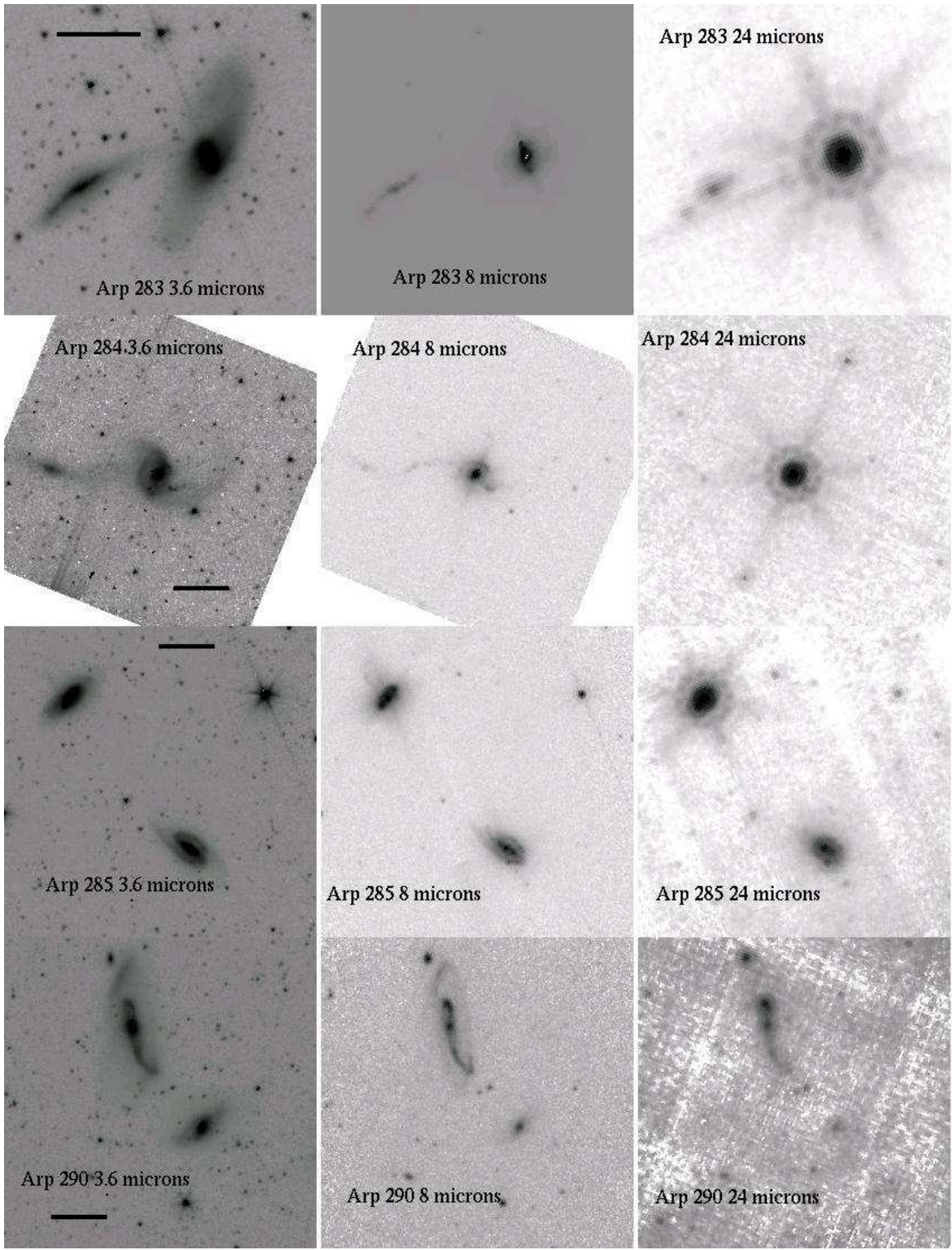}
\caption{
\small 
The 3.6 $\mu$m, 8.0 $\mu$m, and 24 $\mu$m 
images of galaxies in the interacting galaxy sample.
North is up and east to the left.  The scale bar is 60$''$ (1$'$).
In the 24 $\mu$m Arp 283, Arp 284, and Arp 285 images, 
the MIPS point spread function is visible.
The small object in the northeast (upper left) of the Arp 290 images is
a background galaxy.
}
\end{figure*}

\clearpage

\begin{figure*}
\includegraphics[width=6.1in]{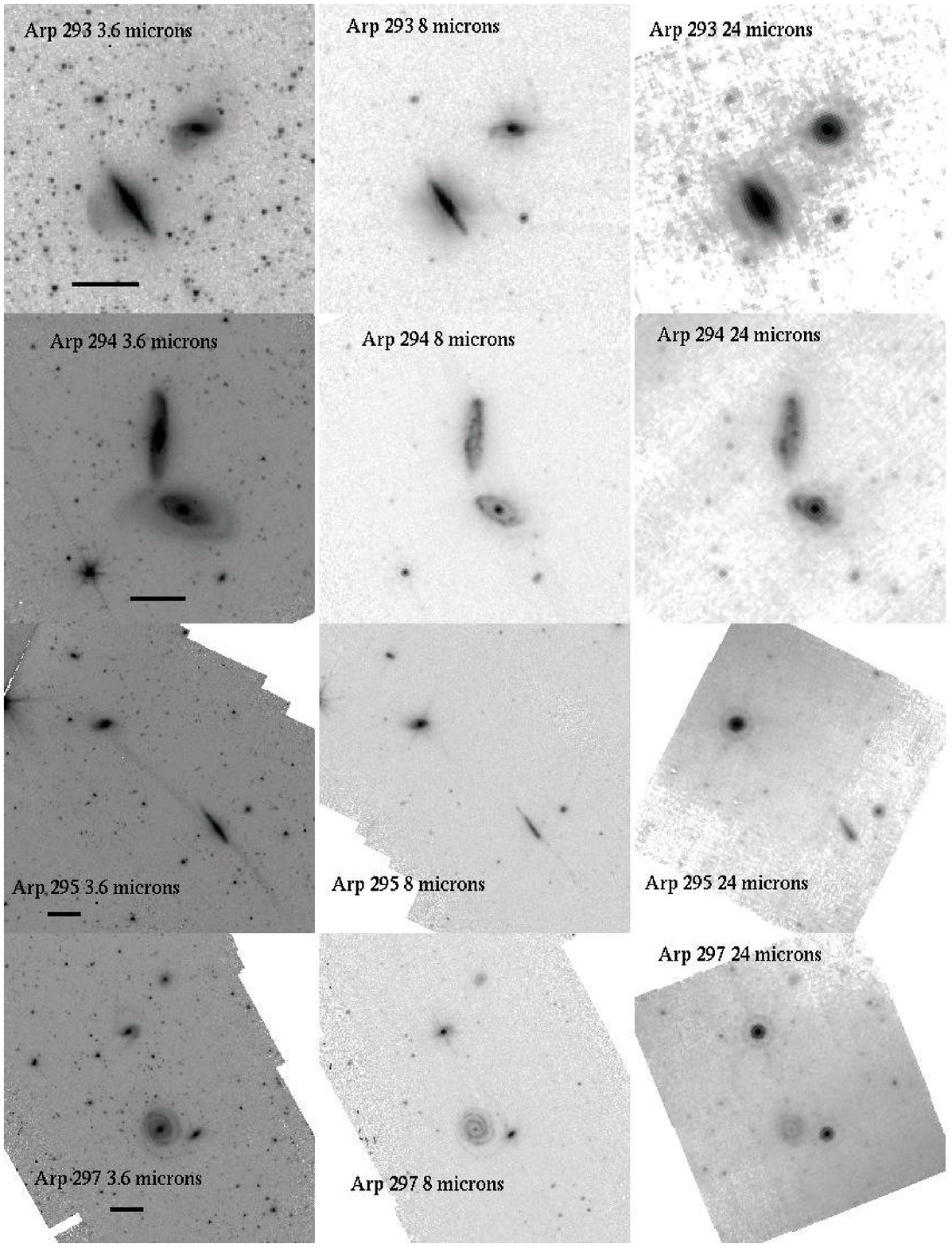}
\caption{
\small 
The 3.6 $\mu$m, 8.0 $\mu$m, and 24 $\mu$m 
images of galaxies in the interacting galaxy sample.
North is up and east to the left.  The scale bar is 60$''$ (1$'$).
In the 24 $\mu$m Arp 295 and Arp 297 images, 
the MIPS point spread function is visible
as a `ring' and `spikes' around the very bright nucleus of the northern
galaxy.
}
\end{figure*}

\clearpage

\begin{figure*}
\includegraphics[width=6.1in]{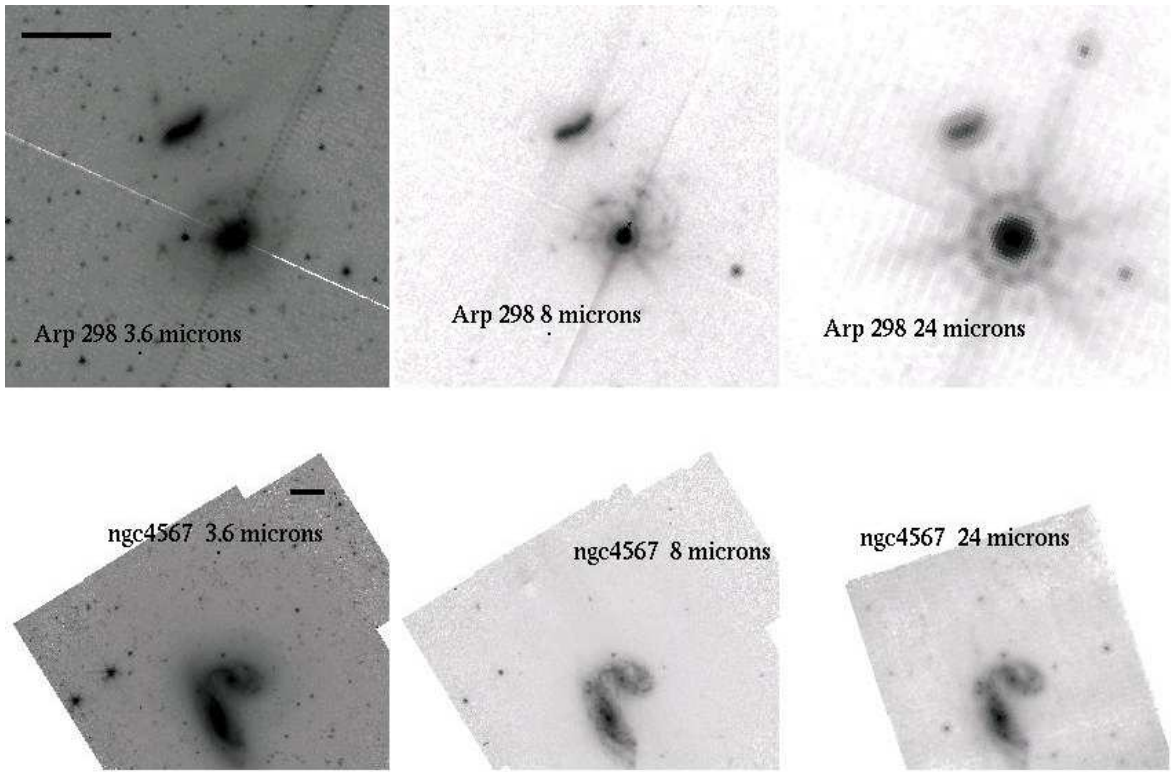}
\caption{
\small 
The 3.6 $\mu$m, 8.0 $\mu$m, and 24 $\mu$m 
images of galaxies in the interacting galaxy sample.
North is up and east to the left.  The scale bar is 60$''$ (1$'$).
The MIPS PSF is visible in the 24 $\mu$m Arp 298 image.
}
\end{figure*}

\begin{figure*}
\includegraphics[width=6.1in]{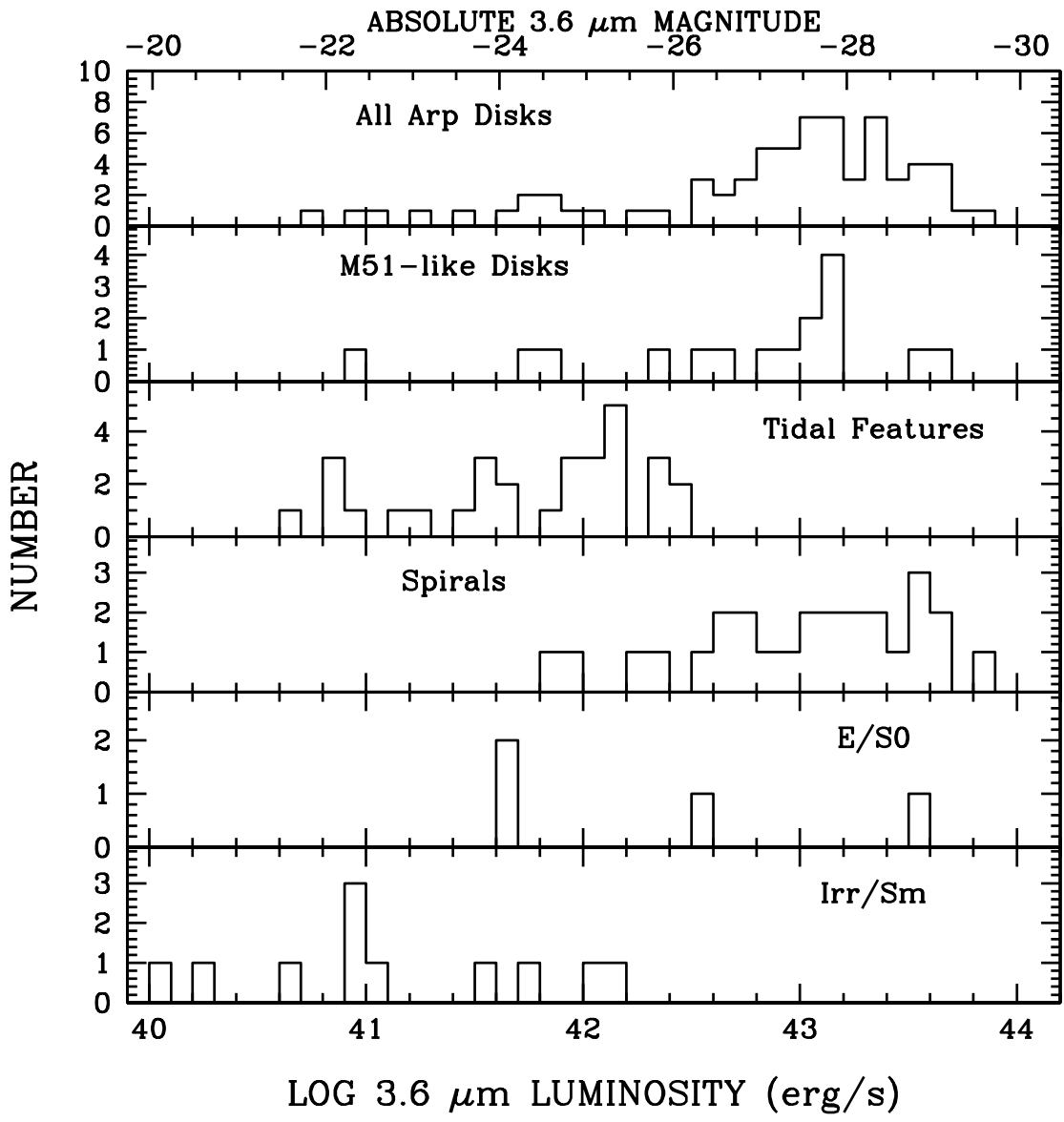}
\caption{
\small 
Histograms of the 3.6 $\mu$m luminosities
for the various samples.  For consistency with previous
work (e.g., \citealp{calzetti05}), we use
the
`monochromatic' luminosity $\nu$L$_{\nu}$, using a
frequency 
of 8.45 $\times$ 10$^{13}$ Hz.
}
\end{figure*}

\clearpage

\begin{figure*}
\includegraphics[width=6.1in]{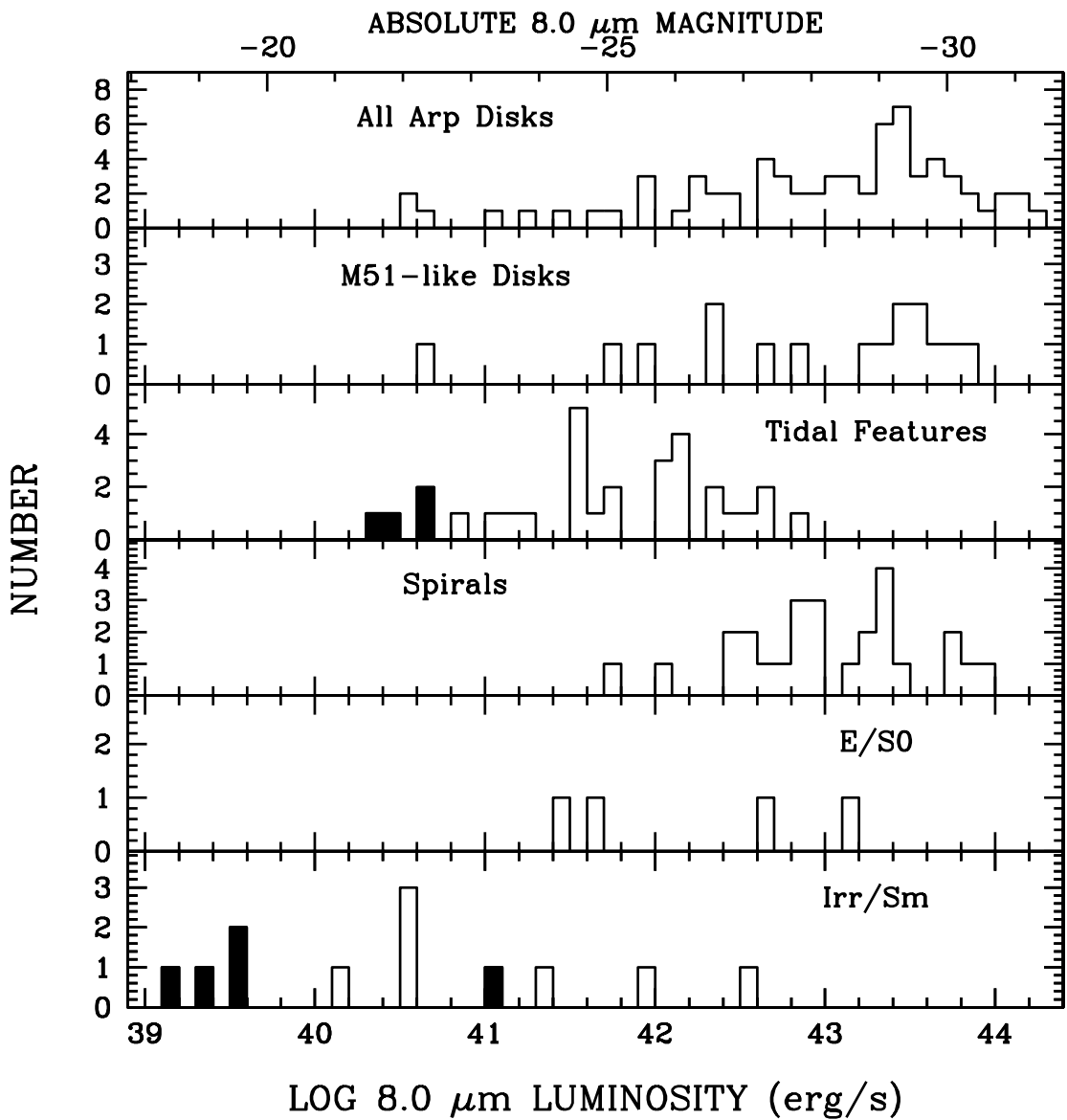}
\caption{
\small 
Histograms of the 8.0 $\mu$m luminosities
for the various samples.  For consistency with previous
work,
we use
the
`monochromatic' luminosity $\nu$L$_{\nu}$, using a
frequency 
of 3.81 $\times$ 10$^{13}$ Hz.
Upper limits are indicated by shaded regions.
}
\end{figure*}

\begin{figure*}
\includegraphics[width=6.1in]{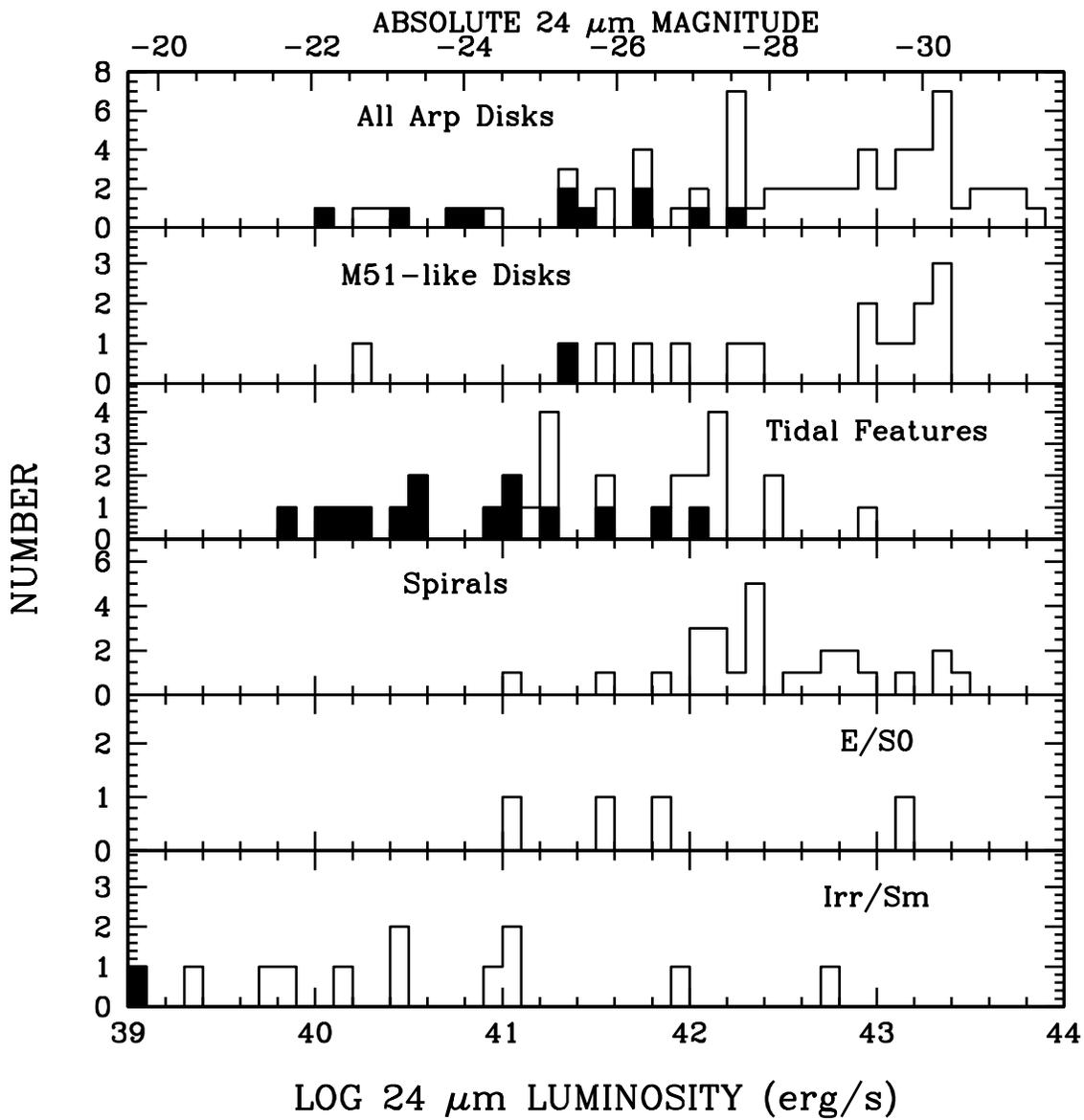}
\caption{
\small 
Histograms of the 24 $\mu$m luminosities
for the various samples.  For consistency with previous
work,
we use
the
`monochromatic' luminosities $\nu$L$_{\nu}$, using a
frequency 
of 1.27 $\times$ 10$^{13}$ Hz.
The Arp galaxy with the highest luminosity is NGC 7469 (Arp 298),
a Seyfert galaxy.
Upper limits are indicated by the filled-in regions.
}
\end{figure*}

\clearpage

\begin{figure}
\plotone{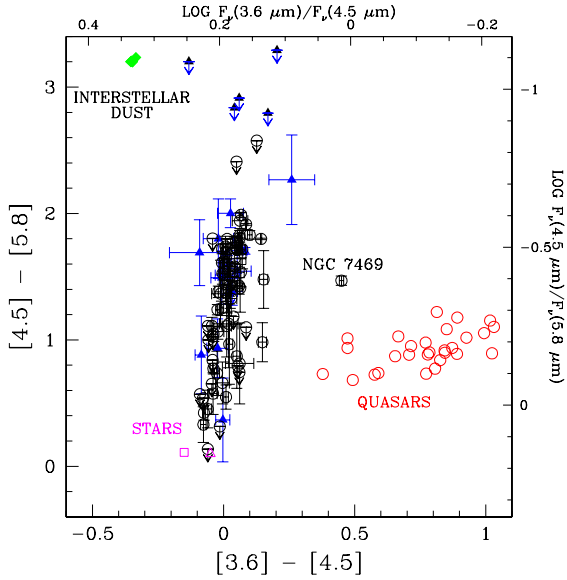}
\caption{
  \small 
The Spitzer IRAC [3.6] $-$ [4.5] vs. [4.5] $-$ [5.8] color-color
plot for the main disks of the interacting galaxy sample (black circles)
and the tidal features of these galaxies (filled blue triangles).
The colors of M0III stars (open magenta square), (M. Cohen 2005, private 
communication), and the mean colors of the 
field stars of \citet{whitney04} (magenta triangle),
and the colors of the \citet{hatz05}
quasars (red circles)
are also shown.
The predicted IRAC colors for interstellar dust \citep{li01}
are also plotted (filled green diamonds), for ISRF 
strengths that vary from 0.3 $-$ 10,000 $\times$ that in the
solar neighborhood.  As the ISRF increases, the IRAC colors
of dust
become redder.  Note that the predicted dust colors vary very little
for this wide range in ISRF.
The Arp disk at [3.6] $-$ [4.5] = 0.4 is the Seyfert galaxy NGC 7469.
}
\end{figure}

\clearpage

 \begin{figure}
\plotone{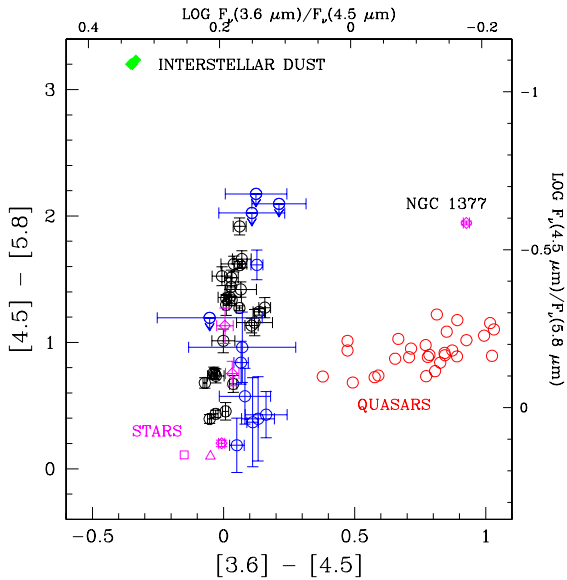}
\caption{
  \small 
The Spitzer IRAC [3.6] $-$ [4.5] vs.\ [4.5] $-$ [5.8] color-color
plot for the spiral galaxies (black circles), ellipticals/S0
(open magenta diamonds), and irregular/Sm galaxies (blue).
The colors of M0III stars (open magenta square), 
field stars (magenta triangle),
quasars (red circles), and 
interstellar dust 
(filled green diamonds) are also shown.  
The elliptical galaxy
NGC 1377, like the Seyfert NGC 7469 (Figure 15), has a very
red [3.6] $-$ [4.5] color, similar to that of quasars, in
spite of not being classified as a Seyfert based on optical
data (e.g., \citealp{kim95}).  Our Spitzer fluxes for NGC 1377 are 
consistent with those of \citet{dale05}.
See \citet{roussel06} for a detailed discussion of NGC 1377.
The bluest elliptical is NGC 4125.
}
\end{figure}

\clearpage

 \begin{figure}
\plotone{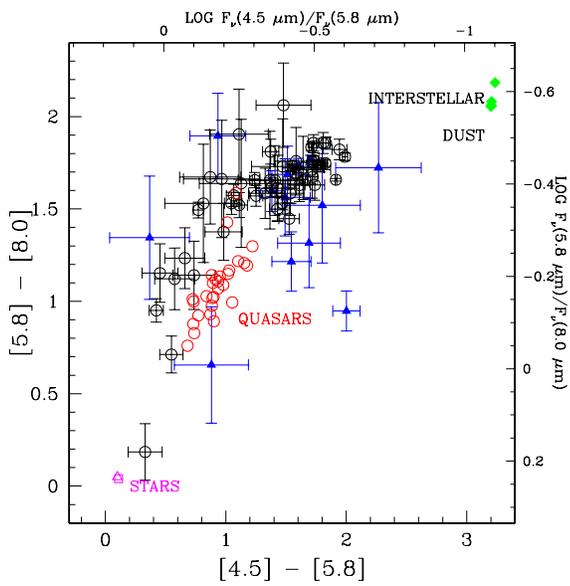}
\caption{
  \small 
The Spitzer 
IRAC [4.5] $-$ [5.8] vs.\ [5.8] $-$ [8.0] color-color
plot for the 
main disks of the interacting galaxy sample (black circles)
and the tidal features of these galaxies (filled blue triangles).
The colors of M0III stars (open magenta square), 
field stars (magenta triangle),
quasars (red circles), and 
interstellar dust 
(filled green diamonds) are also shown.  
The bluest disk is Arp 290 south (IC 195).
}
\end{figure}

\clearpage

 \begin{figure}
\plotone{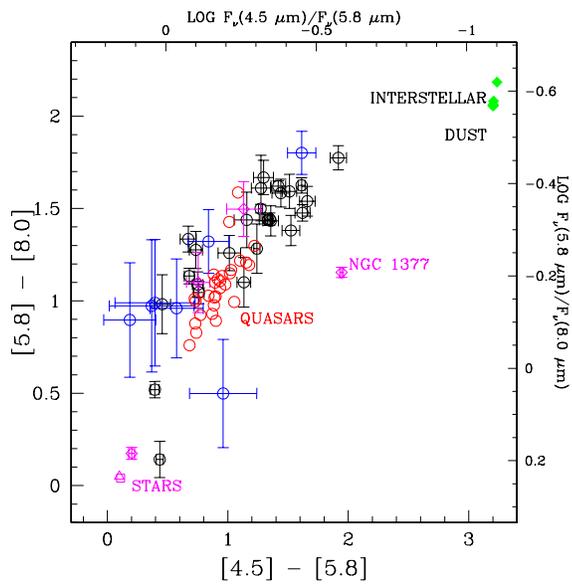}
\caption{
  \small 
The Spitzer 
IRAC [4.5] $-$ [5.8] vs.\ [5.8] $-$ [8.0] color-color
plot for the spiral galaxies (black circles), ellipticals/S0
(open magenta diamonds), and irregular/Sm galaxies (blue).
The colors of M0III stars (open magenta square), 
field stars (magenta triangle),
quasars (red circles), and 
interstellar dust 
(filled green diamonds) are also shown.  
The elliptical NGC 1377 again has peculiar colors.
The bluest elliptical is NGC 4125.
}
\end{figure}

\clearpage

 \begin{figure}
\plotone{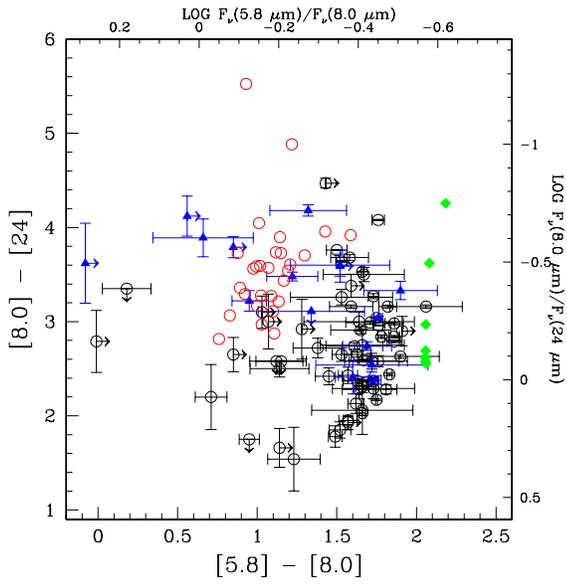}
\caption{
  \small 
The Spitzer 
IRAC [5.8] $-$ [8.0] vs. [8.0] $-$ [24] color-color
plot for the 
main disks of the interacting galaxy sample (black circles)
and the tidal features of these galaxies (filled blue triangles).
The colors of 
quasars (red circles) and 
interstellar dust 
(filled green diamonds) are also shown.  
Stellar photospheres are expected to lie at 0,0 on this plot.
The tidal feature that is the reddest in [8.0] $-$ [24] is the western
tail of Arp 284 (NGC 7714/5), which has a bright star forming
region near its base (see \citealp{smith97}).
}
\end{figure}

\clearpage

 \begin{figure}
\plotone{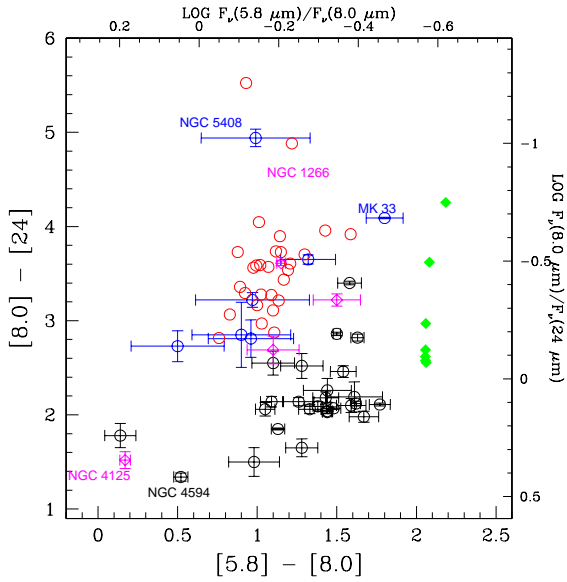}
\caption{
  \small 
The Spitzer 
IRAC [5.8] $-$ [8.0] vs. [8.0] $-$ [24] color-color
plot for the spiral galaxies (black circles), ellipticals/S0
(open magenta diamonds), and irregular/Sm galaxies (blue).
The colors of 
quasars (red circles) and 
interstellar dust 
(filled green diamonds) are also shown.  
Stellar photospheres are expected to lie at 0,0 on this plot.
In [8.0] $-$ [24], the reddest and second reddest Irr/Sm galaxies
are NGC 5408 and Markarian 33, respectively, while the reddest
elliptical is NGC 1266 (which has a high far-infrared luminosity--see Table 2).
The elliptical NGC 1377 is at (1.25, 3.5), with colors similar to those
of quasars.
The bluest spiral galaxy is the SAa galaxy NGC 5494, while the bluest
elliptical is NGC 4125.
}
\end{figure}

\clearpage

 \begin{figure}
\plotone{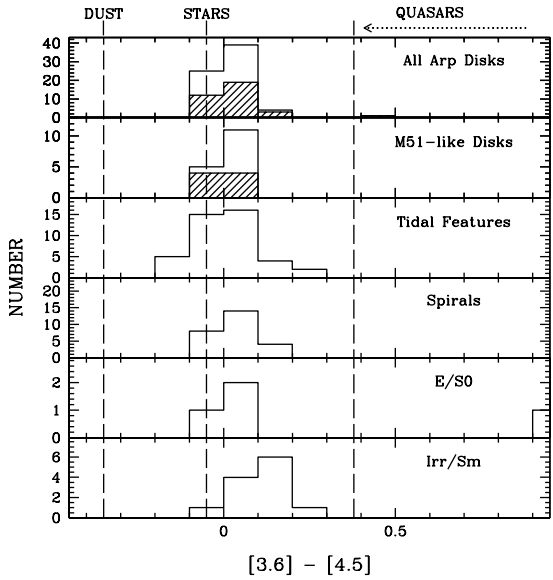}
\caption{
  \small 
Histograms of the Spitzer 
IRAC [3.6] $-$ [4.5] colors for the different samples:
the Arp galaxy main disks (top panel), the subset of M51-like systems
(second panel), tidal features
(third panel),
the spiral galaxies (fourth panel), ellipticals/S0
(fifth panel), and irregular/Sm galaxies (bottom panel).
The hatched galaxies are the more massive in the pair.
The filled regions represent upper limits.
The mean [3.6] $-$ [4.5]
color for the \citet{whitney04}
field stars is $-$0.05, 
while the predicted value
for interstellar dust is $-$0.35
\citep{li01}.
The Arp disk with the very red color of 0.44 is the Seyfert
NGC 7469 (Arp 298S), while the red elliptical is NGC 1377.  
These colors are close to the
colors of quasars \citep{hatz05}.
}
\end{figure}

 \begin{figure}
\plotone{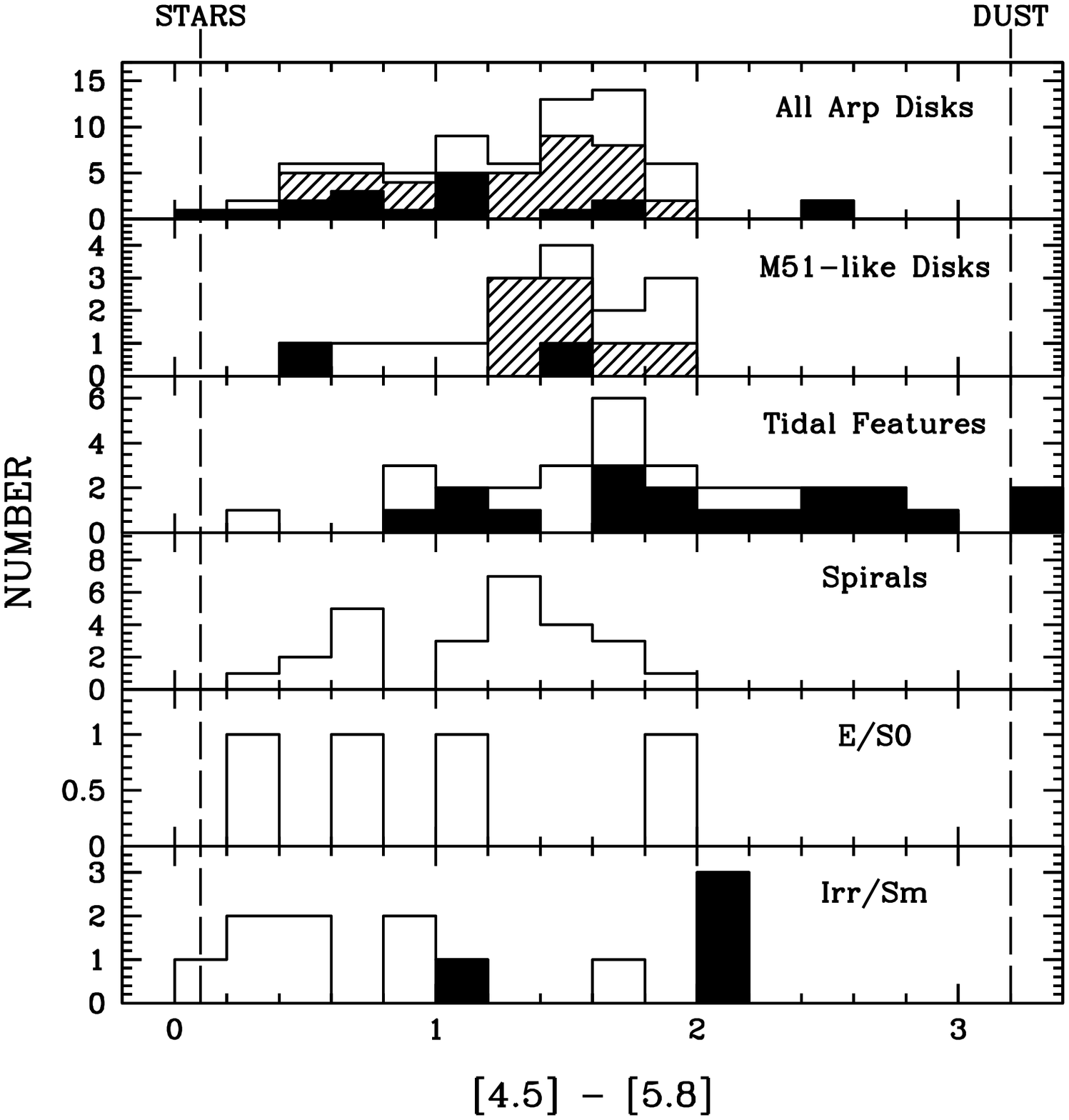}
\caption{
  \small 
Histograms of the Spitzer 
IRAC [4.5] $-$ [5.8] colors for the different samples.
The filled regions represent upper limits.
The hatched galaxies are the more massive in the pair.
The mean 
[4.5] $-$ [5.8]
color for the \citet{whitney04}
field stars is 0.1, 
while the predicted value
for interstellar dust is 3.2
\citep{li01}.
}
\end{figure}

 \begin{figure}
\plotone{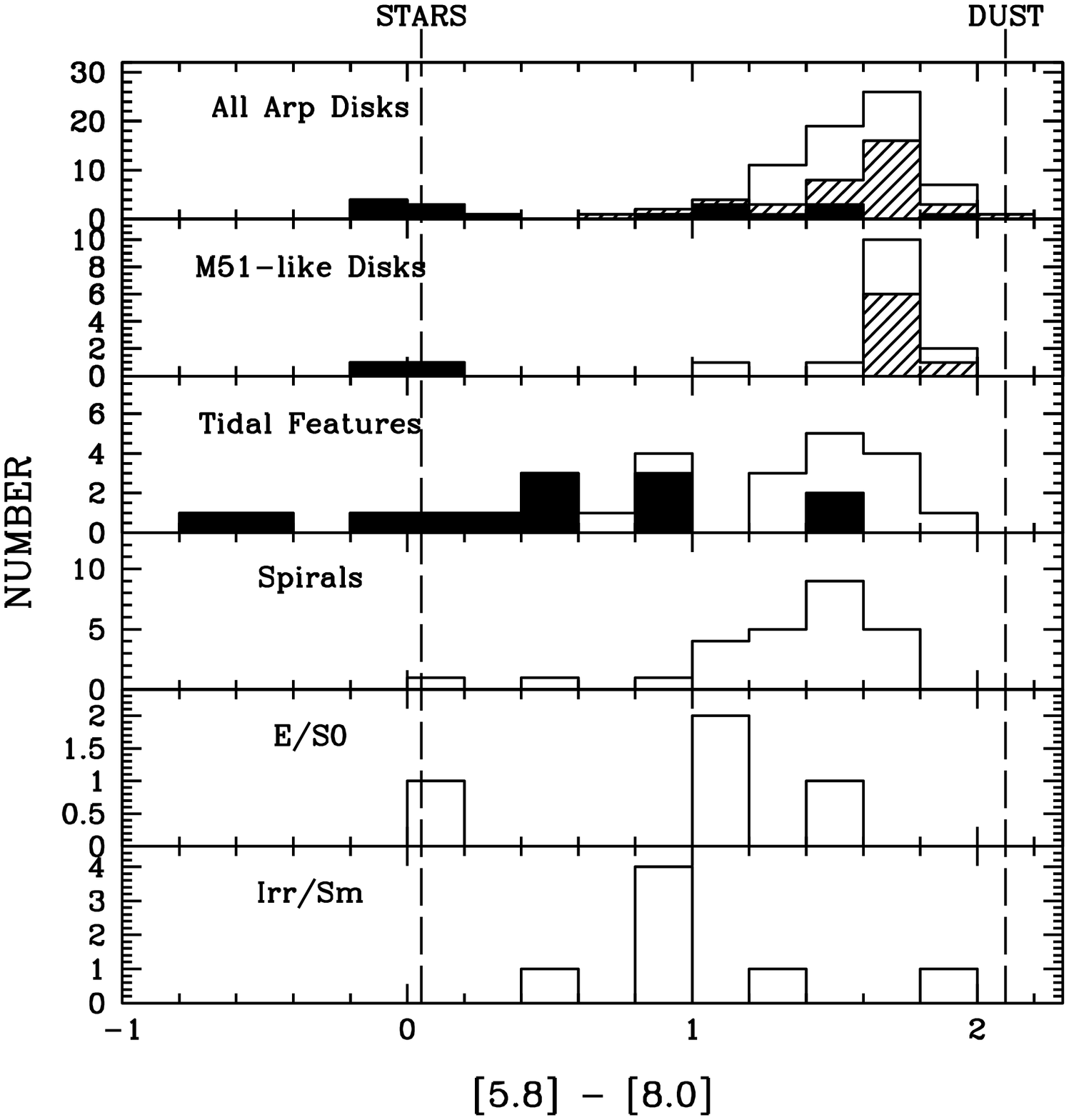}
\caption{
  \small 
Histograms of the Spitzer 
IRAC [5.8] $-$ [8.0] colors for the different samples.
The filled areas represent {\it lower} limits.
The hatched galaxies are the more massive in the pair.
The mean 
[5.8] $-$ [8.0] color for the \citet{whitney04}
field stars is 0.05, 
while the predicted value
for interstellar dust is 2.1
\citep{li01}.
}
\end{figure}

 \begin{figure}
\plotone{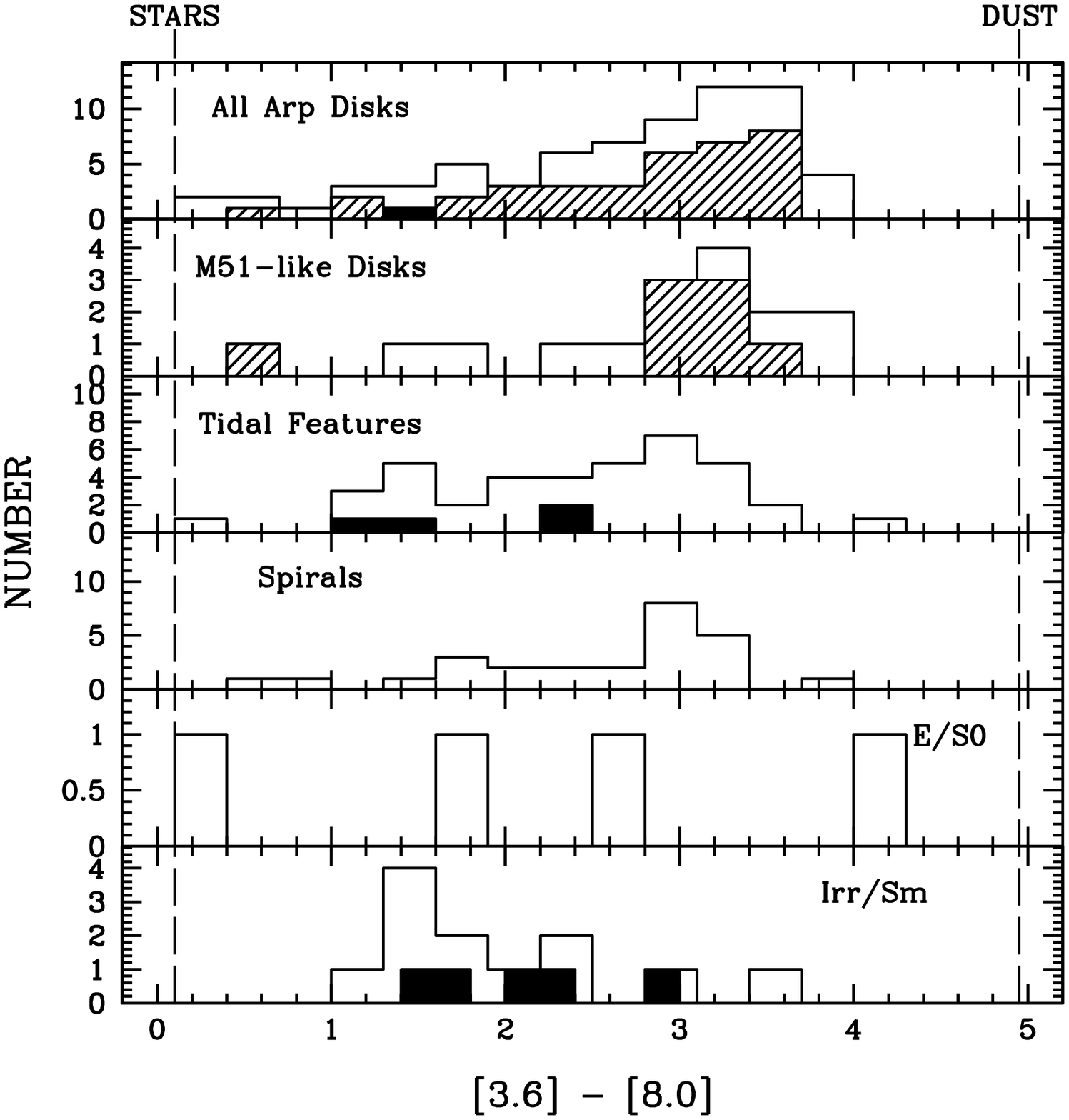}
\caption{
  \small 
Histograms of the Spitzer 
IRAC [3.6] $-$ [8.0] colors for the different samples.
The filled areas represent upper limits.
The hatched galaxies are the more massive in the pair.
The mean 
[3.6] $-$ [8.0] color for the \citet{whitney04}
field stars is 0.1, 
while the predicted value
for interstellar dust is 4.95
\citep{li01}.
}
\end{figure}

\vfill
\eject

\begin{figure}
\plotone{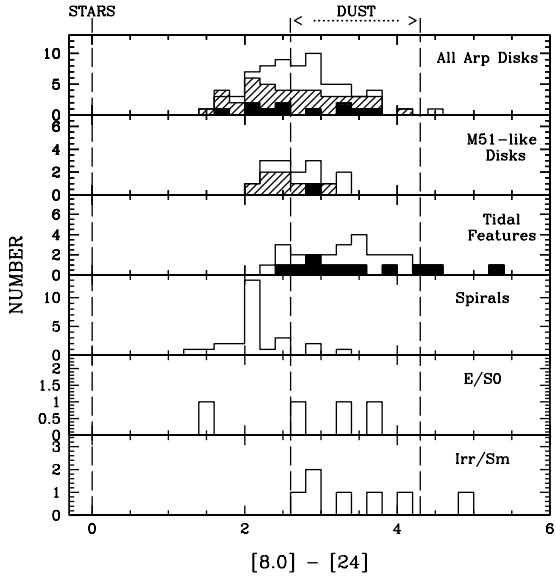}
\caption{
\small 
Histograms of the Spitzer 
IRAC [8.0] $-$ [24] colors for the different samples.
The filled areas represent upper limits.
The hatched galaxies are the more massive in the pair.
The expected [8.0] $-$ [24] color for dust varies from 2.6 $-$ 4.3 \citep{li01},
increasing with increasing interstellar radiation field intensity,
while
stars are expected to be at $\sim$0.0.
}
\end{figure}

\begin{figure}
\plotone{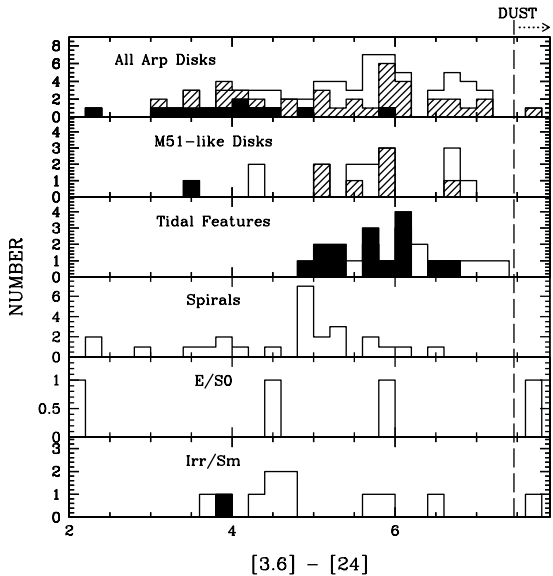}
\caption{
\small 
Histograms of the Spitzer 
IRAC [3.6] $-$ [24] colors for the different samples.
The shaded galaxies are the more massive in the pair.
The expected [3.6] $-$ [24] color for dust varies from 7.5 $-$ 9.3 \citep{li01},
increasing with increasing interstellar radiation field intensity,
while
stars are expected to be at $\sim$0.0.
}
\end{figure}

\begin{figure}
\plotone{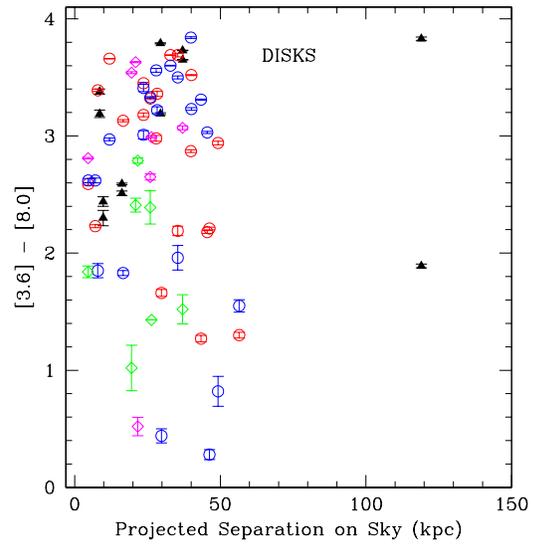}
\caption{
  \small 
Projected pair separation vs.\ [3.6] $-$ [8.0] color, for the main
disks in the galaxy pairs.
Black triangles represent pairs with 3.6 $\mu$m luminosity 
ratios greater than 0.75 (i.e.,
close to equal mass pairs).
Red and blue circles represent the more and less massive galaxy,
respectively, in
a pair with 3.6 $\mu$m luminosity ratios between 0.1 and 0.75 (i.e.,
intermediate mass ratio pairs).
Magenta and green diamonds show the locations of the more
and less massive galaxy, respectively, in pairs with 3.6 $\mu$m
luminosity ratios
less than 0.1 (i.e., highly unequal mass pairs).
}
\end{figure}

\begin{figure}
\plotone{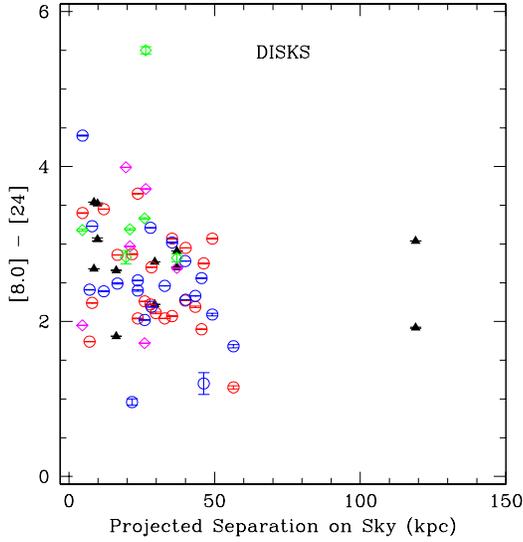}
\caption{
  \small 
Projected pair separation vs.\ [8.0] $-$ [24] color, for the main disks
in the galaxy pairs.
The symbols are the same as in Figure 24.
}
\end{figure}

 \begin{figure}
\plotone{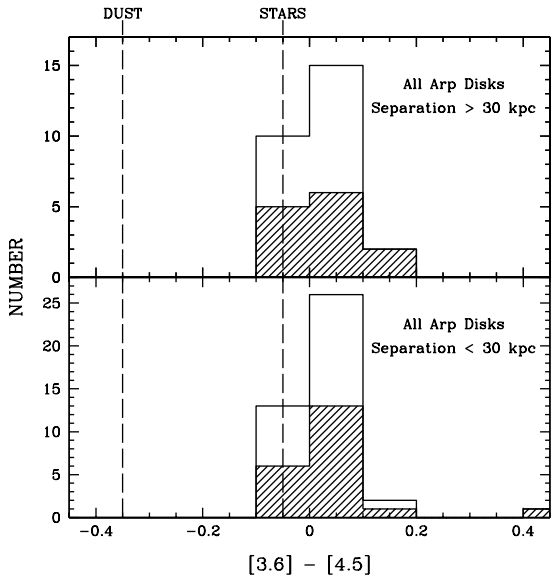}
\caption{
  \small 
Histograms of the [3.6] $-$ [4.5] color for the close and wide pairs
in our sample.
The hatched galaxies are the more massive in the pair.
}
\end{figure}

 \begin{figure}
\plotone{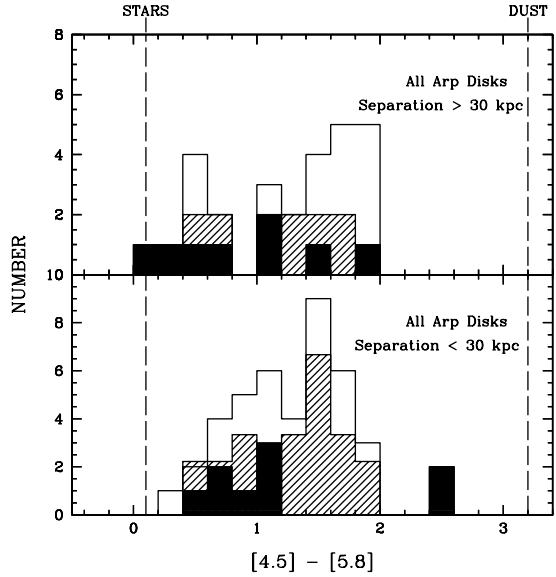}
\caption{
  \small 
Histograms of the [4.5] $-$ [5.8] color for the close and wide pairs
in our sample.
The hatched regions mark galaxies are the more massive in the pair.
The filled-in areas are upper limits.
}
\end{figure}

 \begin{figure}
\plotone{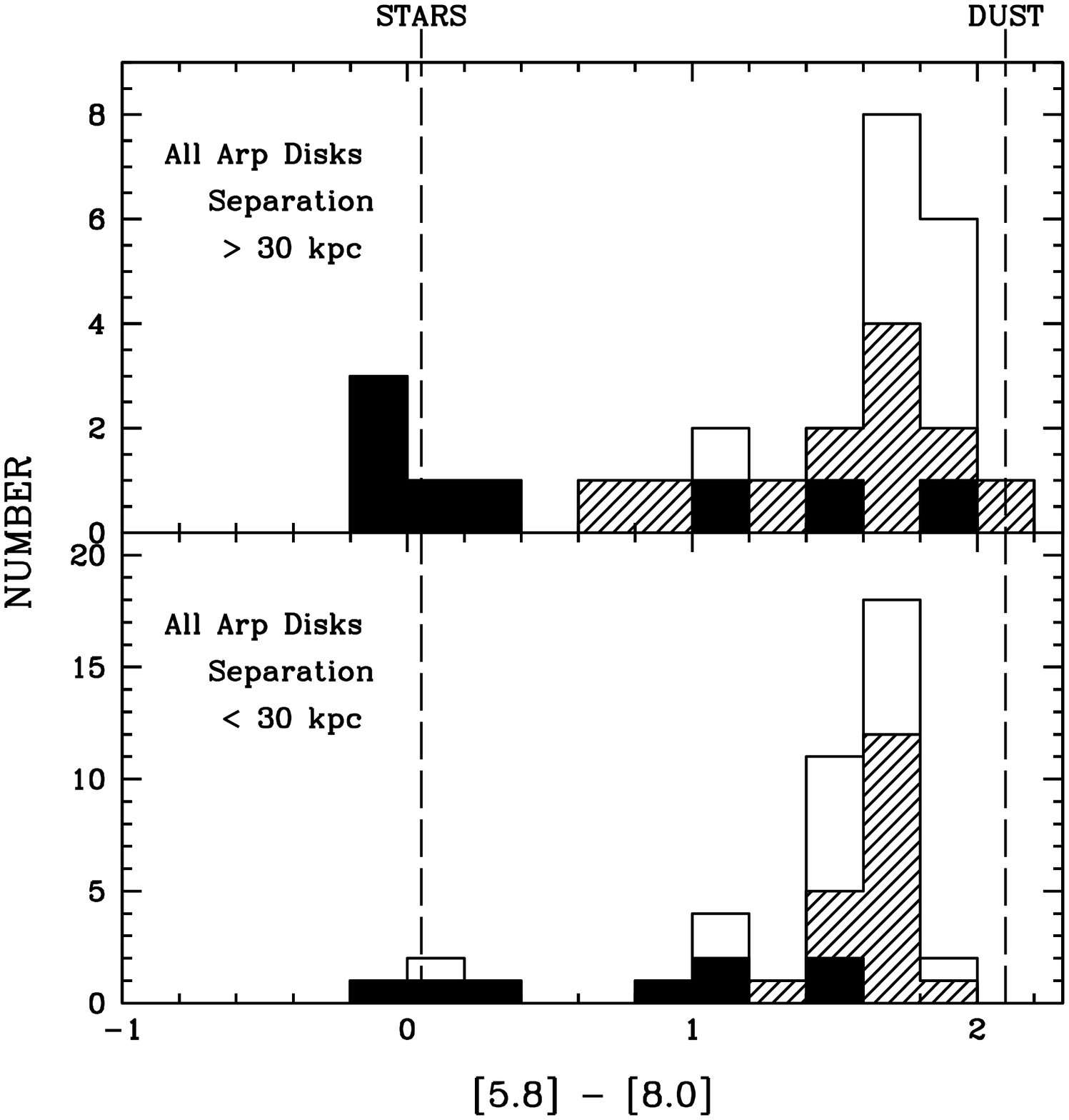}
\caption{
  \small 
Histograms of the [5.8] $-$ [8.0] color for the close and wide pairs
in our sample.
The hatch-marked galaxies are the more massive in the pair.
The filled-in areas indicate {\it lower} limits.
}
\end{figure}

 \begin{figure}
\plotone{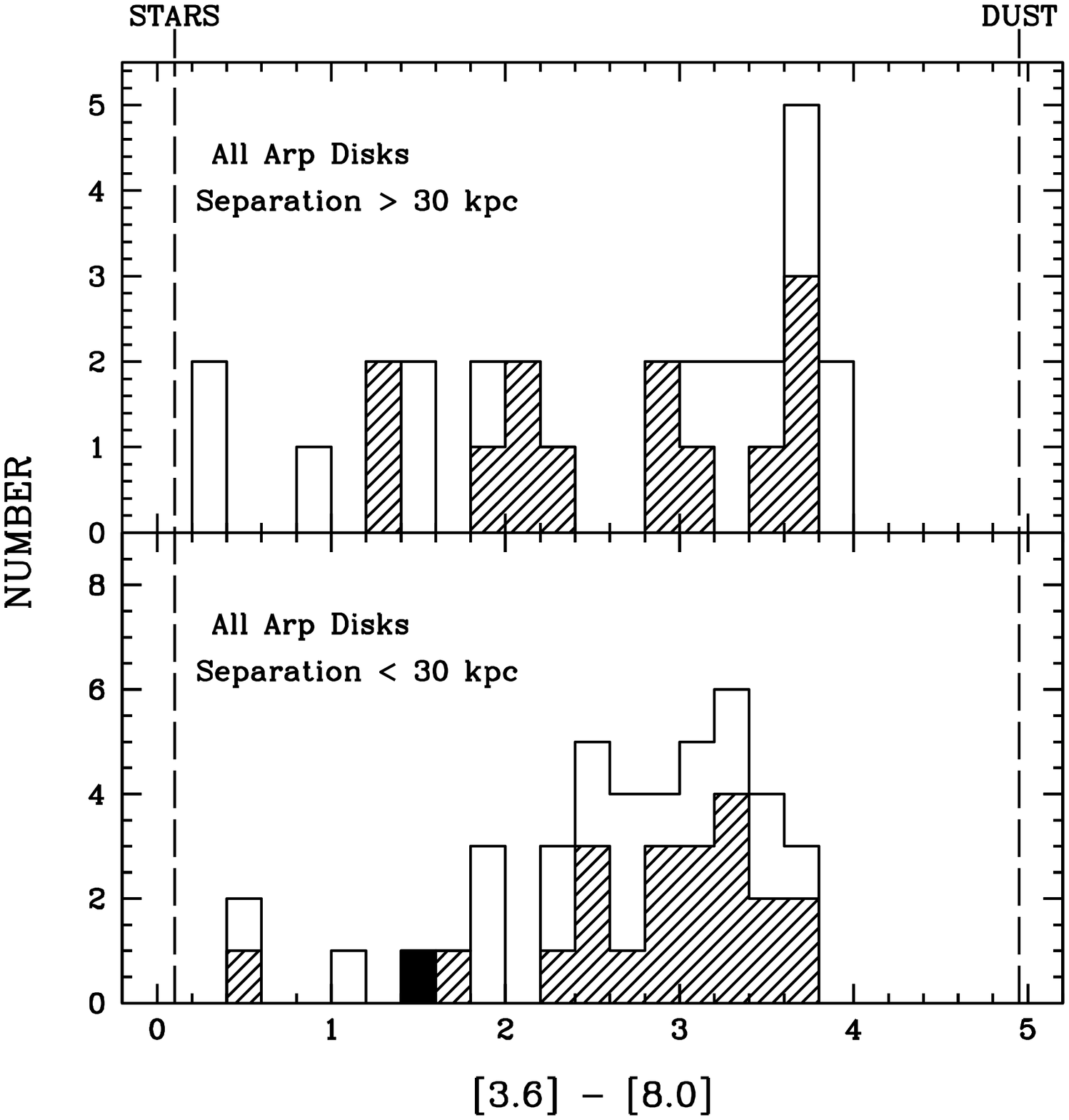}
\caption{
  \small 
Histograms of the [3.6] $-$ [8.0] color for the close and wide pairs
in our sample.
The hatch-marked galaxies are the more massive in the pair.
The filled-in regions indicate upper limits.
}
\end{figure}

 \begin{figure}
\plotone{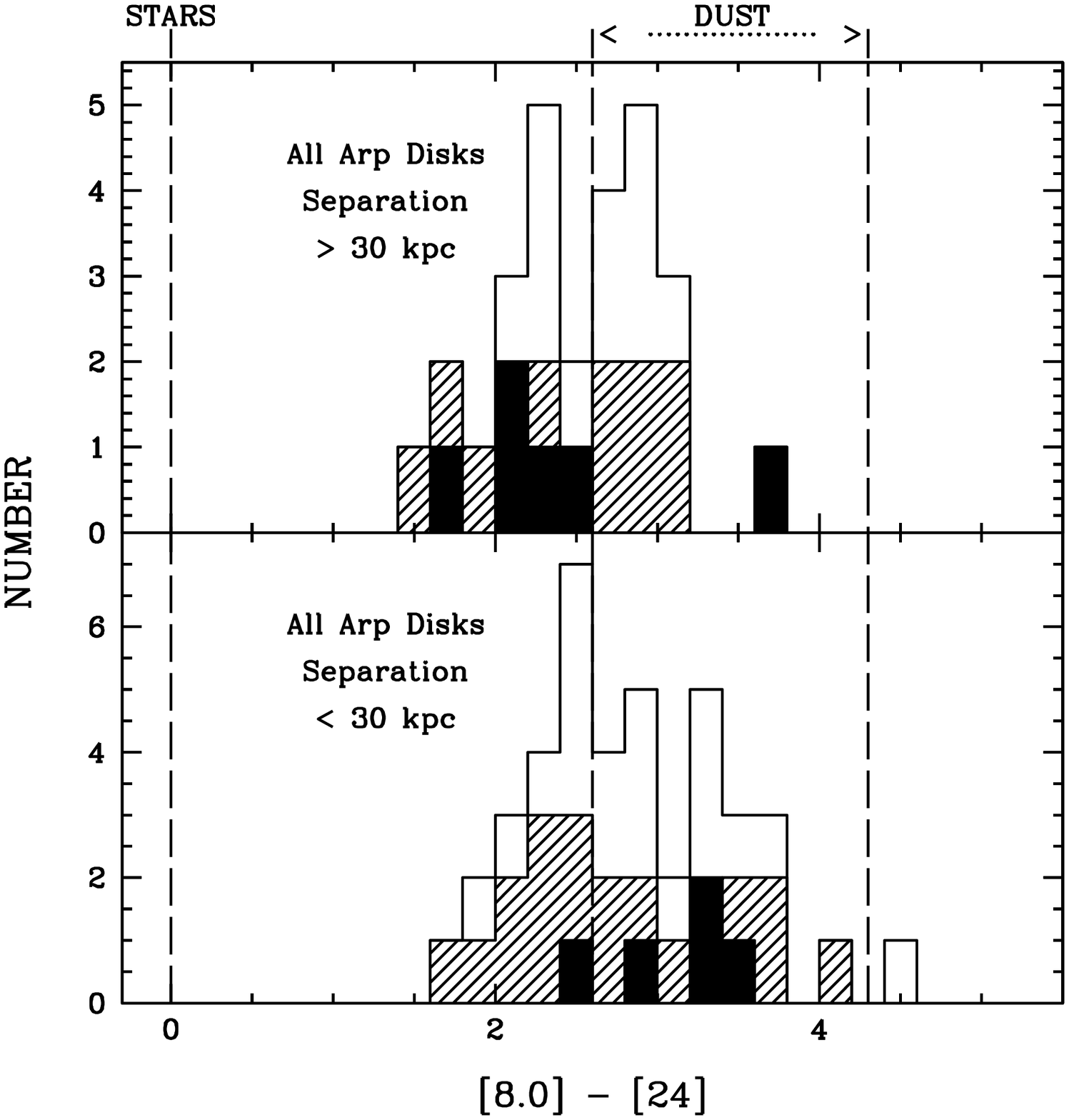}
\caption{
  \small 
Histograms of the [8.0] $-$ [24] color for the close and wide pairs
in our sample.
The hatched regions indicate galaxies are the more massive in the pair.
The filled-in regions indicate upper limits.
}
\end{figure}

 \begin{figure}
\plotone{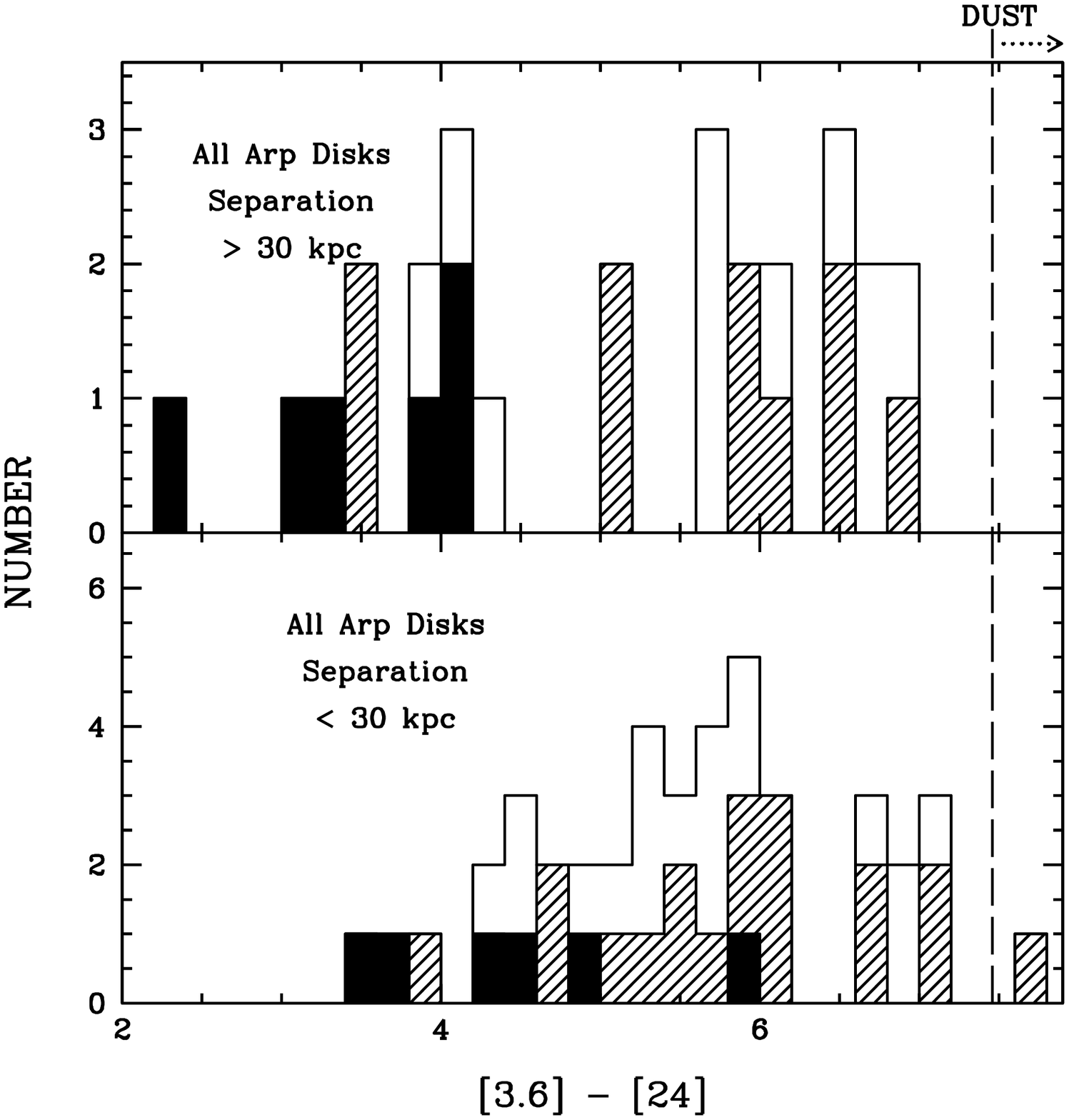}
\caption{
  \small 
Histograms of the [3.0] $-$ [24] color for the close and wide pairs
in our sample.
The hatched regions indicate galaxies are the more massive in the pair.
The filled-in regions indicate upper limits.
}
\end{figure}

 \begin{figure}
\plotone{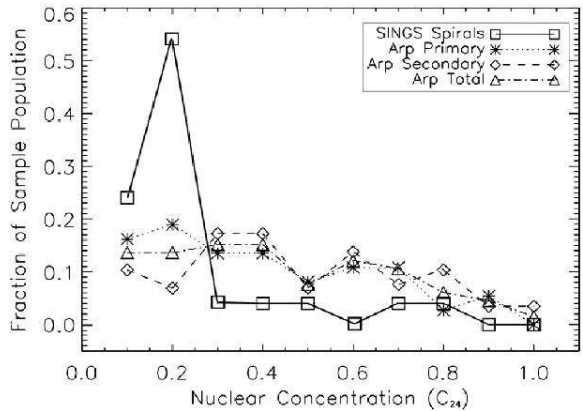}
\caption{
  \small 
The fraction of sample population vs.\ central concentration, for the 
spirals and Arp disks.   See text for more details.
}
\end{figure}

 \begin{figure}
\plotone{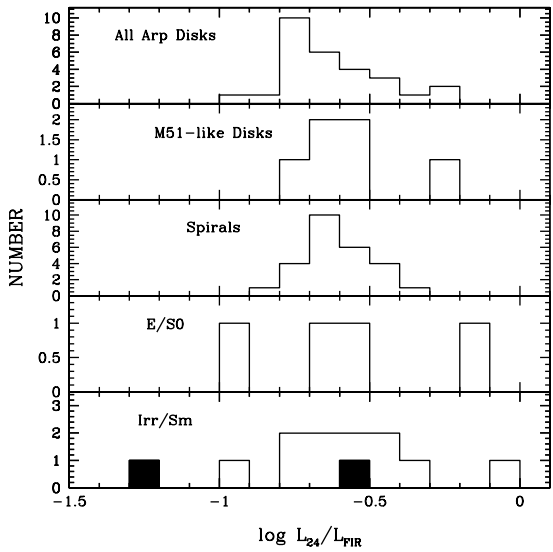}
\caption{
  \small 
Histograms of the L$_{24}$/L$_{FIR}$ ratios for the Arp galaxies (both
disks combined), 
the M51-like systems (both galaxies), the spirals, Irr/Sm, and E/S0 galaxies.
Upper limits are indicated by filled regions.
}
\end{figure}

 \begin{figure}
\plotone{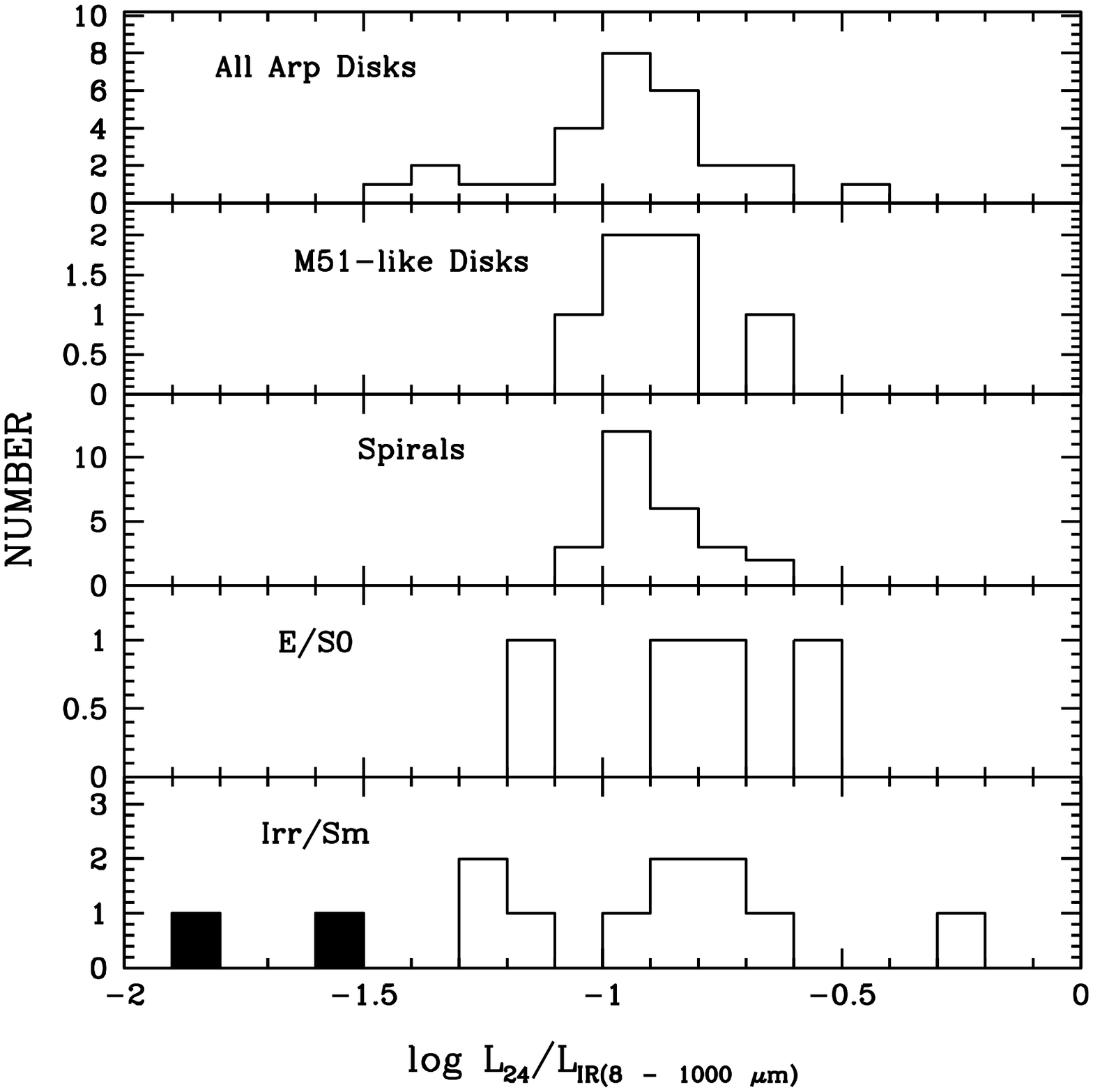}
\caption{
  \small 
Histograms of the L$_{24}$/L$_{IR}$ ratios for the Arp galaxies (both
disks combined), 
the M51-like systems (both galaxies), the spirals, Irr/Sm, and E/S0 galaxies.
The total 8 $-$ 1000 $\mu$m luminosities L$_{IR}$ are calculated
as in \citet{perault87}, \citet{kennicutt98}, and \citet{sanders96}.
Upper limits are indicated by filled regions.
}
\end{figure}



\clearpage

\begin{deluxetable}{ccrrrccrrrrrrrrr}
\rotate
\tabletypesize{\scriptsize}
\def\et#1#2#3{${#1}^{+#2}_{-#3}$}
\tablewidth{0pt}
\tablecaption{Interacting Galaxy Sample\label{tab-1}}
\tablehead{
\multicolumn{1}{c}{System} &
\multicolumn{1}{c}{Other } &
\colhead{Distance} & 
\colhead{Separation} & 
\colhead{LOG L(FIR)$^a$} & 
\colhead{Notes on Morphology} &
\colhead{Nuclear Spectral} 
\\ 
\multicolumn{1}{c}{}
& \multicolumn{1}{c}{Names} 
& \multicolumn{1}{c}{(Mpc)} 
& \multicolumn{1}{c}{(kpc)} 
& \multicolumn{1}{c}{(L$_{\sun}$)} 
& \multicolumn{1}{c}{} 
& \multicolumn{1}{c}{Types} \\
}
\startdata
Arp 24 & NGC 3445 & 29.1 & 9.6 &    9.51 & M51-like, but weak bridge       & HII$^b$\\
Arp 34 & NGC 4613/4/5 & 67.2 & 43.4 &    10.07 & equal mass spirals, short tails, small 3rd galaxy   & $-$\\
Arp 65 & NGC 90/93 & 69.5 & 56.5 &    9.53 & widely separated pair        & $-$\\
Arp 72 & NGC 5994/6 & 46.9 & 21.0 &    10.19 & M51-like          & LINER+HII$^b$\\
Arp 82 & NGC 2535/6 & 57.0& 28.3 &    10.18 & M51-like          & HII/HII$^c$,LINER/HII$^b$\\
Arp 84 & NGC 5394/5 & 50.0 & 28 &    10.66 & M51-like, but bridge from smaller galaxy     & HII/LINER$^{b,c}$\\
Arp 85 & M51,NGC 5194/5 & 6.2 & 8.0 &    9.71 & M51; spiral w/ small companion, bridge     & LINER/HII$^{b,c}$\\
Arp 86 & NGC 7752/3 & 66.5 & 39.9 &    10.66 & M51-like          & HII/LINER$^c$\\
Arp 87 & NGC 3808 & 97.3 & 29.4 &    $-$ & M51-like, but near-equal mass       & LINER/HII$^{b,c}$\\
Arp 89 & NGC 2648 & 31.1 & 21.7 &    $-$ & M51-like          & $-$\\
Arp 104 & NGC 5216/8 & 42.1 & 49.2 &    10.34 & unequal mass spiral pair, long bridge     & no emission/no emission$^b$\\
Arp 107 & UGC 5984 & 142.8 & 46.3 &    $-$ & ring-like spiral w/ small E, bridge,short tail    & Seyfert$^c$\\
Arp 136 & NGC 5820/1 & 47.6 & 50.7 &    8.89 & equal mass, short tail$^d$      & $-$\\
Arp 181 & NGC 3212/5 & 128.9 & 45.5 &    10.60 & two close equal mass spirals, long tail    & $-$\\
Arp 202 & NGC 2719 & 44.0& 4.7 &    9.75 & unequal mass pair        & HII/HII$^c$\\
Arp 205 & NGC 3448/UGC6016 & 23.0& 26.4 &    9.77 & peculiar, tails, small irregular companion      & $-$\\
Arp 240 & NGC 5257/8 & 91.0 & 37.1 &    11.19 & close unequal mass spiral pair, short tail    & HII/HII$^c$,HII+LINER/HII$^b$\\
Arp 242 & NGC 4676 & 91.3 & 16.6 &    10.60 & The Mice; equal mass spiral pair, long tails   & LINER/LINER$^c$,no emission$^b$\\
Arp 244 & NGC 4038/9 & 26.8 & 8.6 &    10.76 & The Antennae; equal mass pair, long tails    & HII/HII$^c$,HII/HII+LINER$^b$\\
Arp 253 & UGC 173/4 & 29.0& 9.8 &    8.68 & two close spirals, short tails      & HII$^b$\\
Arp 271 & NGC 5426/7 & 39.0& 26 &    10.48 & close equal mass spirals, bridge, no tails    & HII/Seyfert$^{b,c}$\\
Arp 279 & NGC 1253 & 22.6 & 26.0 &    9.44 & unequal mass close spiral pair      & HII$^c$\\
Arp 280 & NGC 3769 & 11.5 & 4.6 &    8.85 & unequal mass spirals, short tails      & $-$\\
Arp 282 & NGC 169 & 59.6 & 7.1 &    9.96 & close pair         & no emission/HII+LINER$^c$\\
Arp 283 & NGC 2798/9 & 26.3 & 12.0 &    10.35 & two close spirals; tails+bridge       & HII/HII$^c$,HII/HII+LINER$^b$\\
Arp 284 & NGC 7714/5 & 35.3 & 19.5 &    10.30 & unequal mass pair, partial ring, tails, bridge    & HII/HII$^{b,c}$\\
Arp 285 & NGC 2854/6 & 39.1 & 40.1 &    9.93 & equal mass widely separated spirals      & $-$\\
Arp 290 & IC 195/6 & 47.0& 29.8 &    9.17 & unequal mass separated spirals       & HII/LINER$^c$\\
Arp 293 & NGC 6285/6 & 76.1 & 32.9 &    11.03 & two equal mass separated spirals, tails     & $-$\\
Arp 294 & NGC 3786/8 & 40.2 & 16.2 &    $-$ & equal mass close spiral pair, long tail    & no emission$^{b,c}$\\
Arp 295 & Arp 295 & 87.8 & 119.0 &    10.83 & wide unequal mass pair, long bridge     & LINER/HII$^c$\\
Arp 297N & NGC 5753/5 & 130.8 & 40.0 &    $-$ & spiral with small companion       & HII$^c$\\
Arp 297S & NGC 5752/4 & 63.6 & 35.4 &    10.17 & spiral with small companion       & HII$^c$\\
Arp 298 & NGC 7469/IC5283 & 63.1 & 23.6 &    11.22 & disk galaxies w/ ring, disturbed companion     & Seyfert/HII$^c$\\
NGC 4567 & NGC 4567 & 35.5 & 12.3 &    $-$ & two close spirals, no tails      & $-$\\
\enddata
\tablenotetext{a}{Includes the IRAS flux from both galaxies in the pair.}
\tablenotetext{b}{Nuclear spectral type from \citet{dahari85}.}
\tablenotetext{c}{Nuclear spectral type from \citet{keel85}.}
\tablenotetext{d}{The bright galaxy in the \citet{arp66} 
Atlas photograph of Arp 136 is NGC 5820.  This galaxy has a nearby similar-mass companion NGC 5821 at the same redshift,
outside of the Arp Atlas field of view.  The second small
galaxy in the Arp Atlas image is a background galaxy.}

\end{deluxetable}
%

\clearpage

%
%
\begin{deluxetable}{ccrrrrrrrrrrrrr}
\tabletypesize{\scriptsize}
\def\et#1#2#3{${#1}^{+#2}_{-#3}$}
\tablewidth{0pt}
\tablecaption{Normal Galaxy Sample\label{tab-2}}
\tablehead{
\multicolumn{1}{c}{Name} &
\multicolumn{1}{c}{Type} &
\colhead{Distance} &
\colhead{LOG L(FIR)} 
\\
\multicolumn{1}{c}{}
& \multicolumn{1}{c}{} 
& \multicolumn{1}{c}{(Mpc)} 
& \multicolumn{1}{c}{(L$_{\sun}$)} 
\\
}
\startdata
\multicolumn{4}{c}{Spiral Galaxies}
\\ 
\hline
NGC 24 & SAc  & 8.2 &    8.26 \\
NGC 337 & SBd  & 24.7 &    10.02 \\
NGC 628 & SAc  & 11.4 &    9.75 \\
NGC 925 & SABd  & 10.1 &    9.22 \\
NGC 1097 & SBb  & 16.9 &    10.43 \\
NGC 1291 & SBa  & 9.7 &    8.61 \\
NGC 2403 & SABcd  & 3.5 &    9.00 \\
NGC 2841 & SAb  & 9.8 &    9.00 \\
NGC 3049 & SBab  & 19.6 &    9.24 \\
NGC 3184 & SABcd  & 8.6 &    9.17 \\
NGC 3521 & SABbc  & 9.0 &    9.90 \\
NGC 3621 & Sad  & 6.2 &    9.36 \\
NGC 3938 & SAc  & 12.2 &    9.48 \\
NGC 4254 & SAc  & 20.0 &    10.47 \\
NGC 4321 & SABbc  & 20.0 &    10.33 \\
NGC 4450 & SAab  & 20.0 &    9.21 \\
NGC 4559 & SABcd  & 11.6 &    9.44 \\
NGC 4579 & SABb  & 20.0 &    9.76 \\
NGC 4594 & SAa  & 13.7 &    9.27 \\
NGC 4725 & SABab  & 17.1 &    9.56 \\
NGC 4736 & SAab  & 5.3 &    9.53 \\
NGC 4826 & SAab  & 5.6 &    9.34 \\
NGC 5055 & SAbc  & 8.2 &    9.81 \\
NGC 6946 & SABcd  & 5.5 &    9.87 \\
NGC 7331 & SAb  & 15.7 &    10.34 \\
NGC 7793 & SAd  & 3.2 &    8.61 \\
\hline
\multicolumn{4}{c}{Elliptical/S0 Galaxies}
\\ 
\hline
NGC 855 & E  & 9.6 &    8.42 \\
NGC 1377 & S0  & 24.4 &    9.77 \\
NGC 3773 & SA0  & 12.9 &    8.60 \\
NGC 4125 & E6p  & 21.4 &    8.80 \\
\hline
\multicolumn{4}{c}{Irregular/Sm Galaxies}
\\ 
\hline
DDO 53 & Im  & 3.5 &    6.77 \\
DDO 154 & IBm  & 5.4 & $<$6.72 \\
DDO 165 & Im  & 3.5 & $<$6.35 \\
Holmberg II & Im  & 3.5 &    7.43 \\
IC 4710 & SBm  & 8.5 &    8.16 \\
Markarian 33 & Im  & 21.7 &    9.53 \\
NGC 1705 & Am  & 5.8 &    7.83 \\
NGC 2915 & I0  & 2.7 &    7.06 \\
NGC 4236 & SBdm  & 3.5 &    7.99 \\
NGC 5398 & SBdm  & 15.0 &    8.77 \\
NGC 5408 & IBm  & 4.5 &    7.91 \\
NGC 6822 & IBm  & 0.6 &    6.65 \\
\enddata
\end{deluxetable}
%

%
%
\begin{deluxetable}{c|r|rrr|rrr|rrr|rrr|rrr|}
\tabletypesize{\scriptsize}
\def\et#1#2#3{${#1}^{+#2}_{-#3}$}
\tablewidth{0pt}
\tablecaption{Statistics on IRAS Far-Infrared Luminosities\label{tab-3}}
\tablehead{
\multicolumn{1}{c}{Type} &
\multicolumn{1}{c}{Number} &
\multicolumn{3}{c}{log L(FIR)$^a$ } &
\multicolumn{3}{c}{log L(IR)$^b$ } \\
\multicolumn{1}{c}{}
& \multicolumn{1}{c}{}
& \multicolumn{1}{c}{median} 
& \multicolumn{1}{c}{mean} 
& \multicolumn{1}{c}{rms} 
& \multicolumn{1}{c}{median} 
& \multicolumn{1}{c}{mean} 
& \multicolumn{1}{c}{rms} 
\\
}
\startdata
Arp Galaxies$^c$ & 29 &  10.18 &  10.10 &   0.69 &  10.48 &  10.43 &   0.64 \\
M51-like Disks$^c$ &  7 &  10.18 &  10.14 &   0.43 &  10.43 &  10.43 &   0.44 \\
   Spirals & 26 &   9.44 &   9.50 &   0.57 &   9.72 &   9.78 &   0.57 \\
      E/S0 &  4 &   8.60 &   8.90 &   0.60 &   8.84 &   9.17 &   0.68 \\
    Irr/Sm & 12 &   7.43 &   7.60 &   0.95 &   7.72 &   7.87 &   1.00 \\
\enddata
\tablenotetext{a}{Far-infrared (40 $-$ 120 $\mu$m)
luminosities calculated as in \citet{lonsdale85}
and \citet{persson87}.}
\tablenotetext{b}{Total 8 $-$ 1000 $\mu$m luminosity,
calculated as in \citet{perault87}, \citet{sanders88}, and 
\citet{kennicutt98}.}
\tablenotetext{c}{The luminosities for the Arp galaxies include
the combined fluxes from both galaxies in the pair.}
\end{deluxetable}
%

\begin{deluxetable}{lrrccrrrrrrr}
  \tabletypesize{\scriptsize}
  \setlength{\tabcolsep}{0.01in}
  \tablecaption{IRAC/MIPS24 Observational Parameters\label{tab4}}
  \tablewidth{0pc}
  \tablehead{
    \colhead{Galaxy} & \colhead{RA} & \colhead{Dec} & 
    \multicolumn{2}{c}{IRAC} &  \multicolumn{2}{c}{MIPS24} &\multicolumn{5}{c}{On-source Time (sec)}  \\
    \colhead{} & \colhead{(J2000)} & \colhead{(J2000)} & \colhead{Obs. Date} & \colhead{AORKEY} &
	 \colhead{Obs. Date} & \colhead{AORKEY} & \colhead{3.6$\mu$m} &  \colhead{4.5$\mu$m} & \colhead{5.8$\mu$m} & \colhead{8.0$\mu$m} & \colhead{24$\mu$m}\\
  }
\startdata

Arp 65   &   0:21:59.00 &  22:24:26.0 & 2004-12-14 & 10532352 & 2004-12-24 & 10540544 &  48          & 48	& 48	& 48	& 312  \\   
Arp 282  &   0:36:48.70 &  23:58:47.5 & 2004-12-16 & 10529536 & 2004-12-24 & 10539008 &  276 	& 276	& 276	& 276	& 312  \\    
Arp 290  &   2:03:48.00 &  14:43:34.0 & 2005-01-15 & 10533376 & 2005-01-26 & 10542080 &  48  	& 48	& 48	& 48	& 312  \\    
Arp 279  &   3:14:16.10 &  -2:48:42.0 & 2005-01-17 & 10529024 & 2005-01-31 & 10538496 &  48  	& \nodata 	& 48	& \nodata	& 312  \\    
         &              &              & 2005-01-17 & 10531072 &         &             & \nodata 	& 48	& 	& 48	& \nodata  \\    
Arp 82   &   8:11:13.70 &  25:12:10.0 & 2004-11-01 & 10526464 & 2005-04-02 & 10534656 & 72  	& \nodata	& 72	& \nodata	& 312  \\    
         &              &              & 2004-11-01 & 10532096 &   & & \nodata 	& 72	& \nodata & 72	& \nodata  \\    
Arp 89   &   8:42:39.81 &  14:17:08.0 & 2005-05-06 & 10527232 & 2004-11-05 & 10536704 &  24  	& 	& 24	& 24	& 312  \\    
         &              &              & 2005-05-06 & 10531840 &  & & \nodata  	& 24	& \nodata & 24	& \nodata  \\    
Arp 202  &   9:00:15.64 &  35:43:25.5 & 2005-05-10 & 10528256 & 2004-11-02 & 10537728 &   276 	& 276	& 276	& 276	& 312  \\    
Arp 253  &   9:43:25.00 &  -5:16:52.0 & 2004-12-17 & 10528768 & 2004-11-06 & 10541824 &   48  	& 48	& 48	& 48	& 312  \\    
Arp 285  &   9:24:09.60 &  49:13:34.0 & 2005-05-06 & 10533120 & 2004-12-04 & 10538240 &   276 	& 276	& 276	& 276	& 312  \\    
Arp 181  &  10:28:26.40 &  79:49:18.3 & 2004-12-16 & 10528000 & 2004-11-09 & 10537472 &   276 	& 276	& 276	& 276	& 312  \\    
Arp 107  &  10:52:16.70 &  30:03:55.0 & 2004-12-17 & 10527488 & 2004-12-03 & 10536960 &   276 	& 276	& 276	& 276	& 312  \\    
Arp 205  &  10:54:32.60 &  54:17:54.0 & 2004-12-16 & 10531328 & 2004-11-10 & 10540800 &   48  	& 48	& 48	& 48	& 312  \\    
Arp 24   &  10:54:40.50 &  56:59:04.4 & 2004-12-16 & 10525952 & 2004-11-10 & 10535168 &   276 	& 276	& 276	& 276	& 312  \\    
Arp 280  &  11:37:47.30 &  47:53:10.0 & 2004-12-17 & 10529280 & 2004-12-03 & 10538752 &   276 	& 276	& 276	& 276	& 312  \\    
Arp 294  &  11:39:43.59 &  31:55:12.4 & 2004-12-17 & 10529792 & 2005-05-12 & 10539264 &   276 	& 276	& 276	& 276	& 312  \\    
Arp 87   &  11:40:44.40 &  22:26:16.0 & 2004-12-18 & 10526976 & 2005-05-18 & 10536448 &   276 	& 276	& 276	& 276	& 312  \\    
NGC 4567 &  12:36:32.70 &  11:15:28.3 & 2005-06-10 & 10530816 & 2005-01-28 & 10540032 &   84  	& \nodata & 84	& \nodata & 312  \\    
         &              &              & 2005-06-10 & 10530560 &    & &\nodata 	& 84	& \nodata & 84	&   \nodata\\
Arp 34   &  12:41:34.00 &  26: 3:31.0 & 2005-06-09 & 10532608 & 2005-01-27 & 10535424 &  108 	& 108	& 108	& 108	& 312  \\    
Arp 104  &  13:32:08.90 &  62:44:02.0 & 2004-11-25 & 10531584 & 2004-12-02 & 10540288 &  48  	& 48	& 48	& 48	& 312  \\    
Arp 240  &  13:39:55.20 &   0:50:13.0 &  &   & 2005-02-02 & 10537984 & \nodata & \nodata &\nodata &\nodata & 312  \\
Arp 84   &  13:58:35.80 &  37:26:20.0 &  &   & 2005-01-25 & 10541312 & \nodata & \nodata &\nodata &\nodata & 312  \\
Arp 271  &  14:03:25.50 &  -6:02:60.0 & 2005-07-21 & 10532864 & 2005-08-02 & 10541568 &  36  	& 36	& 36	& 36	& 312  \\    
Arp 297  &  14:45:19.40 &  38:45:10.8 & 2005-01-21 & 10530304 & 2005-01-29 & 10539776 &  276 	& 276	& 276	& 276	& 312  \\    
Arp 136  &  14:58:39.82 &  53:53:09.9 & 2004-12-21 & 10527744 & 2004-12-26 & 10537216 &  48  	& 48	& 48	& 48	& 312  \\    
Arp 72   &  15:46:56.03 &  17:52:43.2 & 2005-03-27 & 10526208 & 2005-03-10 & 10534912 &  276 	& 276	& 276	& 276	& 312  \\    
Arp 295  &  23:41:54.10 &  -3:38:29.0 & 2004-11-26 & 10530048 & 2004-11-29 & 10539520 &  48  	& 48	& 48	& 48	& 312  \\    
Arp 86   &  23:47:01.90 &  29:28:23.7 &  &   & 2004-12-06 & 10536192 & \nodata & \nodata &\nodata &\nodata & 312  \\

\enddata

\end{deluxetable}

\clearpage

%
%
\begin{deluxetable}{ccccccccccccccc}
\tabletypesize{\scriptsize}
\def\et#1#2#3{${#1}^{+#2}_{-#3}$}
\tablewidth{0pt}
\tablecaption{Spitzer Flux Densities for Interacting Galaxy Sample\label{tab-5}}
\tablehead{
\multicolumn{1}{c}{Arp} &
\multicolumn{1}{c}{Component} &
\multicolumn{1}{c}{Other} &
\colhead{F$_{3.6 {\mu}m}$} & 
\colhead{F$_{4.5 {\mu}m}$} & 
\colhead{F$_{5.8 {\mu}m}$} & 
\colhead{F$_{8.0 {\mu}m}$} & 
\colhead{F$_{24 {\mu}m}$} 
\\ 
\multicolumn{1}{c}{Name} &
\multicolumn{1}{c}{}&
\multicolumn{1}{c}{Name} 
& \multicolumn{1}{c}{(mJy)} 
& \multicolumn{1}{c}{(mJy)} 
& \multicolumn{1}{c}{(mJy)} 
& \multicolumn{1}{c}{(mJy)} 
& \multicolumn{1}{c}{(mJy)} 
\\ 
}
\startdata
  Arp 24 &        E &      UGC     6021 &       1.16 $\pm$     0.02 &       0.77 $\pm$     0.04 &  $\le$ 2.0     &       1.1 $\pm$     0.1 &       1.6 $\pm$     0.4   \\
  Arp 24 &        MAIN &      NGC     3445 &      28.3 $\pm$     0.2 &      18.6 $\pm$     0.4 &      44.5 $\pm$     7.3 &     109.4 $\pm$     1.4 &     151.9 $\pm$     4.7   \\
  Arp 34 &       NE &      NGC     4615 &      12.49 $\pm$     0.04 &       8.14 $\pm$     0.04 &      21.7 $\pm$     0.6 &      59.5 $\pm$     0.2 &      58.8 $\pm$     2.0   \\
  Arp 34 &    NE E TAIL &      NGC 4615 EAST TAIL &       3.54 $\pm$     0.01 &       2.44 $\pm$     0.01 &       7.5 $\pm$     0.2 &      21.23 $\pm$     0.06 &      38.5 $\pm$     0.7   \\
  Arp 34 &    NE W TAIL &      NGC 4615 WEST TAIL &       2.36 $\pm$     0.01 &       1.54 $\pm$     0.01 &       4.3 $\pm$     0.2 &      11.88 $\pm$     0.05 &      11.8 $\pm$     0.5   \\
  Arp 34 & NW SMALL &      NGC     4613 &       6.49 $\pm$     0.03 &       4.04 $\pm$     0.03 &       6.7 $\pm$     0.4 &      15.4 $\pm$     0.1 &      19.8 $\pm$     1.3   \\
  Arp 34 &        S &      NGC     4614 &      27.8 $\pm$     0.1 &      18.0 $\pm$     0.1 &      19.1 $\pm$     1.7 &      20.5 $\pm$     0.5 &      17.3 $\pm$     5.5   \\
  Arp 65 &        E &      NGC       93 &      61.5 $\pm$     0.2 &      36.5 $\pm$     0.1 &      34.7 $\pm$     1.9 &      46.4 $\pm$     1.1 &  $\le$25.9       \\
  Arp 65 &       N TAIL &      NGC      90 TAIL &       0.74 $\pm$     0.07 &       0.45 $\pm$     0.05 &  $\le$ 1.9     &       1.7 $\pm$     0.4 &  $\le$ 9.6       \\
  Arp 65 &       S TAIL &      NGC      90 TAIL &       0.73 $\pm$     0.05 &       0.47 $\pm$     0.03 &  $\le$ 1.4     &       1.8 $\pm$     0.3 &  $\le$ 6.9       \\
  Arp 65 &        W &      NGC       90 &      20.5 $\pm$     0.2 &      12.4 $\pm$     0.1 &      12.0 $\pm$     1.7 &      19.5 $\pm$     0.9 &  $\le$23.2       \\
  Arp 72 &         BRIDGE &      NGC  5994/6 BRIDGE &       6.9 $\pm$     0.2 &       4.5 $\pm$     0.2 &      11.6 $\pm$     1.6 &      30.7 $\pm$     0.6 &      42.6 $\pm$     1.6   \\
  Arp 72 &        E &      NGC     5996 &      30.8 $\pm$     0.1 &      20.8 $\pm$     0.1 &      71.2 $\pm$     1.2 &     199.0 $\pm$     0.4 &     364.5 $\pm$     1.0   \\
  Arp 72 &       E TAIL &      NGC    5996 TAIL &       4.6 $\pm$     0.1 &       3.1 $\pm$     0.1 &       8.1 $\pm$     1.2 &      13.9 $\pm$     0.4 &      38.2 $\pm$     1.2   \\
  Arp 72 &        W &      NGC     5994 &       2.45 $\pm$     0.08 &       1.65 $\pm$     0.06 &       2.2 $\pm$     0.7 &       5.1 $\pm$     0.2 &      11.5 $\pm$     0.7   \\
  Arp 82 &         BRIDGE &      NGC  2535/6 BRIDGE &       0.83 $\pm$     0.07 &       0.51 $\pm$     0.06 &  $\le$ 4.7     &       3.7 $\pm$     0.4 &  $\le$ 5.3       \\
  Arp 82 &        N &      NGC     2535 &      43.7 $\pm$     0.6 &      28.5 $\pm$     0.5 &      78.1 $\pm$    13.1 &     220.2 $\pm$     3.2 &     261.6 $\pm$    13.4   \\
  Arp 82 &        S &      NGC     2536 &      10.6 $\pm$     0.2 &       6.9 $\pm$     0.2 &      17.1 $\pm$     4.0 &      46.9 $\pm$     1.0 &      43.0 $\pm$     4.2   \\
  Arp 84 &        N &      NGC     5394 &      46.2 $\pm$     0.3 &      30.9 $\pm$     0.3 &     102.0 $\pm$     2.7 &     279.8 $\pm$     3.8 &     633.4 $\pm$    12.8   \\
  Arp 84 &       N TAIL &      NGC    5394 TAIL &       1.86 $\pm$     0.09 &       1.2 $\pm$     0.1 &  $\le$ 2.7     &  $\le$ 3.9     &  $\le$13.4       \\
  Arp 84 &        S &      NGC     5395 &     154.4 $\pm$     0.9 &      97.0 $\pm$     1.1 &     221.2 $\pm$     8.9 &     548.1 $\pm$    12.4 &     434.2 $\pm$    40.2   \\
  Arp 85 &        N &      NGC     5195 &    1077.8 $\pm$     3.1 &     676.9 $\pm$     5.6 &     848.6 $\pm$   136.5 &    1351.6 $\pm$    75.6 &    1505.6 $\pm$    89.2   \\
  Arp 85 &        S &      NGC     5194 &    2707.4 $\pm$     4.4 &    1732.0 $\pm$     7.9 &    5403.8 $\pm$   191.8 &   14023.6 $\pm$   106.2 &   16970.6 $\pm$   125.1   \\
  Arp 86 &         BRIDGE &      NGC  7752/3 BRIDGE &       5.45 $\pm$     0.08 &       3.59 $\pm$     0.04 &       8.2 $\pm$     0.8 &      19.9 $\pm$     0.5 &      20.2 $\pm$     3.3   \\
  Arp 86 &        N &      NGC     7753 &     112.9 $\pm$     0.5 &      70.2 $\pm$     0.3 &     140.8 $\pm$     5.3 &     362.1 $\pm$     3.5 &     338.6 $\pm$    19.5   \\
  Arp 86 &        S &      NGC     7752 &      22.5 $\pm$     0.1 &      15.37 $\pm$     0.06 &      61.5 $\pm$     1.2 &     176.5 $\pm$     0.7 &     269.0 $\pm$     4.2   \\
  Arp 87 &         BRIDGE &      NGC    3808 BRIDGE &       1.47 $\pm$     0.02 &       0.93 $\pm$     0.03 &       2.4 $\pm$     0.4 &       5.5 $\pm$     0.2 &       0.0       \\
  Arp 87 &        N &      NGC   3808B &      14.88 $\pm$     0.04 &      10.44 $\pm$     0.08 &      36.0 $\pm$     1.1 &     111.4 $\pm$     0.4 &     169.7 $\pm$     2.0   \\
  Arp 87 &        S &      NGC    3808A &      15.02 $\pm$     0.05 &       9.70 $\pm$     0.09 &      21.9 $\pm$     1.3 &      64.7 $\pm$     0.5 &      58.8 $\pm$     2.5   \\
  Arp 87 &  SMALL N &                &       0.32 $\pm$     0.01 &       0.20 $\pm$     0.02 &  $\le$ 0.7     &       1.20 $\pm$     0.09 &       2.0 $\pm$     0.6   \\
  Arp 87 &       S TAIL &      NGC   3808A TAIL &       0.37 $\pm$     0.01 &       0.25 $\pm$     0.02 &  $\le$ 1.0     &       0.9 $\pm$     0.1 &  $\le$ 0.0       \\
  Arp 89 &        E &      KPG      168 &       7.74 $\pm$     0.09 &       5.0 $\pm$     0.1 &       9.1 $\pm$     2.9 &      23.1 $\pm$     0.4 &      40.7 $\pm$     3.7   \\
  Arp 89 &       E TAIL &                   &       0.77 $\pm$     0.05 &       0.44 $\pm$     0.09 &  $\le$ 5.3     &  $\le$ 0.64    0 &  $\le$ 5.4       \\
  Arp 89 &        W &      NGC     2648 &     130.8 $\pm$     0.9 &      78.4 $\pm$     1.4 &  $\le$81.8     &      48.6 $\pm$     3.5 &  $\le$83.5       \\
  Arp 104 &         BRIDGE &      NGC  5216/8 BRIDGE &       4.2 $\pm$     0.4 &       3.2 $\pm$     0.4 &  $\le$18.2     &       4.0 $\pm$     1.1 &    $-$   \\
  Arp 104 &        N &      NGC     5218 &      86.4 $\pm$     0.9 &      56.8 $\pm$     0.9 &     122.7 $\pm$    15.3 &     295.7 $\pm$     2.7 &     604.9 $\pm$     8.1   \\
  Arp 104 &       N TAIL &      NGC    5218 TAIL &       2.1 $\pm$     0.3 &       1.6 $\pm$     0.3 &  $\le$13.0     &  $\le$ 2.3     &    $-$   \\
  Arp 104 &        S &      NGC     5216 &      47.0 $\pm$     0.9 &      28.7 $\pm$     0.9 &  $\le$46.2     &      22.8 $\pm$     2.7 &  $\le$23.2       \\
  Arp 107 &         BRIDGE &      UGC    5984 BRIDGE &       1.39 $\pm$     0.02 &       0.84 $\pm$     0.04 &  $\le$ 1.6     &       1.36 $\pm$     0.05 &  $\le$ 4.1       \\
  Arp 107 &        N &      UGC    5984N &      10.29 $\pm$     0.06 &       6.2 $\pm$     0.1 &  $\le$ 4.5     &       3.0 $\pm$     0.1 &  $\le$10.4       \\
  Arp 107 &       N TAIL &      UGC   5984N TAIL &       1.23 $\pm$     0.04 &       0.84 $\pm$     0.07 &  $\le$ 2.6     &       0.38 $\pm$     0.07 &  $\le$ 6.2       \\
  Arp 107 &        S &      UGC    5984S &      17.9 $\pm$     0.1 &      11.3 $\pm$     0.2 &  $\le$ 9.7     &      31.5 $\pm$     0.2 &      50.8 $\pm$     7.3   \\
  Arp 107 &       W TAIL &      UGC   5984S TAIL &       0.71 $\pm$     0.02 &       0.42 $\pm$     0.04 &  $\le$ 1.5     &       0.51 $\pm$     0.05 &  $\le$ 3.8       \\
  Arp 136 &       NW &      NGC     5821 &      15.6 $\pm$     0.2 & $-$ &      30.2 $\pm$     7.3 & $-$  &      15.6 $\pm$     2.0   \\
  Arp 136 &       SW &      NGC     5820 &      83.3 $\pm$     0.5 &      49.0 $\pm$     0.6 &  $\le$53.1     &      25.5 $\pm$     1.0 &  $\le$18.1       \\
  Arp 181 &        E &      NGC     3215 &      29.1 $\pm$     0.2 &      17.8 $\pm$     0.3 &  $\le$20.9     &      49.5 $\pm$     0.5 &      32.3 $\pm$     2.2   \\
  Arp 181 &        W &      NGC     3212 &      20.7 $\pm$     0.1 &      13.4 $\pm$     0.2 &      23.8 $\pm$     5.3 &      77.1 $\pm$     0.4 &      96.8 $\pm$     1.8   \\
  Arp 181 &       W TAIL &      NGC    3212 TAIL &       1.2 $\pm$     0.1 &       0.6 $\pm$     0.1 &  $\le$10.8     &       1.4 $\pm$     0.3 &    $-$   \\
  Arp 202 &        N &      NGC     2719 &       9.05 $\pm$     0.05 &       6.06 $\pm$     0.07 &       8.6 $\pm$     2.0 &      22.4 $\pm$     0.2 &      62.8 $\pm$     4.3   \\
  Arp 202 &        S &      NGC    2719A &       3.88 $\pm$     0.04 &       2.69 $\pm$     0.06 &  $\le$ 4.8     &       9.9 $\pm$     0.1 &      67.6 $\pm$     3.1   \\
  Arp 205 &         BRIDGE &      NGC 3448/UGC6016 BRIDGE &       1.8 $\pm$     0.2 &       1.3 $\pm$     0.2 &  $\le$10.0     &       3.0 $\pm$     0.6 &  $\le$ 4.7       \\
  Arp 205 &        E &      NGC     3448 &      74.6 $\pm$     0.3 &      50.0 $\pm$     0.4 &     119.3 $\pm$     7.1 &     264.8 $\pm$     1.2 &     941.2 $\pm$     3.1   \\
  Arp 205 &       E TAIL &      NGC    3448 TAIL &       0.8 $\pm$     0.1 &       0.6 $\pm$     0.2 &  $\le$ 8.5     &  $\le$ 1.5     &  $\le$ 3.8       \\
  Arp 205 &        W &      UGC     6016 &       2.0 $\pm$     0.2 &       1.4 $\pm$     0.2 &  $\le$ 9.8     &  $\le$ 1.7     &  $\le$ 4.3       \\
  Arp 240 &         BRIDGE &      NGC  5257/8 BRIDGE &       3.26 $\pm$     0.07 &       2.1 $\pm$     0.1 &  $\le$14.7     &      18.0 $\pm$     0.6 &      65.7 $\pm$     6.4   \\
  Arp 240 &        E &      NGC     5258 &      52.55 $\pm$     0.06 &      35.53 $\pm$     0.09 &     111.9 $\pm$     4.6 &     345.8 $\pm$     0.5 &     503.2 $\pm$     5.9   \\
  Arp 240 &       E TAIL &      NGC    5258 TAIL &       2.71 $\pm$     0.05 &       1.72 $\pm$     0.08 &  $\le$10.4     &      12.2 $\pm$     0.4 &  $\le$13.9       \\
  Arp 240 &        W &      NGC     5257 &      46.62 $\pm$     0.08 &      31.5 $\pm$     0.1 &     106.8 $\pm$     5.6 &     330.3 $\pm$     0.6 &     572.4 $\pm$     7.5   \\
  Arp 240 &       W TAIL &      NGC    5257 TAIL &       0.90 $\pm$     0.02 &       0.57 $\pm$     0.03 &  $\le$ 4.6     &       3.0 $\pm$     0.2 &  $\le$ 6.6       \\
  Arp 242 &        N &      NGC    4676A &      28.6 $\pm$     0.2 &      19.3 $\pm$     0.1 &      45.4 $\pm$     7.9 &     116.6 $\pm$     0.5 &     187.8 $\pm$     6.2   \\
  Arp 242 &       N TAIL &      NGC   4676A TAIL &       6.1 $\pm$     0.2 &       3.8 $\pm$     0.1 &  $\le$18.7     &       9.6 $\pm$     0.4 &    $-$   \\
  Arp 242 &        S &      NGC    4676 BRIDGE &      20.3 $\pm$     0.2 &      12.5 $\pm$     0.1 &  $\le$20.7     &      25.1 $\pm$     0.4 &      32.0 $\pm$     5.3   \\
  Arp 242 &       S TAIL &      NGC   4676 BRIDGE TAIL &       2.3 $\pm$     0.1 &       1.31 $\pm$     0.09 &  $\le$14.7     &       1.9 $\pm$     0.3 &  $\le$ 0.0       \\
  Arp 244 &        N &      NGC     4038 &     299.6 $\pm$     1.2 &     191.7 $\pm$     1.6 &     609.1 $\pm$    45.3 &    1523.6 $\pm$    29.4 &    1948.5 $\pm$    26.6   \\
  Arp 244 &       N TAIL &      NGC    4038 TAIL &       8.7 $\pm$     0.5 &       5.9 $\pm$     0.7 &  $\le$55.0     &  $\le$36.2     &    $-$   \\
  Arp 244 &        S &      NGC     4039 &     291.4 $\pm$     1.5 &     198.9 $\pm$     2.0 &     520.8 $\pm$    54.9 &    1255.6 $\pm$    36.4 &    4108.6 $\pm$    31.8   \\
  Arp 244 &       S TAIL &      NGC    4039 TAIL &       7.5 $\pm$     1.3 &  $\le$ 5.14    0 &  $\le$143.4     &  $\le$92.7     &  $\le$ 0.0       \\
  Arp 253 &        E &     UGCA      174 &       2.38 $\pm$     0.05 &       1.57 $\pm$     0.07 &  $\le$ 3.0     &       4.5 $\pm$     0.3 &       7.9 $\pm$     2.1   \\
  Arp 253 &        W &     UGCA      173 &       3.11 $\pm$     0.05 &       2.10 $\pm$     0.06 &  $\le$ 2.8     &       6.7 $\pm$     0.2 &      16.8 $\pm$     1.8   \\
  Arp 271 &        N &      NGC     5427 &     155.8 $\pm$     0.9 &     103.4 $\pm$     0.7 &     283.6 $\pm$     6.8 &     749.6 $\pm$     3.6 &     734.0 $\pm$    17.4   \\
  Arp 271 &        S &      NGC     5426 &      84.1 $\pm$     0.4 &      54.3 $\pm$     0.3 &     160.2 $\pm$     2.8 &     412.0 $\pm$     1.5 &     297.6 $\pm$     7.2   \\
  Arp 279 &       NE &      NGC    1253A &      10.6 $\pm$     0.4 &       7.1 $\pm$     0.2 &  $\le$41.8     &      21.8 $\pm$     2.7 &  $\le$62.2       \\
  Arp 279 &       SW &      NGC     1253 &     108.8 $\pm$     1.0 &      71.5 $\pm$     0.5 &     110.8 $\pm$    32.4 &     285.0 $\pm$     6.4 &     213.6 $\pm$    50.4   \\
  Arp 280 &        E &      NGC    3769A &       4.58 $\pm$     0.06 &       3.09 $\pm$     0.07 &  $\le$ 3.9     &       5.6 $\pm$     0.3 &      10.9 $\pm$     1.7   \\
  Arp 280 &        W &      NGC     3769 &      90.5 $\pm$     0.3 &      57.9 $\pm$     0.3 &     117.1 $\pm$     6.1 &     277.2 $\pm$     1.2 &     186.1 $\pm$     7.5   \\
  Arp 282 &       E TAIL &      NGC     169 TAIL &       4.1 $\pm$     0.3 &       2.7 $\pm$     0.2 &  $\le$ 3.6     &       7.9 $\pm$     0.7 &  $\le$14.0       \\
  Arp 282 &        N &      NGC      169 &      73.2 $\pm$     0.5 &      45.1 $\pm$     0.4 &      58.7 $\pm$     2.2 &     129.1 $\pm$     1.3 &      74.1 $\pm$     7.7   \\
  Arp 282 &        S &      NGC     169A &      12.5 $\pm$     0.2 &       7.8 $\pm$     0.1 &      13.3 $\pm$     0.7 &      31.5 $\pm$     0.4 &      32.9 $\pm$     2.6   \\
  Arp 283 &         BRIDGE &      NGC  2798/9 BRIDGE &       1.15 $\pm$     0.03 &       0.72 $\pm$     0.03 &       2.4 $\pm$     0.7 &       5.5 $\pm$     0.2 &       0.0       \\
  Arp 283 &        E &      NGC     2799 &      14.34 $\pm$     0.08 &       9.09 $\pm$     0.09 &      23.8 $\pm$     1.8 &      50.4 $\pm$     0.4 &      52.2 $\pm$     4.3   \\
  Arp 283 &       N TAIL &      NGC 2798 NORTH TAIL &       5.35 $\pm$     0.05 &       3.18 $\pm$     0.06 &       4.6 $\pm$     1.3 &       4.6 $\pm$     0.3 &       0.0       \\
  Arp 283 &       S TAIL &      NGC 2798 SOUTH TAIL &       2.21 $\pm$     0.06 &       1.27 $\pm$     0.06 &  $\le$ 4.1     &       3.8 $\pm$     0.3 &       0.0       \\
  Arp 283 &        W &      NGC     2798 &     111.83 $\pm$     0.08 &      77.01 $\pm$     0.09 &     288.9 $\pm$     1.8 &     742.8 $\pm$     0.4 &    2136.5 $\pm$     4.5   \\
  Arp 284 &         BRIDGE &      NGC  7714/5 BRIDGE &       3.7 $\pm$     0.3 &       2.2 $\pm$     0.6 &  $\le$11.1     &       5.7 $\pm$     1.3 &    $-$   \\
  Arp 284 &        E &      NGC     7715 &       3.77 $\pm$     0.10 &       2.3 $\pm$     0.2 &  $\le$ 3.2     &       2.2 $\pm$     0.4 &  $\le$ 5.6       \\
  Arp 284 &       E TAIL &      NGC    7715 TAIL &       1.5 $\pm$     0.1 &       0.8 $\pm$     0.2 &  $\le$ 2.9     &  $\le$ 1.1     &    $-$   \\
  Arp 284 &        W &      NGC     7714 &      63.8 $\pm$     0.4 &      42.7 $\pm$     0.7 &     134.5 $\pm$     4.9 &     382.7 $\pm$     1.7 &    1826.8 $\pm$     8.1   \\
  Arp 284 &       W TAIL &      NGC    7714 TAIL &       7.8 $\pm$     0.3 &       4.6 $\pm$     0.5 &      14.0 $\pm$     3.0 &      26.0 $\pm$     1.1 &    $-$   \\
  Arp 285 &        N &      NGC     2856 &      61.5 $\pm$     0.2 &      41.55 $\pm$     0.08 &     128.4 $\pm$     3.6 &     362.1 $\pm$     0.8 &     616.2 $\pm$     2.1   \\
  Arp 285 &       N TAIL &      NGC    2856 TAIL &       0.47 $\pm$     0.03 &       0.39 $\pm$     0.02 &       2.0 $\pm$     0.6 &       5.4 $\pm$     0.2 &       6.3 $\pm$     0.4   \\
  Arp 285 &        S &      NGC     2854 &      41.4 $\pm$     0.3 &      26.2 $\pm$     0.1 &      73.9 $\pm$     6.1 &     184.9 $\pm$     1.4 &     184.4 $\pm$     3.6   \\
  Arp 290 &        N &       IC      196 &      49.3 $\pm$     0.3 &      30.4 $\pm$     0.4 &      33.2 $\pm$     5.0 &      51.9 $\pm$     1.4 &  $\le$62.2       \\
  Arp 290 &       N TAIL &       IC     196 TAIL &       4.70 $\pm$     0.06 &       2.98 $\pm$     0.06 &       2.7 $\pm$     0.8 &       5.2 $\pm$     0.2 &  $\le$10.1       \\
  Arp 290 &        S &       IC      195 &      29.4 $\pm$     0.1 &      17.6 $\pm$     0.2 &      15.3 $\pm$     2.0 &      10.1 $\pm$     0.6 &  $\le$24.5       \\
  Arp 293 &         BRIDGE &      NGC  6285/6 BRIDGE &       2.09 $\pm$     0.05 &       1.35 $\pm$     0.08 &  $\le$ 2.3     &       5.2 $\pm$     0.3 &      15.9 $\pm$     2.3   \\
  Arp 293 &        N &      NGC     6285 &      17.40 $\pm$     0.03 &      11.66 $\pm$     0.06 &      36.0 $\pm$     0.5 &     108.4 $\pm$     0.2 &     115.3 $\pm$     1.4   \\
  Arp 293 &       N TAIL &      NGC    6285 TAIL &       1.46 $\pm$     0.02 &       0.91 $\pm$     0.03 &       1.4 $\pm$     0.3 &       4.4 $\pm$     0.1 &      10.5 $\pm$     0.9   \\
  Arp 293 &        S &      NGC     6286 &      71.21 $\pm$     0.06 &      52.3 $\pm$     0.1 &     175.7 $\pm$     1.0 &     486.2 $\pm$     0.4 &     399.7 $\pm$     2.8   \\
  Arp 293 &       S TAIL &      NGC    6286 TAIL &       2.19 $\pm$     0.03 &       1.44 $\pm$     0.06 &       5.8 $\pm$     0.6 &       7.8 $\pm$     0.2 &      16.8 $\pm$     1.6   \\
  Arp 294 &        N &      NGC     3788 &      76.0 $\pm$     0.3 &      46.8 $\pm$     0.3 &      83.3 $\pm$     5.2 &     188.3 $\pm$     1.1 &     115.3 $\pm$     9.2   \\
  Arp 294 &        S &      NGC     3786 &      76.7 $\pm$     0.6 &      56.3 $\pm$     0.7 &      88.9 $\pm$    12.5 &     176.5 $\pm$     2.8 &     240.8 $\pm$    22.8   \\
  Arp 295 &         BRIDGE &      ARP     295 BRIDGE &       2.4 $\pm$     0.6 &  $\le$ 2.22    0 &  $\le$18.4     &       2.8 $\pm$     0.7 &  $\le$ 0.0       \\
  Arp 295 &        N &      ARP    295B &      31.4 $\pm$     0.4 &      21.4 $\pm$     0.5 &      81.8 $\pm$     4.2 &     243.7 $\pm$     0.5 &     498.5 $\pm$     7.1   \\
  Arp 295 &        S &      ARP     295A &      39.9 $\pm$     0.5 &      25.3 $\pm$     0.5 &      29.7 $\pm$     4.5 &      51.9 $\pm$     0.5 &      23.9 $\pm$     7.4   \\
  Arp 295 &       S TAIL &      ARP    295A TAIL &       2.8 $\pm$     0.4 &       2.1 $\pm$     0.5 &  $\le$13.0     &       2.1 $\pm$     0.5 &    $-$   \\
  Arp 297 &       NE &      NGC     5755 &      13.7 $\pm$     0.1 &      10.1 $\pm$     0.1 &      25.2 $\pm$     5.3 &      93.5 $\pm$     1.0 &     191.3 $\pm$     2.5   \\
  Arp 297 &    NE N TAIL &      NGC    5755 TAIL &       0.98 $\pm$     0.08 &       0.65 $\pm$     0.08 &  $\le$11.8     &       2.7 $\pm$     0.7 &       0.0       \\
  Arp 297 &       NW &      NGC     5753 &       4.25 $\pm$     0.07 &       2.77 $\pm$     0.06 &  $\le$ 9.3     &       5.9 $\pm$     0.6 &  $\le$ 4.7       \\
  Arp 297 &       SE &      NGC     5754 &      37.4 $\pm$     0.3 &      22.6 $\pm$     0.3 &  $\le$40.2     &      64.1 $\pm$     2.3 &      32.6 $\pm$     6.1   \\
  Arp 297 &    SE TAIL S &      NGC    5754 TAIL &       1.14 $\pm$     0.05 &       0.73 $\pm$     0.05 &  $\le$ 6.7     &       2.8 $\pm$     0.4 &  $\le$ 0.0       \\
  Arp 297 &       SW &      NGC     5752 &       9.47 $\pm$     0.06 &       6.46 $\pm$     0.06 &      20.0 $\pm$     2.8 &      54.3 $\pm$     0.5 &      95.0 $\pm$     1.3   \\
  Arp 298 &        N &       IC     5283 &      31.4 $\pm$     0.3 &      20.8 $\pm$     0.8 &      60.4 $\pm$     9.4 &     165.5 $\pm$     4.5 &     196.7 $\pm$    14.4   \\
  Arp 298 &        S &      NGC     7469 &     187.3 $\pm$     0.5 &     181.4 $\pm$     1.3 &     449.5 $\pm$    15.0 &    1025.3 $\pm$     7.1 &    3263.6 $\pm$    22.5   \\
  Arp 298 &       W TAIL &       IC    5283 TAIL &       3.0 $\pm$     0.2 &       2.0 $\pm$     0.5 &  $\le$17.4     &  $\le$ 8.4     &    $-$   \\
 NGC 4567 &        N &      NGC    4567N &     143.4 $\pm$     3.0 &      90.9 $\pm$     2.1 &     203.6 $\pm$    49.3 &     523.4 $\pm$    13.0 &     734.0 $\pm$   115.6   \\
 NGC 4567 &        S &      NGC    4567S &     429.1 $\pm$     4.4 &     277.0 $\pm$     3.1 &     732.3 $\pm$    71.5 &    1831.7 $\pm$    18.7 &    1666.1 $\pm$   150.7   \\
\enddata
\end{deluxetable}
%

\clearpage

%
%
\begin{deluxetable}{cccccccccccccc}
\tabletypesize{\scriptsize}
\def\et#1#2#3{${#1}^{+#2}_{-#3}$}
\tablewidth{0pt}
\tablecaption{Spitzer Flux Densities for Normal Galaxy Sample\label{tab-6}}
\tablehead{
\colhead{Galaxy} &
\colhead{F$_{3.6 {\mu}m}$} & 
\colhead{F$_{4.5 {\mu}m}$} & 
\colhead{F$_{5.8 {\mu}m}$} & 
\colhead{F$_{8.0 {\mu}m}$} & 
\colhead{F$_{24 {\mu}m}$} 
\\ 
\multicolumn{1}{c}{}
& \multicolumn{1}{c}{(mJy)} 
& \multicolumn{1}{c}{(mJy)} 
& \multicolumn{1}{c}{(mJy)} 
& \multicolumn{1}{c}{(mJy)} 
& \multicolumn{1}{c}{(mJy)} 
\\
}
\startdata
\multicolumn{6}{c}{Spiral Galaxies}
\\
\hline
 NGC 24 &     112.9 $\pm$     0.6 &      74.2 $\pm$     0.8 &      88.9 $\pm$     5.4 &     168.6 $\pm$     3.8 &     125.2 $\pm$     4.6   \\
 NGC 337 &     104.8 $\pm$     0.5 &      70.9 $\pm$     0.5 &     201.7 $\pm$     5.1 &     499.9 $\pm$    14.3 &     747.6 $\pm$    12.8   \\
 NGC 628 &    1010.5 $\pm$     9.8 &     749.1 $\pm$    13.8 &    1544.2 $\pm$   109.9 &    3791.9 $\pm$   557.0 &    3204.0 $\pm$    40.1   \\
 NGC 925 &     360.2 $\pm$     6.3 &     255.0 $\pm$     4.0 &     462.1 $\pm$    23.2 &     709.3 $\pm$    79.8 &     827.4 $\pm$    26.4   \\
 NGC 1097 &    1356.9 $\pm$     8.7 &     917.4 $\pm$     8.3 &    1891.0 $\pm$    25.9 &    4196.2 $\pm$    69.0 &    6511.8 $\pm$    19.8   \\
 NGC 1291 &    2293.8 $\pm$    11.7 &    1427.4 $\pm$     8.9 &    1369.9 $\pm$    36.3 &     868.7 $\pm$    75.3 &     498.5 $\pm$    40.6   \\
 NGC 2403 &    1943.4 $\pm$    36.0 &    1414.3 $\pm$    14.5 &    2835.9 $\pm$    60.2 &    5138.8 $\pm$   622.9 &    5830.4 $\pm$    53.2   \\
 NGC 2841 &    1382.1 $\pm$     5.0 &     829.0 $\pm$     4.6 &     992.4 $\pm$    38.0 &    1580.7 $\pm$    10.6 &     967.6 $\pm$     7.0   \\
 NGC 3049 &      44.9 $\pm$     0.4 &      29.5 $\pm$     0.3 &      71.2 $\pm$     4.7 &     170.2 $\pm$     3.0 &     434.2 $\pm$     1.6   \\
 NGC 3184 &     570.9 $\pm$    19.7 &     372.0 $\pm$     8.2 &     833.1 $\pm$    45.9 &    1733.2 $\pm$    88.9 &    1437.8 $\pm$    44.1   \\
 NGC 3521 &    2150.6 $\pm$    12.4 &    1388.5 $\pm$    12.1 &    2942.4 $\pm$   221.3 &    7635.9 $\pm$   303.4 &    5268.6 $\pm$   185.9   \\
 NGC 3621 &    1128.6 $\pm$    21.6 &     749.1 $\pm$    30.9 &    2131.6 $\pm$    85.1 &    4601.1 $\pm$    47.9 &    3449.0 $\pm$    24.8   \\
 NGC 3938 &     344.0 $\pm$     7.3 &     234.7 $\pm$    12.0 &     555.5 $\pm$    12.7 &    1376.8 $\pm$    15.8 &    1080.7 $\pm$    22.9   \\
 NGC 4254 &     738.8 $\pm$    11.8 &     499.5 $\pm$     8.0 &    1873.7 $\pm$   111.9 &    5331.6 $\pm$    55.8 &    4185.0 $\pm$    35.5   \\
 NGC 4321 &    1068.0 $\pm$    18.9 &     702.3 $\pm$    11.5 &    1805.9 $\pm$    95.9 &    4353.7 $\pm$   290.9 &    3386.1 $\pm$    30.9   \\
 NGC 4450 &     570.9 $\pm$     3.9 &     368.6 $\pm$     3.1 &     360.3 $\pm$    22.8 &     495.3 $\pm$    65.9 &     217.6 $\pm$     9.8   \\
 NGC 4559 &     449.3 $\pm$    18.6 &     321.0 $\pm$    15.3 &     598.0 $\pm$    50.1 &    1255.6 $\pm$   138.7 &    1121.2 $\pm$    41.1   \\
 NGC 4579 &     938.8 $\pm$     4.5 &     578.8 $\pm$     5.3 &     745.9 $\pm$    29.7 &    1093.6 $\pm$    43.0 &     812.3 $\pm$    43.8   \\
 NGC 4594 &    4135.7 $\pm$    16.8 &    2503.5 $\pm$    15.0 &    2315.8 $\pm$    75.0 &    2083.8 $\pm$    52.6 &     797.4 $\pm$    25.1   \\
 NGC 4725 &    1160.3 $\pm$    26.0 &     722.0 $\pm$    13.0 &     905.1 $\pm$    42.7 &    1640.1 $\pm$   131.4 &     827.4 $\pm$    28.9   \\
 NGC 4736 &    3737.2 $\pm$    75.2 &    2390.8 $\pm$    83.7 &    3878.8 $\pm$   307.5 &    6900.1 $\pm$   263.5 &    5466.4 $\pm$    66.5   \\
 NGC 4826 &    2633.6 $\pm$    19.3 &    1623.9 $\pm$    10.8 &    2073.5 $\pm$    83.0 &    3154.0 $\pm$   155.0 &    2521.7 $\pm$    41.6   \\
 NGC 5055 &    2658.0 $\pm$    14.9 &    1732.0 $\pm$    12.2 &    3808.0 $\pm$    74.0 &    8069.7 $\pm$   189.6 &    5830.4 $\pm$    34.4   \\
 NGC 6946 &    3986.1 $\pm$    61.6 &    2719.9 $\pm$    79.9 &    8029.7 $\pm$   416.7 &   18486.6 $\pm$   981.9 &   20030.8 $\pm$   435.5   \\
 NGC 7331 &    1822.0 $\pm$    13.5 &    1198.3 $\pm$     9.8 &    2610.3 $\pm$    61.7 &    5481.0 $\pm$    68.8 &    3960.0 $\pm$    40.9   \\
 NGC 7793 &     825.2 $\pm$    13.5 &     527.9 $\pm$    16.4 &    1369.9 $\pm$    83.2 &    2721.8 $\pm$   121.8 &    2078.3 $\pm$    19.5   \\
\hline
\multicolumn{6}{c}{Elliptical/S0 Galaxies}
\\ 
\hline
 NGC 855 &      48.8 $\pm$     0.7 &      32.4 $\pm$     0.6 &      41.4 $\pm$     3.6 &      63.5 $\pm$     7.8 &      84.3 $\pm$     5.0   \\
 NGC 1377 &      60.9 $\pm$     0.3 &      91.7 $\pm$     0.9 &     350.5 $\pm$     3.9 &     568.7 $\pm$    12.9 &    1760.8 $\pm$     6.0   \\
 NGC 3773 &      24.9 $\pm$     0.3 &      15.9 $\pm$     0.4 &      28.9 $\pm$     3.7 &      64.1 $\pm$     3.4 &     138.6 $\pm$     3.7   \\
 NGC 4125 &     732.1 $\pm$     2.7 &     468.3 $\pm$     4.9 &     360.3 $\pm$     7.6 &     234.9 $\pm$     5.0 &     106.1 $\pm$     8.4   \\
\hline
\multicolumn{6}{c}{Irregular/Sm Galaxies}
\\ 
\hline
 DDO 53 &       5.7 $\pm$     0.4 &       4.4 $\pm$     0.3 &  $\le$19.4     &  $\le$20.1     &      29.2 $\pm$     2.8   \\
 DDO 154 &       4.3 $\pm$     0.4 &       3.0 $\pm$     0.2 &  $\le$12.5     &  $\le$7.0     &  $\le$3.7       \\
 DDO 165 &      15.7 $\pm$     0.9 &      11.3 $\pm$     1.0 &  $\le$53.5     &  $\le$14.3     &      11.6 $\pm$     2.3   \\
 Holmberg II &      94.7 $\pm$     4.6 &      70.2 $\pm$     3.9 &      66.8 $\pm$    10.6 &  $\le$167.1     &     169.7 $\pm$    24.5   \\
 IC 4710 &      76.0 $\pm$     3.5 &      52.3 $\pm$     4.1 &      57.1 $\pm$    10.7 &      77.1 $\pm$    12.4 &     114.2 $\pm$    10.1   \\
 Markarian 33 &      28.6 $\pm$     0.3 &      20.6 $\pm$     0.4 &      58.2 $\pm$     6.2 &     170.2 $\pm$     1.2 &     827.4 $\pm$     2.3   \\
 NGC 1705 &      28.1 $\pm$     0.2 &      19.9 $\pm$     0.4 &      17.9 $\pm$     5.8 &      24.4 $\pm$     1.4 &      52.7 $\pm$     2.2   \\
 NGC 2915 &      65.5 $\pm$     1.0 &      43.9 $\pm$     0.9 &      33.2 $\pm$     6.5 &      42.4 $\pm$     8.7 &      65.1 $\pm$    15.8   \\
 NGC 4236 &     280.9 $\pm$    23.4 &     171.6 $\pm$    28.2 &  $\le$328.6     &  $\le$293.0     &     485.0 $\pm$    28.9   \\
 NGC 5398 &      45.3 $\pm$     0.7 &      30.7 $\pm$     0.6 &      42.5 $\pm$     6.6 &      80.7 $\pm$     3.0 &     259.2 $\pm$     6.5   \\
 NGC 5408 &      39.9 $\pm$     1.5 &      28.7 $\pm$     1.3 &      26.3 $\pm$     8.0 &      36.5 $\pm$     3.1 &     388.8 $\pm$     5.0   \\
 NGC 6822 &    2293.8 $\pm$   361.8 &    1565.1 $\pm$   160.2 &    2424.9 $\pm$   571.1 &    2142.2 $\pm$   280.7 &    2949.2 $\pm$   222.5   \\
\enddata
\end{deluxetable}
%

%
%
\begin{deluxetable}{c|r|rrr|rrr|rrr|}
\tabletypesize{\scriptsize}
\def\et#1#2#3{${#1}^{+#2}_{-#3}$}
\tablewidth{0pt}
\tablecaption{Statistics on Spitzer Luminosities$^{\rm a}$\label{tab-7}}
\tablehead{
\multicolumn{1}{c}{Type} &
\multicolumn{1}{c}{Number} &
\multicolumn{3}{c}{LOG L$_{3.6 {\mu}m}$} &
\multicolumn{3}{c}{LOG L$_{8.0 {\mu}m}$} &
\multicolumn{3}{c}{LOG L$_{24 {\mu}m}$} 
\\
\multicolumn{1}{c}{}
& \multicolumn{1}{c}{}
& \multicolumn{1}{c}{median} 
& \multicolumn{1}{c}{mean} 
& \multicolumn{1}{c}{rms} 
& \multicolumn{1}{c}{median} 
& \multicolumn{1}{c}{mean} 
& \multicolumn{1}{c}{rms} 
& \multicolumn{1}{c}{median} 
& \multicolumn{1}{c}{mean} 
& \multicolumn{1}{c}{rms} 
\\
}
\startdata
All Arp Disks & 53 &  43.06 &  42.86 &   0.71 &  43.11 &  42.92 &   0.88 &  42.62 &  42.51 &   0.97 \\
M51-like Disks & 16 &  43.00 &  42.74 &   0.71 &  43.29 &  42.90 &   0.91 &  42.92 &  42.51 &   0.91 \\
Tails/Bridges & 30 &  41.92 &  41.76 &   0.54 &  41.71 &  41.73 &   0.69 &  41.29 &  41.39 &   0.80 \\
Spirals & 26 &  43.08 &  43.01 &   0.54 &  42.95 &  43.02 &   0.55 &  42.36 &  42.46 &   0.56 \\
Irr/Sm & 12 &  40.91 &  41.01 &   0.73 &  40.52 &  40.52 &   1.07 &  40.42 &  40.57 &   1.08 \\
\enddata
\tablenotetext{a}{These are `monochromatic luminosities', $\nu$L$_{\nu}$,
in units of erg~s$^{-1}$.
For the 3.6 $\mu$m, 8.0 $\mu$m, and 24 $\mu$m bands, 
the effective
frequencies used were 8.45 $\times$ 10$^{13}$ Hz, 3.81 $\times$ 10$^{13}$ Hz,
and 1.27 $\times$ 10$^{13}$ Hz, respectively.  Upper limits were included
in calculating the means and medians.}
\end{deluxetable}
%

\clearpage
\thispagestyle{empty}
%
%
\begin{deluxetable}{c|r|rrr|rrr|rrr|rrr|rrr|rrr|}
\rotate
\tabletypesize{\scriptsize}
\def\et#1#2#3{${#1}^{+#2}_{-#3}$}
\tablewidth{0pt}
\tablecaption{Statistics on Spitzer Colors$^{a,b}$\label{tab-8}}
\tablehead{
\multicolumn{1}{c}{Type} &
\multicolumn{1}{c}{N} &
\multicolumn{3}{c}{[3.6] $-$ [4.5] } &
\multicolumn{3}{c}{[4.5] $-$ [5.8] } &
\multicolumn{3}{c}{[5.8] $-$ [8.0] } &
\multicolumn{3}{c}{[3.6] $-$ [8.0] } &
\multicolumn{3}{c}{[8.0] $-$ [24] } &
\multicolumn{3}{c}{[3.6] $-$ [24] } \\
\multicolumn{1}{c}{}
& \multicolumn{1}{c}{}
& \multicolumn{1}{c}{median} 
& \multicolumn{1}{c}{mean} 
& \multicolumn{1}{c}{rms} 
& \multicolumn{1}{c}{median} 
& \multicolumn{1}{c}{mean} 
& \multicolumn{1}{c}{rms} 
& \multicolumn{1}{c}{median} 
& \multicolumn{1}{c}{mean} 
& \multicolumn{1}{c}{rms} 
& \multicolumn{1}{c}{median} 
& \multicolumn{1}{c}{mean} 
& \multicolumn{1}{c}{rms} 
& \multicolumn{1}{c}{median} 
& \multicolumn{1}{c}{mean} 
& \multicolumn{1}{c}{rms} 
& \multicolumn{1}{c}{median} 
& \multicolumn{1}{c}{mean} 
& \multicolumn{1}{c}{rms} 
\\
}
\startdata
 Arp Disks & 53 &   0.02 &   0.03 &   0.08 &   1.47 &   1.35 &   0.44 &   1.66 &   1.58 &   0.31 &   2.94 &   2.66 &   0.94 &   2.72 &   2.73 &   0.60 &   5.76 &   5.68 &   0.98 \\
                  Wide Disks & 27 &   0.01 &   0.01 &   0.06 &   1.55 &   1.38 &   0.51 &   1.73 &   1.61 &   0.35 &   2.94 &   2.56 &   1.12 &   2.74 &   2.58 &   0.48 &   5.81 &   5.56 &   1.12 \\
                Close Disks  & 42 &   0.02 &   0.03 &   0.08 &   1.43 &   1.34 &   0.40 &   1.64 &   1.56 &   0.30 &   2.97 &   2.73 &   0.80 &   2.65 &   2.81 &   0.65 &   5.67 &   5.75 &   0.89 \\
  M51-like & 16 &   0.02 &   0.02 &   0.04 &   1.43 &   1.45 &   0.38 &   1.67 &   1.66 &   0.17 &   3.07 &   2.87 &   0.94 &   2.74 &   2.70 &   0.36 &   5.78 &   5.73 &   0.78 \\
     Tidal & 30 &   0.01 &   0.01 &   0.10 &   1.54 &   1.48 &   0.50 &   1.56 &   1.46 &   0.36 &   2.59 &   2.48 &   0.82 &   3.48 &   3.33 &   0.59 &   6.19 &   6.26 &   0.58 \\
   Spirals & 26 &   0.03 &   0.03 &   0.06 &   1.27 &   1.15 &   0.42 &   1.43 &   1.32 &   0.36 &   2.80 &   2.50 &   0.78 &   2.10 &   2.17 &   0.43 &   4.95 &   4.67 &   1.08 \\
      E/S0 &  4 &   0.01 &   0.24 &   0.46 &   0.75 &   1.01 &   0.73 &   1.10 &   0.98 &   0.57 &   1.89 &   2.23 &   1.52 &   2.69 &   2.76 &   0.91 &   4.58 &   4.99 &   2.41 \\
    Irr/Sm & 12 &   0.11 &   0.10 &   0.07 &   0.43 &   0.67 &   0.46 &   0.97 &   1.06 &   0.37 &   1.53 &   1.86 &   0.81 &   3.22 &   3.47 &   0.82 &   4.62 &   5.09 &   1.20 \\
\enddata
\tablenotetext{a}{Upper/lower limits are not included in
these statistics.  No color corrections were included.
}
\tablenotetext{b}
{
For comparison with these numbers, the mean [3.6] $-$ [4.5], [4.5] $-$ [5.8],
[5.8] $-$ [8.0], and [3.6] $-$ [8.0] colors in the \citet{whitney04}
field stars are $-$0.05, 0.1, 0.05, and 0.1, respectively,
and the predicted values
for interstellar dust are $-$0.35, 3.2, 2.1, and 4.95,
respectively \citep{li01}.
The expected [8.0] $-$ [24] color for dust varies 
from 2.6 $-$ 4.3 \citep{li01},
increasing with increasing interstellar radiation field intensity,
while
stars are expected to be at [8.0] $-$ [24] $\sim$0.0.
}
\end{deluxetable}
%

\clearpage

%
%
\begin{deluxetable}{c|r|r|r|r|r|r}
\tabletypesize{\scriptsize}
\def\et#1#2#3{${#1}^{+#2}_{-#3}$}
\tablewidth{0pt}
\tablecaption{Kolmogorov-Smirnov Test Results: Probability Same Parent Sample$^a$\label{tab-9}}
\tablehead{
\multicolumn{1}{c}{Samples} &
\multicolumn{1}{c}{[3.6] $-$ [4.5] } &
\multicolumn{1}{c}{[4.5] $-$ [5.8] } &
\multicolumn{1}{c}{[5.8] $-$ [8.0] } &
\multicolumn{1}{c}{[3.6] $-$ [8.0] } &
\multicolumn{1}{c}{[8.0] $-$ [24] } &
\multicolumn{1}{c}{[3.6] $-$ [24] } \\
\\
}
\startdata
    Arp Disks vs. Spirals  &  90.6  &   3.8$^b$  &   0.004$^b$/ 0.3$^c$/ 0.2$^d$  &   7.6$^b$  &     0.001$^b$/   0.001$^c$/   0.094$^e$  &      0.01$^b$/    0.30$^c$   \\
Tails/Bridges vs. Spirals  &  18.7     &   3.3$^b$ &  19.9$^b$ &  81.5$^b$ &  $^f$  &  $^f$   \\
     M51-like vs. Spirals  &  91.7  &   7.2$^b$ &   0.01$^b$/ 0.09$^c$/ 0.04$^e$  &   7.9  &     0.01$^b$/   0.004$^c$/   0.02$^e$  &   0.4$^b$/  1.0$^c$   \\
     Close vs. Wide Pairs  &  56.2  &  61.5$^b$  &  10.2$^b$  &  14.5$^b$  &       21.7$^b$  &  38.9$^b$   \\
       Irr/Sm vs. Spirals  &   0.11  &  $^f$  &  $^f$ &  2.1$^b$/  4.1$^c$ & $^f$ &    46.1$^b$   \\
Arp Disks vs. Arp Tails/Bridges  &  20.2  &     57.7$^b$ &  37.9$^b$ &     23.4$^b$ &  $^f$  &  $^f$   \\
          Arp Disks vs. Irr/ISm  &   0.005  &  $^f$     &  $^f$     &  $^f$     &  $^f$  &  $^f$   \\
      Tails/Bridges vs. Irr/ISm  &   0.003  &  $^f$     &  $^f$     &  $^f$     &  $^f$  &  $^f$   \\
\enddata
\tablenotetext{a}{The quoted numbers
are percentages.  Probabilities $<$1$\%$ mean it is highly probable that
the two samples have different parent samples.  Probabilities $>$ 1$\%$ are
inconclusive;  we cannot rule out the hypothesis that the colors
came from the same parent population.  
}
\tablenotetext{b}
{Upper/lower limits are not included in
these statistics.
}
\tablenotetext{c}
{Upper/lower limits are included as detections.
}
\tablenotetext{d}
{Half of the lower limits are placed at the spiral peak,
and half 0.2 dex lower.
}
\tablenotetext{e}
{Upper limits are placed at the peak of the spiral distribution.
}
\tablenotetext{f}
{Too few detections.}
\end{deluxetable}
%



\end{document}